\documentstyle[12pt,epsf]{article}
\newcommand{\la}[1]{\label{#1}}

\newcommand{\pp}[1]{\langle\phi^\dagger\phi(#1)\rangle}

\newlength{\numlen}
\newcommand{\n}{\settowidth{\numlen}{0}\makebox[\numlen]{}}
\newcommand{\cen}[1]{\multicolumn{1}{|c|}{#1}}

\newlength{\indexlength}

\newcommand{\itsep}{
  \setlength{\parsep}{0mm}
  \setlength{\topsep}{0mm}
  \setlength{\itemsep}{0mm}}

\newcommand{\be}{\begin{equation}}
\newcommand{\ee}{\end{equation}}
\newcommand{\ba}{\begin{eqnarray}}
\newcommand{\ea}{\end{eqnarray}}

\newcommand{\rmi}[1]{{\mbox{\scriptsize #1}}}
\newcommand{\pdp}{\langle\phi^\dagger\phi\rangle}

\newcommand{\etal}{{et al.\ }}
\newcommand{\eq}{eq.~}
\newcommand{\eqs}{eqs.~}
\newcommand{\fig}{fig.~}
\newcommand{\figs}{figs.~}

\newcommand{\nr}[1]{(\ref{#1})}
\newcommand{\h}{{\hspace{0.5 cm}}}

\newcommand{\tr}{{\rm Tr\,}}

\newcommand{\bfx}{\mbox{\bf x}}

\newcommand{\fr}[2]{{\frac{#1}{#2}}}

\newcommand{\msbar}{\overline{\mbox{\rm MS}}}

\def\lsim{\raise0.3ex\hbox{$<$\kern-0.75em\raise-1.1ex\hbox{$\sim$}}}
\def\gsim{\raise0.3ex\hbox{$>$\kern-0.75em\raise-1.1ex\hbox{$\sim$}}}
\makeatletter \@addtoreset{equation}{section} \makeatother
\renewcommand{\theequation}{\arabic{section}.\arabic{equation}}

\begin{document}
\begin{titlepage}
\begin{flushright}
CERN-TH/95-263\\
HD-THEP-95-44\\
HU-TFT-95-57\\
IUHET-318\\
hep-lat/9510020\\
October 12, 1995
\end{flushright}
\begin{centering}
\vfill

{\bf THE ELECTROWEAK PHASE TRANSITION:\\
A NON-PERTURBATIVE ANALYSIS}
\vspace{0.8cm}

K. Kajantie$^{\rm a,b,}$\footnote{kajantie@phcu.helsinki.fi},
M. Laine$^{\rm a,c,}$\footnote{mlaine@rock.helsinki.fi},
K. Rummukainen$^{\rm d,}$\footnote{kari@trek.physics.indiana.edu} and
M. Shaposhnikov$^{\rm b,}$\footnote{mshaposh@nxth04.cern.ch} \\

\vspace{0.3cm}
{\em $^{\rm a}$Department of Physics,
P.O.Box 9, 00014 University of Helsinki, Finland\\}
\vspace{0.3cm}
{\em $^{\rm b}$Theory Division, CERN, CH-1211 Geneva 23, Switzerland\\}
\vspace{0.3cm}
{\em $^{\rm c}$Institut f\"ur Theoretische Physik,
Philosophenweg 16,\\
D-69120 Heidelberg, Germany\\}
\vspace{0.3cm}
{\em $^{\rm d}$Indiana University, Department of Physics,
Swain Hall West 117,\\Bloomington IN 47405 USA}

\vspace{0.7cm}
{\bf Abstract}

\end{centering}

\vspace{0.3cm}\noindent
We study on the lattice the 3d SU(2)+Higgs model, which is an
effective theory of a large class of 4d high temperature gauge
theories. Using the exact constant physics curve, continuum
($V\to\infty, a\to0$) results for the properties of the phase
transition (critical temperature, latent heat, interface tension) are
given. The 3-loop correction to the effective potential of the scalar
field is determined. The masses of scalar and vector excitations are
determined and found to be larger in the symmetric than in the broken
phase. The vector mass is considerably larger than the scalar one,
which suggests a further simplification to a scalar effective theory
at large~$m_H$. The use of consistent 1-loop relations between 3d
parameters and 4d physics permits one to convert the 3d simulation
results to quantitatively accurate numbers for different physical
theories, such as the Standard Model -- excluding possible
nonperturbative effects of the U(1) subgroup -- for Higgs masses up to
about 70~GeV. The applications of our results to cosmology are
discussed.

\vfill \vfill
\noindent

\end{titlepage}
\section{Introduction}

The 3d SU(2)$\times$U(1)+Higgs model is a universal theory for
the description of the electroweak phase transition in
the standard electroweak theory
and many extensions thereof, including the
MSSM~[1--5] (for a motivation of the study of the electroweak
phase transition, see~[1--4]).
In the present paper, we study with lattice simulations
the dominant SU(2)+Higgs part of the theory, defined by the Lagrangian
\be
L={1\over4} F_{ij}^aF_{ij}^a +
(D_{i}\phi)^\dagger(D_{i}\phi) + m_3^2\phi^\dagger\phi
+\lambda_3(\phi^\dagger\phi)^2.
\la{lagr}
\ee
The procedure of dimensional
reduction~\cite{fkrs1}, [6--13] allows one to compute
perturbatively the
relationship between the temperature $T$ and the physical parameters of the
underlying 4d electroweak theory or its extensions, and the
parameters of the 3d theory. Concrete
formulae for the SU(2)+Higgs model and MSM can be found
in~\cite{klrs} (see also below).

The aim of the present paper is to study
the 3d SU(2)+Higgs model, especially its phase diagram,
on the lattice. We confine
ourselves to a small ratio $\lambda_3/g_3^2 < 1/8$, which in 4d
terms corresponds to the case of small Higgs masses, $m_H <
m_W\approx 80$ GeV. This case seems to be the most interesting one for
cosmological applications, because the phase transition at least in this
region is of first order.

First Monte Carlo results on the electroweak phase transition using a
3d effective theory have already been given in~\cite{krs,fkrs2}, see
also~\cite{knp,ikps}. In comparison with~\cite{krs,fkrs2}
we considerably extend numerical calculations. This
makes it possible to determine for the first time or more accurately
than previously a number of finite $T$ quantities
such as latent heat, correlation lengths, interface
tension, magnitude of the higher order perturbative terms, etc.
An essential ingredient in the increased accuracy is that the
continuum-lattice mapping formulae,
which are exact in 3d, are now known~\cite{fkrs3,mlaine2}. Thus the continuum
limit $V\to\infty,a\to0$ can be carried out under controlled conditions.
Preliminary results of the simulations described here were published
in~\cite{fkrslatproc}.

Lattice Monte Carlo studies of the 4d SU(2)+Higgs model have been reported
in [18--23]. Whenever comparison can be made,
the results are in agreement within errorbars. However, the 3d
approach used in this paper gives much smaller errors than
the 4d one. Monte Carlo simulations of the $O(4)$ pure scalar
theory in 3d with a
non-analytic cubic term have been performed in \cite{knp}. The spectrum of
excitations in this theory, however, is very different from that of
the SU(2) model, so that comparison is not possible.

The paper is organized as follows. In Section~\ref{3dtheory}
we present the basic relations for the 3d
SU(2)+Higgs model in continuum and on lattice.
In Section~\ref{MC} we describe the Monte Carlo update algorithm.
In Section~\ref{sec:broken} we study the properties of the broken phase
and the convergence of perturbation theory there.
Section~\ref{pt} is a lattice investigation of the phase
transition for different values of the scalar self-coupling.
In Section~\ref{sec:correlation} we measure the different correlation
lengths, and in Section~\ref{sec:metastab} we study the metastability
region and the properties of the symmetric phase.
Section~\ref{sec:80} contains a short account
of simulations with a larger Higgs mass, and in
Section~\ref{sec:largemH} we discuss
the problem of studying large Higgs masses more generally.
In Section~\ref{sec:comparison} we summarize the
information acquired on lattice about the properties
of the 3d SU(2)+Higgs theory, and compare to perturbation
theory and to some non-perturbative approaches.
In Section~\ref{sec:phys} we relate the 3d lattice results
to 4d continuum physics.
Section~\ref{sec:appl} is a discussion of some implications
of our results to cosmology.
The conclusions and proposals for
future work are in Section~\ref{sec:conclusions}.

Readers not interested in details of lattice simulations could check
tables 5-7, and go directly to Section~\ref{sec:comparison},
which contains a summary of the non-perturbative results.

In this paper we use results from~\cite{fkrs1} and~\cite{fkrs3}.
All references to specific formulae from these papers are indicated by
I and II followed by the number of the corresponding expression.

\section{3d theory in continuum and on lattice\la{3dtheory}}

To make the paper self-contained, we summarize here the essential
properties of the 3d theory in continuum and on lattice. A more
detailed discussion can be found in~\cite{fkrs3,klrs,mlaine2}.

The two couplings $g_3^2$ and
$\lambda_3$ of the theory in eq.~\nr{lagr} have the
dimensionality of mass. They do not possess ultraviolet
renormalization and, say, in the $\msbar$ scheme are scale ($\mu$)
independent. The mass squared of the scalar field has a linear and
logarithmic divergence on the 1- and 2-loop levels,
respectively. In the $\msbar$ scheme the relation between the scalar
mass and the renormalization-group invariant parameter $\Lambda_m$
is
\be
m_3^2(\mu)=  {f_{2m}\over16\pi^2}
\log\frac{\Lambda_m}{\mu},
\la{m3}
\ee
where, in the SU(2)+Higgs theory,
\be
f_{2m} ={51\over16}g_3^4+9\lambda_3 g_3^2-12\lambda_3^2.
\la{f2m}
\ee

Since all the three parameters of the 3d theory are dimensionful,
the theory is uniquely fixed by giving three parameters, the
gauge coupling of dimension mass and two dimensionless
ratios:
\be
g_3^2,\qquad x\equiv {\lambda_3\over g_3^2},\qquad
y\equiv {m_3^2(g_3^2)\over g_3^4}=\frac{1}{g_3^4}
{f_{2m}\over16\pi^2}\log{\Lambda_m\over g_3^2}.
\la{3dvariables}
\ee
Renormalization introduces an intermediate mass scale on which the
physics does not depend.

The 3d theory of eq.~\nr{lagr} is described
on a lattice with the lattice constant $a$
by the action
\begin{eqnarray}
S&=& \beta_G \sum_x \sum_{i<j}(1-\fr12 \tr P_{ij}) +\nonumber \\
 &-& \beta_H \sum_x \sum_i
\fr12\tr\Phi^\dagger(\bfx)U_i(\bfx)\Phi(\bfx+i) +
\la{lagrangian} \\
 &+& \sum_x
\fr12\tr\Phi^\dagger(\bfx)\Phi(\bfx) + \beta_R\sum_x
 \bigl[ \fr12\tr\Phi^\dagger(\bfx)\Phi(\bfx)-1 \bigr]^2.
\nonumber
\end{eqnarray}
The three dimensionless parameters $\beta_G,
\beta_H,\beta_R$ of eq.~\nr{lagrangian}
are in the continuum limit $a \rightarrow 0$
related to the three dimensionless
parameters $g_3^2a,x,y$ by the following equations:
\begin{eqnarray}
g_3^2a & = & {4\over \beta_G}, \la{betag}\\
x & = & {1\over4}\lambda_3 a \beta_G =
{\beta_R\beta_G\over\beta_H^2}, \la{betar}\\
y & = &
{\beta_G^2\over8}\biggl({1\over\beta_H}-3-
{2x\beta_H\over\beta_G}\biggr)+{3\Sigma\beta_G\over32\pi}
(1+4x)+
\nonumber\\
&&+{1\over16\pi^2}\biggl[\biggl({51\over16}+9x-12x^2\biggr)
\biggl(\ln{3\beta_G\over2}+\zeta\biggr)+5.0+5.2x\biggr].
\la{y}
\end{eqnarray}
Eq.~(\ref{y}) depends on several constants arising from lattice
perturbation theory; $\Sigma=3.17591$, $\zeta=0.09$
and the two numbers 5.0 and 5.2, specific for SU(2)+Higgs theory,
and computed in \cite{mlaine2}. This reference also gives the
analogous numbers for some other relevant 3d theories.
Note that the logarithmic 2-loop term on the second line
in eq.~\nr{y} is absolutely necessary with the accuracy
which we have in our lattice simulations.
For instance, changing the number 5.0 in the 2-loop
part by 0.05 changes the critical temperature by an amount equal
to the statistical uncertainty in one of our
lattice simulations (138.38$\pm$0.05 GeV).

When the lattice constant $a$ is varied,
eqs.~(\ref{betag}--\ref{y}) define for the fixed parameters
$g_3^2$, $x$, $y$ of eq.~\nr{3dvariables} a curve,
the constant physics curve (CPC),
in the space of $\beta_G,\beta_H,\beta_R$.
All the curves end in the point ($\infty,1/3,0$) for $a\to0$.

The above discussion was entirely confined to the 3d theory. As
explained in \cite{klrs}, a single 3d theory is the effective
theory of a large class of finite $T$ field theories and we shall
later give several quantitative examples of this. For each 4d theory
one separately has to establish the transformation from the physical
parameters of the 4d theory to $g_3^2,x,y$. For the Standard Model
these are given in \cite{klrs}.

Since the use of the 3d parameters $g_3^2,x,y$ is rather unilluminating,
we shall in the presentation of lattice results
replace them by a ``Higgs mass'' $m_H^*$
and ``temperature'' $T^*$ using the following equations:
\ba
g_3^2&\equiv&0.44015T^*,
\la{g3eff}\\
x&=&{\lambda_3\over g_3^2}= -0.00550+0.12622 h^2,
\la{lameff}\\
y&=&{m_3^2(g_3^2)\over g_3^4}
\la{yeff}\\
&=& 0.39818+0.15545 h^2-
0.00190 h^4 - 2.58088{(m_H^*)^2\over (T^*)^2},\nonumber
\ea
where
\be
h\equiv{m_H^*\over 80.6{\rm GeV}}.
\la{def1h}
\ee
These equations follow from the tree-level formulae
\be
g=2/3,\,\,\lambda=\fr18 g^2\frac{(m_H^*)^2}{m_W^2},\quad
m_W=80.6\, {\rm GeV},\quad m_D=\sqrt{\fr56} g T=0.60858T,
\la{nprms}
\ee
from
a formula relating $m_3^2(\mu)$ to 4d quantities (I.66),
and from a subsequent
integration over the $A_0$-field~\cite{fkrs1}. Each quoted set of
($m_H^*,T^*$) can with eqs.~(\ref{g3eff}--\ref{yeff}) be converted to a set
of $g_3^2,x,y$, which for a given theory can, as discussed in
Sec.~\ref{sec:phys}, in turn be converted
to a precise set of values for the zero-temperature pole Higgs mass
and the physical temperature. For the SU(2)+Higgs
theory without fermions the difference between
the simplified parameters $m_H^*$, $T^*$ and
the physical parameters is relatively small.

In studying the phase structure of the 3d theory we search for a
critical curve in the $(x,y)$-plane. On the tree level this curve is
the line
$y=0$: for $y>0$ the theory is in a symmetric, for $y<0$ in a
broken phase. In 1- or 2-loop perturbation theory the line splits
in three: a critical curve $y=y_c(x)$ and upper and lower ends
$y=y_\pm(x)$ of metastability branches. On the critical curve the
system can exist in two different phases with the same vacuum energy
$\epsilon_\rmi{vac}$ but different values of various gauge invariant
condensates like $\pdp$. The broken phase exists for $y<y_+(x)$,
the symmetric phase for $y>y_-(x)$. One knows that perturbation theory
can never conclusively determine the curve $y=y_c(x)$
or the jumps of various gauge invariant
condensates like $\pdp$ across the curve, and
the main purpose of the lattice Monte Carlo study is to do this.
For instance, one is interested in knowing whether the curve
$y=y_c(x)$ continues to large ($>1/8$) values of~$x$ or whether it
terminates.
Our present simulations do not provide an answer to
the last question.

When performing simulations with the action (\ref{lagrangian}),
the procedure is somewhat
different depending on whether one is performing simulations at
some fixed $(x,y)$ or whether one is searching for the critical
curve $y=y_c(x)$.

Assume first that one is studying the system, e.g., determining correlation
lengths or the value of $\pdp$, at some fixed $(x,y)$.
The procedure then is as follows. Choose some $\beta_G$ which then
gives the lattice constant $a$. The value of $\beta_G$
on an $N^3$ lattice should satisfy the
constraints that the smallest correlation length is larger than $a$
and the largest correlation
length is smaller than $Na$.
The former requirement gives a lower limit for $\beta_G$ and the
latter a lower limit for $N$. Since $(x,y)$
are fixed, eq.~(\ref{y}) gives the value of $\beta_H$ and
eq.~(\ref{betar}) the value of $\beta_R$. Simulate the system with
these values for larger and larger $N$ and perform an extrapolation
to $N\to\infty$. Choose then larger and larger
values of $\beta_G$ (smaller and smaller $a$) doing always the same.
The set of different extrapolations to $N\to\infty$ can then
be extrapolated to $\beta_G\to\infty$ ($a\to0$), which is the final
continuum limit.

When searching for the critical curve $y_c(x)$, one can
in the beginning only
fix the value of $x$. Then one again first chooses
$\beta_G$ subject to the constraints given above, simulates the
system for various $\beta_H$ with $\beta_R=x\beta_H^2/\beta_G$ and finds
the value of $\beta_{H,c}$. Extrapolating at fixed $\beta_G$ to
$N=\infty$ gives a value $\beta_{H,c}(N=\infty)$ which using
eq.~(\ref{y}) can be converted to a value of $y(x)$ at this $a$. The
extrapolation to $a\to0$ is carried out as before and gives the
final $y=y_c(x)$.

\newcommand{\figysize}{14.5cm}
\newcommand{\figtopspace}{\vspace*{-1.3cm}}
\newcommand{\figbottomspace}{\vspace*{-5.5cm}}

\section{The Monte Carlo update algorithm}
\la{MC}

The lattice spacing $a$ and the
linear lattice size $Na$ are constrained by the length scales set by
the $W$ and Higgs masses: $a \ll 1/m_W(T) < 1/m_H(T)\ll Na$ (assuming
$m_H(T) < m_W(T)$).  Even though this requirement is much milder than
the 4d one ($a\ll 1/T$)~\cite{fkrs3}, in many cases it still
mandates quite large lattice sizes (our largest volume is $50^2\times
200$).  Therefore, it is important that the update
algorithm be as efficient as possible.

The gauge field update is not qualitatively different from the
standard SU(2) pure gauge update, in spite of the hopping term
$\tr\Phi^\dagger(\bfx) U_i(\bfx)\Phi(\bfx+i)$ in the action.  To
update the gauge links we use the conventional reflection
overrelaxation and Kennedy-Pendleton heat bath \cite{Kennedy85}
methods.  All the gauge field modes are much `faster' than the Higgs
modes, i.e.,\@ they have much shorter autocorrelation times.

Due to the flatness of the Higgs potential, the `slow' modes of the
system are associated with the radial sector of the Higgs field $\Phi
= R\,V$, $R\ge 0$, $V\in\mbox{SU(2)}$.  In what follows we summarize
the more non-standard methods we use to increase the efficiency of the
update program.

\paragraph{Global radial update.}
First improvement comes from multiplying the radial part of the Higgs
field at all locations simultaneously by the {\em same\,} factor:
$R(\bfx)\rightarrow e^{\xi} R(\bfx)$, where $\xi$ is randomly chosen
from a constant distribution around zero: $\xi\in [-\epsilon,\epsilon]$.
Under this update, the action \nr{lagrangian} changes as $\Delta
S(\xi) = a e^{2\xi} + b e^{4\xi} - a - b$, where $a$ and $b$ are the
sums of the terms proportional to $R^2$ and $R^4$ in the action.
Accounting for the measure factors, the update is accepted with the
Metropolis probability $p(\xi) = \min(1,\exp[4V\xi - \Delta S(\xi)])$,
where $V$ is the volume of the system.

The multiplication factor $e^{\xi}$ has to be very close to unity in order
for $p(\xi)$ to be non-negligible.  In practice, we choose $\xi$ from a
window of width $\sim 0.05$ -- 0.001 around zero, depending on the
volume.  The width of the window is chosen so that the acceptance is
approximately 60--70\%.

In \fig\ref{fig:autocorr} we show the autocorrelation functions for
an $m_H^*=60$, $\beta_G=8$ system on a $14^3$ lattice with both (a) local
Metropolis and (b) local + global update.  Even though the change in
$R(x)$ in each global multiplication is very small, the gain over only
local Metropolis/heat bath is about a factor of 5.  The additional
cost of the global update in terms of cpu time is negligible,
since it involves only one accept/reject step for the whole volume.

\paragraph{Higgs field overrelaxation.}
Let us parametrize the Higgs field as $\Phi = \alpha_\mu \tau_\mu$,
$\mu=0\ldots 3$, $\alpha_\mu\in{\bf R}$, where $\tau_i$ are Pauli
matrices and $\tau_0$ is the $2\times 2$ unit matrix.  From
\eq\nr{lagrangian} we see that the local potential of the Higgs field
at location $\bfx$ is
\be
  V[\Phi(\bfx)] = - \alpha_\mu(\bfx) F_\mu(\bfx) +
        R(\bfx)^2 + \beta_R (R(\bfx)^2 - 1)^2
  \la{local-act}
\ee
where $R^2 = \alpha_\mu \alpha_\mu$ and $F_\mu$ is the sum of the
hopping terms
\be
F_\mu(\bfx) = \beta_H \fr12\tr \tau_\mu \bigl[\sum_{i=1,2,3}
\Phi^\dagger(\bfx-i)U_i(\bfx-i) +
\Phi^\dagger(\bfx+i)U^\dagger_i(\bfx) \big].
\ee
This form seems to suggest separate update steps for the radial and
SU(2)-components of the Higgs field.  However, this is {\em not\,} the
optimal method: even though the overrelaxation for the SU(2) direction
can be readily performed, updating the radial component becomes quite
complicated.  In this case the overrelaxation step $R\rightarrow R'$
would require finding $R'$ so that
\be
  [dG(R)/dR]^{-1} \exp[-G(R)]\,,\h
        \mbox{where}\h G(R) = V(R) - \log R^3\,,
  \la{radialoverrelax}
\ee
remains invariant (here we use a notation where the $\bfx$-dependence
of the variables is suppressed) .  This can be approximated by finding
a solution $R'$ to the equation $G(R')=G(R)$, and performing a
Metropolis accept/reject step using the probability weight
$[dG(R)/dR]/[dG(R')/dR']$.  This update has been used in 4d Higgs
model simulations \cite{Fodor94,desylattice1}.  However, the
acceptance rate is only $\sim 80$\%, and the algorithm behaves
dynamically rather like a heat bath or Metropolis update.

A more efficient method is to update the Higgs variables in the plane
defined by 4-dimensional vectors $\alpha_\mu$ and $F_\mu$, using the
Cartesian components of $\alpha_\mu$ parallel and perpendicular to
$F_\mu$:
\be
  X = \alpha_\mu f_\mu\,, \h\h
  Y_\mu = \alpha_\mu - X f_\mu\,,
\ee
where $f_\mu = F_\mu/F$ and $F = \sqrt{F_\mu F_\mu}$.  In terms of $X$
and $Y_\mu$ \eq\nr{local-act} becomes
\be
  V[\Phi] = - X F + (1 + 2\beta_R(Y^2-1)) X^2 + \beta_R (X^4 + Y^4)\,.
  \la{xy-act}
\ee
Now we can update the $X$ and $Y$ components of $\Phi$ separately:
overrelaxation in $Y$ is simply the reflection $Y_\mu\rightarrow Y'_\mu =
-Y_\mu$, or $\alpha'_\mu = -\alpha_\mu + 2X f_\mu$ (this is exactly
equivalent to the conventional SU(2) reflection overrelaxation
procedure).  In order to perform an exact overrelaxation to the
$X$-component, we need to find $X'$ so that
\be
  [dV(X')/dX']^{-1}\exp[-V(X')] = [dV(X)/dX]^{-1}\exp[-V(X)]\,.
  \la{overrelaxstep}
\ee
To solve this equation we would have to resort to iterative numerical
methods, which can be costly in terms of cpu-time.  Instead, we used
the following approximation to the overrelaxation \nr{overrelaxstep}:
we find the solution to equation $V(X') = V(X)$ and accept $X'$ with the
probability
\be
  p(X') = \min(p_0,1)\,,\h\h p_0 = \fr{dV(X)/dX}{dV(X')/dX'}\,.
  \la{prob-accept}
\ee
Since $V(X)$ is a fourth order polynomial, solving the equation
$V(X')=V(X)$ boils down to finding zeros to a third order polynomial
(we already know one zero $X'=X$, which can be factored out).  In all
realistic cases the parameters of $V(X)$ are such that there always is
only one other real root, and it is straightforward to write a closed
expression for $X'$.  This update is an almost perfect overrelaxation: in
our simulations the acceptance rate varies between 99.7\% -- 99.98\%,
depending on the $\beta_G$ used.  The acceptance is high enough so
that the ``diffusive'' update dynamics inherent in the Metropolis acceptance
step does not play any role, and the evolution of the
field configurations is almost deterministic.  A different but related
overrelaxation schema to the one described here has also been used in
4d simulations \cite{desylattice2}.

\paragraph{Wavefront update.}
Let us next consider the order in which the lattice variables are
traversed.  In the conventional even-odd update, the lattice is
decomposed into even and odd sectors -- in our 3d case, to two sets
with $x+y+z$ even and odd.  The variables are first updated on all the
even points before updating the odd points.  Using this schema it
takes $L/2$ whole lattice updates before any kind of signal from a given
location on the lattice can propagate through the whole volume.

In the wavefront traversal we pick an arbitrary 2-dimensional plane
from the system, defined by one of the conditions $\pm x\pm y \pm
z=\mbox{const.}$ (modulo periodic boundary conditions).  We select one
of the directions perpendicular to the plane to be the positive
direction.  The update proceeds in two stages: first, we update the
Higgs variables $\Phi$ on this plane, and second, the gauge matrices
$U$ on the links emanating from this plane to the positive direction.
Both updates are performed with the overrelaxation algorithms.  After
the gauge field update we move up to the next plane to the positive
direction, and start the updates again.  With this method, a planar
wave of overrelaxation updates propagates through the volume in a
single update sweep.  In our implementation we keep the same
orientation of the plane until the volume has been swept through 4--9
times.

The particular diagonal orientation of the plane is chosen in order to
simplify the update: the spins and gauge links on the plane can be
updated independently of each other, and all the gauge links have an
equal footing with respect to the plane --- there are no link variables
within the plane, as would be the case if the plane was oriented along
the principal axes.

In \fig\ref{fig:autocorr} we compare the autocorrelation functions
of the standard even-odd overrelaxation (c) and the wavefront
overrelaxation (d).  In both cases we perform four overrelaxation sweeps
through the volume, followed by one heat bath/Metropolis update and
one global radial update.  For both cases the individual Higgs
variable overrelaxation step is the $XY$ overrelaxation described in
paragraph II above.  The overrelaxation methods perform much better
than the pure heat bath algorithms (a and b).  The wavefront
overrelaxation has much better initial decorrelation, as seen from the
very rapid decrease in the autocorrelation function, even though it
seems to have roughly the same exponential autocorrelation time as the
conventional even-odd method.  However, this rapid initial decrease
means that the {\em integrated autocorrelation time\,} $\tau_{\rm
int}$ is small, giving correspondingly small statistical errors for
the observables.  In our tests the wavefront overrelaxation had
typically 1.5--3 times smaller $\tau_{\rm int}$ than the even-odd
overrelaxation.

\begin{figure}[tb]
\figtopspace
\epsfysize=\figysize
\centerline{\epsffile{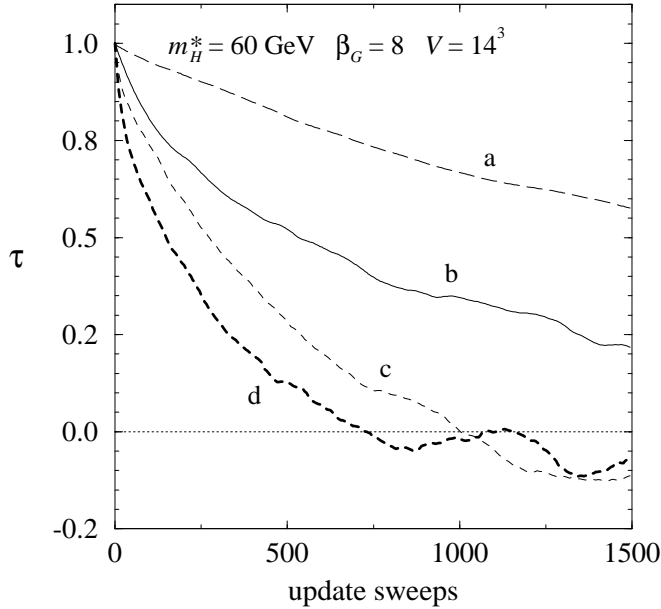}}
\figbottomspace
\caption[a]{The autocorrelation function of the observable $L=
V^\dagger(\bfx) U_i(x)V(\bfx+i)$ calculated from
an $m_H^*=60$\,GeV, $V=14^3$ lattice.  (a) Heat bath/Metropolis, (b) Heat
bath/Metropolis with global $R$-update, (c) $4\times$\,(overrelaxation
with even-odd traversal) + $1\times$\,(heat bath + global update), and
(d) $4\times$\,(wavefront overrelaxation) + $1\times$\,(heat bath +
global update).  In all the cases, one update means going once
through all the lattice points.}
\la{fig:autocorr}
\end{figure}

\paragraph{Multicanonical update.}
Multicanonical update \cite{multicanonical} is essential for the
interface tension calculations with the histogram method (see
Sec.~\ref{sec:tension}).  At the transition temperature the probability
distribution of some order parameter, say $R^2$, has
two distinct peaks corresponding to the two pure phases.  The
probability that the system resides in a mixed state, consisting of
domains of the two phases separated by phase interfaces, is suppressed
by the interface tension times the area of the interface (see,
for example, \fig\ref{m60-all-Rhg}).

To enhance the probability of the mixed states the action
\eq\nr{lagrangian} is modified with the multicanonical weight function $W$:
\be
  S_{\rm MC} = S + W(R_S^2)\,,\h\h
  R_S^2 = \sum_x R(\bfx)^2
\ee
The weight function $W(R_S^2)$ is chosen so that the resulting
distribution $p(R_S^2)$ is approximately constant in the interval
$R_{S,1}^2 \le R_S^2 \le R_{S,2}^2$, where $R_{S,1}^2$ and $R_{S,2}^2$
denote the pure phase peak locations.  This is the main disadvantage
of the multicanonical method: a priori, the weight function is not
known; an exact knowledge of the weight function is
equivalent to knowing the probability distribution of the order
parameter, which is one of the quantities we attempt to calculate with
the Monte Carlo simulation.

The canonical expectation value of an operator ${\cal O}$ can be
calculated by reweighting the individual multicanonical measurements
${\cal O}_k$ with the weight function:
\be
  \langle{\cal O}\rangle = \fr{\sum_k {\cal O}_k e^{-W(R_{S,k}^2)}}
        {\sum_k e^{-W(R_{S,k}^2)}}
\ee
where the sums go over all measurements of ${\cal O}$ and $R_S^2$.

The choice of $R_S^2$ for the argument of the weight function is by no
means unique; equally well one could use, for example, the hopping
term $\sum_{x,i}\fr12\tr\Phi^\dagger(\bfx) U_i(\bfx)\Phi(\bfx+i)$.
The advantage of $R_S^2$ is that in this case the weight function does
not modify the update of the gauge fields $U$ and the SU(2)-direction
of the Higgs field $V$.  However, the $XY$ overrelaxation described
above has to be modified.

We parametrize $W$ with a continuous piecewise linear function:
\be
  W(R_S^2) = w_i + (w_{i+1}-w_i)\fr{R_S^2 - r_i}{r_{i+1} - r_i}\,,
  \h r_i\le R_S^2 < r_{i+1}\,.
\ee
An initial guess for the parameters $w_i$ can be obtained by finite
size scaling the probability distributions obtained from simulations
using smaller lattice sizes.  If deemed necessary, the parameters are
further adjusted after preliminary runs.

Since we perform most of the simulations with vector supercomputers,
it is important to vectorize the multicanonical update.  This is
achieved with the following steps:

\begin{itemize}

\item[(i)] If $r_i\le R_S^2<r_{i+1}$ initially, the weight function
is fixed to the linear form $W'(R_S^2) = w_i + (w_{i+1}-w_i)(R_S^2 -
r_i)/(r_{i+1} - r_i)$ for {\em all\,} $R_S^2$.

\vspace*{-2mm}
\item[(ii)] A number of Higgs variables are updated with the action
$S' = S + W'$.  Since $W'$ is linear in $R^2$, it is straightforward
to do this with a fully vectorized algorithm.  We update $\Phi$ in
$\sim $100--200 points in one vector.

\vspace*{-2mm}
\item[(iii)] The whole vector of updates is accepted with
the probability $p_{\rm mc} = \min(1,$\ $\exp[W'(R_S^2) - W(R_S^2)])$.
\end{itemize}

Obviously, the acceptance in the step (iii) decreases when the number
of variables updated in a single vector increases.  Here the vector
length $\sim 100$ is short enough so that $R_S^2$ in practice often
remains between the limits $r_i$ and $r_{i+1}$ after the update, and
the acceptance is exceptionally good: in our runs the rejection rate
was only $\sim 10^{-6}$!  At the same time the vector length is long
enough so that increasing it does not give any significant gain in
computational speed.  Thus, the performance hit caused by the
implementation of the multicanonical algorithm is negligible.

The simulations were performed with Cray C-90 and X-MP supercomputers
and, for the smaller volumes, with IBM RS6000 and HP 9000/735
workstations. The total amount of computing power used was about
$5\cdot10^{15}$ flop = 160 Mflops year.

\section{Properties of the broken phase:
the 3-loop effective potential\la{sec:broken}}

We shall first study the 3d SU(2)+Higgs system when it is
in the broken phase. This will permit us to show the accuracy
of the method and, in particular, to
determine the size of the so far uncomputed
3-loop term in the effective potential (see (I.73),
(II.151-152)), and verify that it is linear in $\phi$.

Although the object of study is the 3d theory, we use
the simple parametrisation in eqs.~(\ref{g3eff}--\ref{yeff}) to
permit one to use the 4d quantities
$m_H^*,T^*$ in fixing the 3d parameters $x,y$. Whenever $m_H^*,T^*$
are quoted, the values of $g_3^2,x,y$ are to be computed from
eqs.~(\ref{g3eff}--\ref{yeff}).
The same values of $m_H^*,T^*$ can correspond to
different physical parameters in different theories.

We study the effective potential in the broken phase
with the help of the scalar condensate $\pdp$~\cite{fkrs3}.
Eq. (II.139) relates the value of the scalar condensate
in the $\msbar$ scheme, where
\be
\pp{\mu_1}-\pp{\mu_2}={3g_3^2\over16\pi^2}\log{\mu_1\over\mu_2},
\la{scaledep}
\ee
to the corresponding lattice
quantity $\langle R^2\rangle$. The relation, in which the
correction term vanishes in the continuum limit, is
\be
{\pp{\mu}\over g_3^2}=\fr18\beta_G\beta_H\biggl(\langle R^2\rangle-
{\Sigma\over\pi\beta_H}\biggr)-
{3\over(4\pi)^2}\biggl(\log{3\beta_G g_3^2\over2\mu}+\zeta
+\fr14\Sigma^2-\delta\biggr)+O(\frac{1}{\beta_G}).\la{rl2}
\ee
Here numerically $\zeta+\fr14\Sigma^2-\delta=0.67$.
Thus, the extrapolation of lattice measurements of the quantity
$\langle R^2\rangle$ to the limit $\beta_G\rightarrow \infty$
allows one to determine an "exact" value of the scalar
condensate which then may be
confronted with the perturbative expression.

We use the strategy appropriate for simulations at $T<T_c$
as explained in
Sec.~\ref{3dtheory}. We
take $m_H^*=80$ GeV and select several temperatures for which the
system is in the broken phase: $T^*= 110, 145, 155, 165, 166, 167$ GeV.
For each temperature we choose different values of $\beta_G$ (usually,
$6\leq\beta_G\leq 32$). These numbers completely define the
parameters of simulation through the constant physics curve
(\ref{betag}-\ref{y}). For every simulation,
the volume of the system was large
enough to make finite size effects smaller
than the statistical uncertainty in $\langle R^2\rangle$.

Examples of measurements at $T^*=110, 145$ GeV and 165 GeV are
shown in \figs\ref{beta1}--\ref{beta2}, and the values of $\pp{T^*}/T^*$
in the continuum limit are collected in table 1. The range of the
temperatures used is quite wide, the expectation value
of the Higgs field varies (in 4d units)
from $\phi_b/T^* \approx 2$ to $\phi_b/T^*\approx 0.6$.

\begin{table}
\center
\begin{tabular}{|c|c|c|c|c|c|}
\hline
 $T^*$ & 110 & 145 &155& 165&167\\
\hline
$\pp{T^*}/T^*$
&1.9553(13)&0.7061(7)&0.4718(6)&0.2506(10)&0.2052(36)\\
\hline
$\mu=1.2m_T$&1.9642&0.7184&0.4861&0.2707&0.2235\\
$\mu=4.7m_T$&1.9650&0.7207&0.4898&0.2793&0.2359\\
\hline
CW\,$\mu=1.2m_T$ &1.9649&0.7206&0.4902&0.2818&0.2397\\
CW\,$\mu=4.7m_T$ &1.9660&0.7223&0.4922&0.2847&0.2433\\
\hline
\end{tabular}
\caption[1]{Measured and computed values of
$\pp{T^*}/T^*$ for $m_H^*=80$ GeV
and $T^*=110,145,155,165,167$ GeV. 
The first row gives the lattice value in the limit
$V\to\infty, a\to0$,
obtained using eq.~(\ref{rl2}).
The following two rows with $\mu$-values refer to the result
in $\hbar$ expansion including the 2-loop term and the
known part of the 3-loop term (Appendix~A,
eq.\,(\ref{sp_for_pdp}) with $\beta(h)=0$), and the last two rows to a
numerical CW calculation using the
RG-improved 2-loop effective potential.}
\la{pdpmu}
\end{table}

\begin{figure}[p]
\figtopspace \hspace{1cm}
\epsfysize=\figysize
\centerline{\epsffile{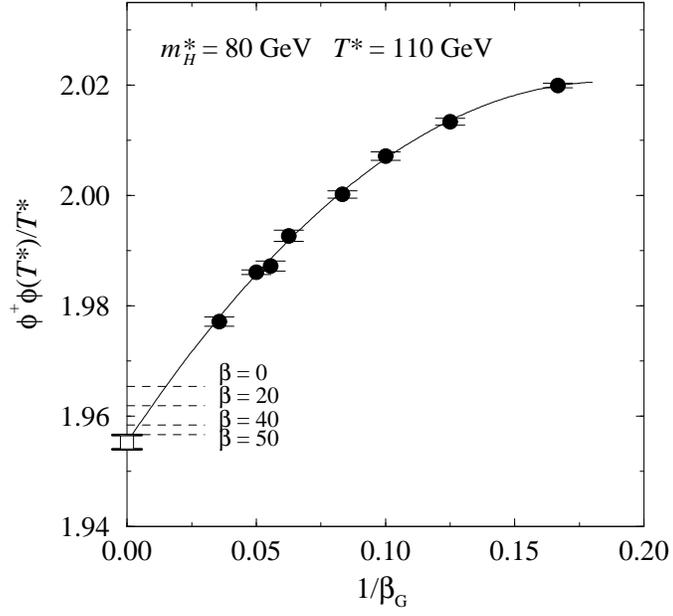}}
\figbottomspace
\caption[a]{Data for $\pp{T^*}/T^*$ as a function of $1/\beta_G$
for $m_H^*=80$, $T^*=110$ GeV computed from measured values of
$\langle R^2\rangle$ using eq.~(\ref{rl2}) with $\mu=T^*$,
$g_3^2=0.44015T^*$ (parametrisation in eq.~(\ref{g3eff})).
The perturbative values corresponding
to $\beta=0,20,40,50$, calculated with the CW-method
at the scale $\mu=2.37 m_T$ and then run to $\mu=T^*$
with eq.~\nr{scaledep}, are shown on the vertical axis.}
\la{beta1}
\end{figure}

\begin{figure}[p]
\vspace*{-1.2cm}
\centerline{\hspace{-2mm}
\epsfxsize=9.5cm\epsfbox{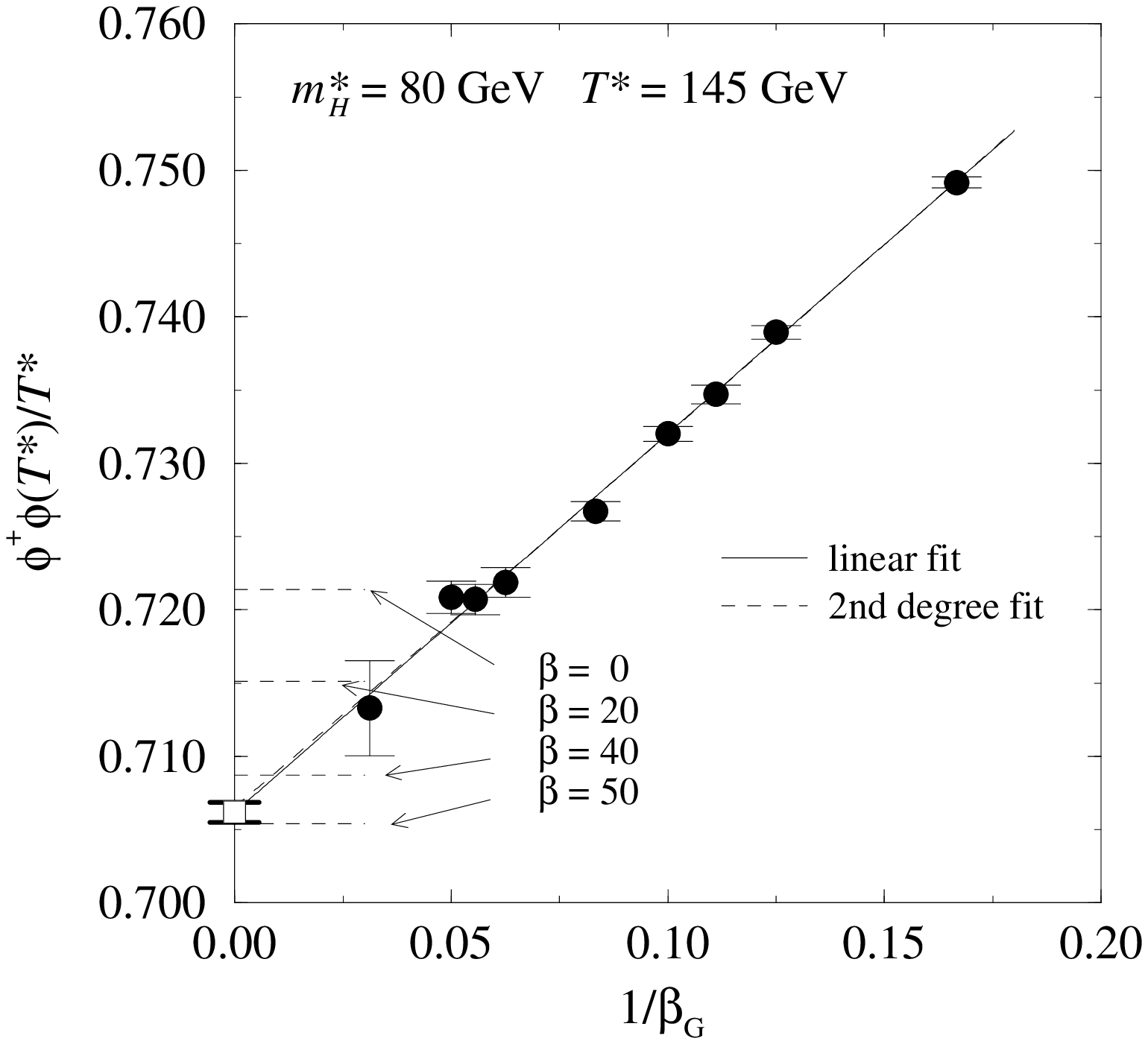}
\hspace{-1.5cm}
\epsfxsize=9.5cm\epsfbox{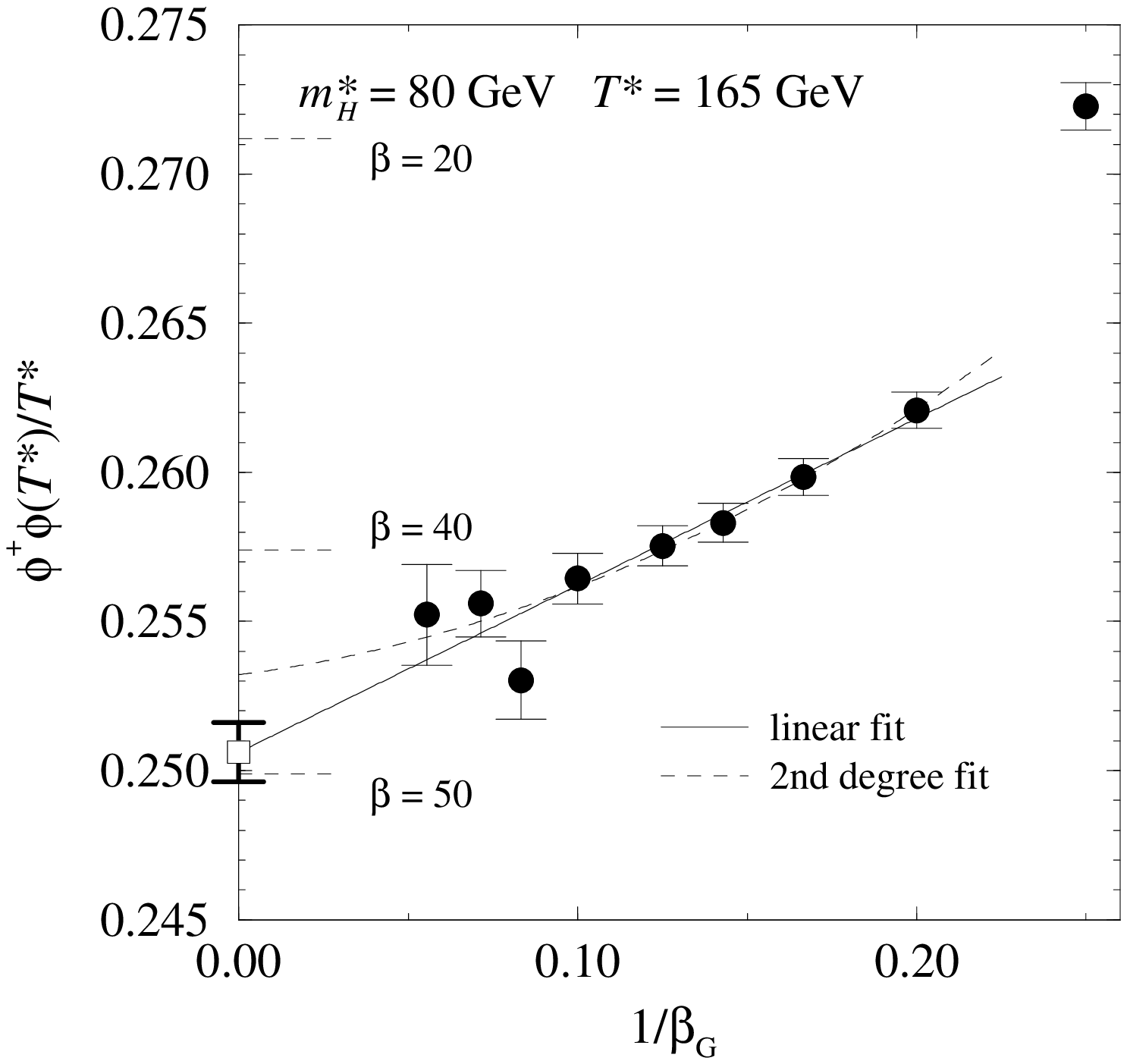}}
\vspace*{-4.5cm}
\caption[a]{
As \fig\ref{beta1},
but for $T^*=145$ GeV and $T^*=165$ GeV\@.}
\la{beta2}
\end{figure}

\begin{figure}[tb]
\vspace*{1.0cm}
\hspace{1cm}
\epsfysize=8.5cm
\centerline{\epsffile{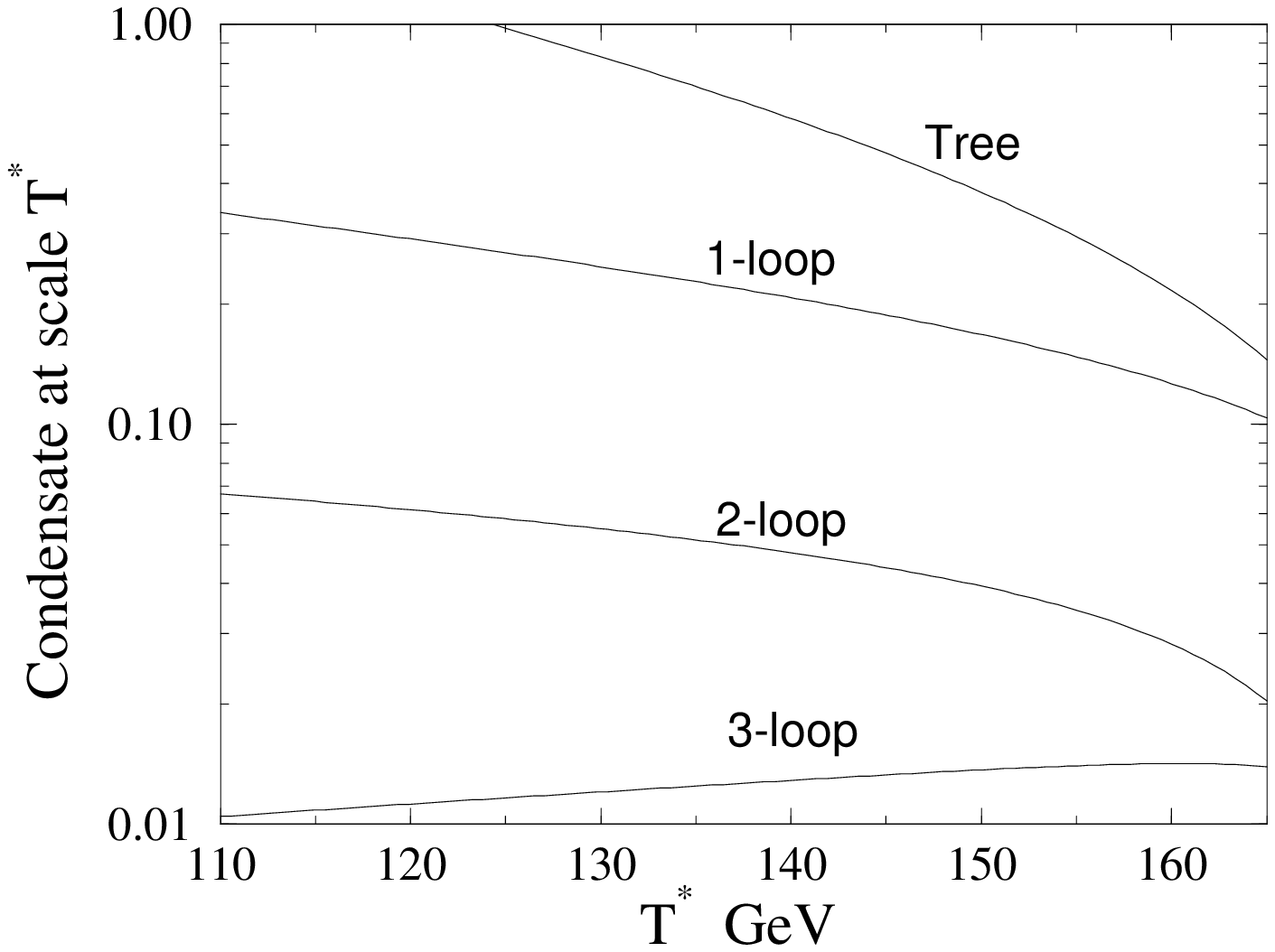}}
\vspace*{-2.5cm}
\caption[a]{The values
of $\pp{T^*}/T^*$ in the $\hbar$ expansion
as a function of temperature for $m_H^*=80$ GeV
from eq.~\nr{sp_for_pdp}.
The 3-loop curve contains the known part thereof,
with $\beta(h)=0$.
It is seen that the $\hbar$-calculation becomes increasingly
unreliable as one approaches the critical temperature.}
\la{cond5}
\end{figure}

The perturbative 2-loop computations were done with two methods,
described in Sec.~5 of II. The first method is based on a
straightforward $\hbar$ expansion of the condensate.
The result is given in Appendix A (eq.(\ref{sp_for_pdp}))
and shown in \fig\ref{cond5}.  Fig.~\ref{cond5}
contains the tree, 1-loop and 2-loop terms
(eqs. (II.76-78)) but also that part of the 3-loop term which
is known because it is related to lower order potentials. In
the second method (which we call the CW -- Coleman-Weinberg -- type of
computation), one numerically finds the location of the
broken minimum of the 2-loop effective
potential, determines the ground state energy,
and then computes the condensate with the help of (II.42)
as $\partial V(\phi_b)/\partial m_3^2$.
To avoid infrared divergences, the straightforward
CW-method has to be somewhat improved,
see~\ref{oldB}.

The expansion parameter of the $\hbar$ expansion is
$\sim g_3^2/\sqrt{-m_3^2}$ and thus
it converges the better the deeper one is in the broken phase.
This also means that the signal for the 3-loop term vanishes
if one goes too deep into the broken phase. Since one cannot
be too close to $m_3^2=0$ (or $T^*=173.5$ GeV in table 1) either,
there is an
optimal region in between. The numerical CW method works also
at $T_c$ and is thus more accurate. We include the $\hbar$ results
mainly since the formulas are very explicit.

When all orders of perturbation theory are summed,
$\pp{\mu}$ depends on $\mu$ according to eq.~\nr{scaledep}.
At a finite order in perturbation theory, however,
there is extra $\mu$-dependence which can be used
as an indication of the accuracy of the calculation.
In table~\ref{pdpmu} the parameter $\mu$ used
in the calculation of $\pp{\mu}$ was varied
within the limits $0.5\cdot 2.37 <\mu/m_T<2\cdot 2.37$, the central
value 2.37 (table 2 of I) being determined by the requirement of the
"best" convergence of
perturbation theory for the effective potential in the vicinity
of the minimum. Then $\pp{\mu}$ was run to $\mu=T^*$ according
to the exact running in eq.~\nr{scaledep}.

Inspection of table~\ref{pdpmu} shows that the CW
values of the condensate are practically independent of $\mu$
so that the convergence of CW
type perturbation theory is better than
that of the ordinary one. The
difference between these two methods is due to higher order
corrections; the CW method sums a subset of them and can be
used at the phase transition, where the first method
fails\footnote{The CW method provides an automatic summation of one-particle
reducible diagrams, whereas in ordinary perturbation theory
they have to explicitly calculated to the desired order, see Appendix~A.}.
One can see that
the CW values of the condensate are larger than the corresponding
lattice values, and that the difference between the two is statistically
significant. This proves the necessity of higher order effects
and singles out their sign: the ``exact'' vev of the Higgs field at
some fixed temperature is {\em smaller} than that given by 2-loop
perturbation theory.

The study of the difference between the 2-loop and lattice values of
the scalar condensate allows one to determine the magnitude and the
structure of the 3-loop term. In CW perturbation theory
for the scalar condensate we add to the
2-loop effective potential a linear term, expected on the 3-loop
level (see (I.73)):\footnote{
The structure of the 3-loop effective potential is discussed
in more detail in Appendix A.}
\be
\Delta V_3 = \frac{\beta}{(4 \pi)^3}g_3^4 m_T(\phi).
\la{3loop}
\ee
Here $\beta$ is a constant to be determined by fitting the prediction
to the lattice number at each temperature (for constant
$m_H^*$)\footnote{In~\cite{fkrs3} a rough estimate of the
parameter $\beta$ was given, $\beta=-15(20)$ with only statistical
errors quoted. After the analytical computation of the constant
physics curve~\cite{mlaine2} the systematic errors can be removed,
while the higher statistics allows to reduce the statistical errors
considerably.}.  In the $\hbar$ expansion method the procedure is
extremely simple: one takes $\pp{\mu}$ from eq.~(\ref{sp_for_pdp}) of
Appendix~A,
runs it to the scale $T^*$
using eq.(\ref{scaledep}) and fits the constant $\beta$ to
get agreement with data. In other words, $\beta$ is linearly
proportional to the difference of the perturbative
and the lattice value in table~\ref{pdpmu}. The
outcome of this procedure is in table~\ref{betamu}.

\begin{table}
\center
\begin{tabular}{|c|c|c|c|c|c|}
\hline
 $T^*$  & 110 & 145 &155& 165&167  \\
\hline
$\beta$&53.0(7.3)(2.7) & 41.8(2.1)(5.1) & 38.3(1.3)(6.7) &
35.1(1.1)(12.0)  & 28.8(2.8)(14.4)\\
$\beta_\rmi{CW}$&58.0(7.4)(2.3)&48.1(2.3)(2.3) & 46.6(1.8)(3.6) &
49.1(1.3)(2.6)\n & 46.2(3.6)(2.5)\n \\
\hline
\end{tabular}
\caption[1]{The values of the 3-loop coefficient $\beta$ at different
temperatures using the $\hbar$ expansion
and the numerical CW computation. 
The first number in brackets is statistical
uncertainty and the second is an estimate of the systematic error
associated with the change of the parameter $\mu$ within the interval
in table~\ref{pdpmu}.}
\la{betamu}
\end{table}

Within the $T^*$ range discussed the vev of the scalar field
varies from 0.6$T^*$ to 2$T^*$. Since
the value of $\beta$ does not depend on $T^*$
within error bars, one may conclude on the basis of
the lattice data that for $m_H^*$ around 80 GeV there
exists in the effective potential a 3-loop linear term with
positive sign and $\beta \simeq 50$.

We estimated $\beta$ at different Higgs masses also by another method.
We computed on the lattice the continuum limit of the critical
temperature and the value of the condensate at $T_c^*$ (the methods
are described in detail in the next section). We found for $m_H^*= 60$ GeV
that $T_c^*= 138.38(5)$ GeV and $\pp{T_c^*}/T_c^* = 0.227(6)$;
and for $m_H^*= 70$ GeV that
$T_c^*= 154.52(10)$ GeV and $\pp{T_c^*}/T_c^* = 0.162(12)$.
These numbers may be reproduced
with the 3-loop effective potential with $\beta \simeq 49(2)$ and
with $\beta \simeq 46(4)$, respectively.
The small Higgs mass $m_H^*=35$ GeV is not
informative since higher order corrections at the critical
temperature are numerically small and the extraction of $\beta$ with
any reasonable accuracy is not possible with the data we have.

Results at $m_H^*=60,70$ GeV indicate that within errorbars,
$\beta$ does not depend on $m_H^*$. This is what one expects:
in the abelian U(1)+Higgs model there is no linear
term~\cite{hebecker}, so that its coefficient
should be proportional to the non-abelian gauge coupling.

\section{The phase transition\la{pt}}

Since we are mostly interested in the properties of the phase
transition, most of our simulations are performed at and immediately
around the transition temperature.  We study three different Higgs mass
parameters $m_H^* = 35$, 60 and 70~GeV\@.  We have mostly concentrated
on 60~GeV Higgs, since it is close to the physically allowed mass
range, but the transition is not yet too weakly first order to be
studied with moderately sized lattices.

\begin{table}[ht]
\newcommand{\tube}[2]{~~$#1^2\times #2$}
\newcommand{\cube}[1]{~~$#1^3$}
\newcommand{\mc}{_m}
\center
\begin{tabular}{|c|c|llll|} \hline
 $m_H^*$ & $\beta_G$ & \multicolumn{4}{c|}{volumes}  \\ \hline\hline
 35    &  8    & \tube{6}{18}&  \tube{8}{24\mc}& \tube{10}{30\mc}&
        \tube{12}{36\mc}, \\
       &       & \tube{14}{42\mc}& \tube{8}{80\mc}& \tube{10}{80\mc}&  \\
        \cline{2-6}
       & 12    & \cube{12}& \cube{16\mc}& \tube{12}{24\mc}& \tube{12}{48\mc}\\
       &       & \tube{16}{32\mc}& \tube{18}{36\mc}& \tube{20}{40\mc}&
        \tube{22}{44\mc} \\
        \cline{2-6}
       & 20    & \cube{10}& \tube{10}{30}& \tube{12}{36\mc}& \\
       &       & \tube{16}{48\mc}& \tube{20}{60\mc}& \tube{24}{72\mc}& \\
        \hline\hline
 60    &  5    & \tube{12}{72\mc}& \tube{16}{80\mc}& &
        \\ \cline{2-6}
       &  8    &  \multicolumn{4}{l|}{
        \cube{12}\,\,\, \cube{16}\,\,\, \cube{24\mc}\,\, \cube{32\mc}} \\
       &       & \tube{20}{140\mc}& \tube{24}{120\mc}& \tube{30}{120\mc}&
        \\ \cline{2-6}
       & 12    &  \multicolumn{4}{l|}{
        \cube{16}\,\,\, \cube{24\mc}\,\, \cube{32\mc}\,\, \cube{40\mc}} \\
       &       & \tube{26}{156\mc}& \tube{30}{150\mc}& \tube{36}{144\mc}&
        \\  \cline{2-6}
       & 20    &  \multicolumn{4}{l|}{
        \cube{16}\,\,\, \cube{24}\,\,\, \cube{32}\,\,\, \cube{40}} \\
       &       & \tube{40}{200\mc}& \tube{50}{200\mc}& &
        \\  \hline\hline
 70    &  8    & \multicolumn{4}{l|}{
        \cube{12}\, \cube{16}\, \cube{24}\, \cube{32}}
        \\  \cline{2-6}
       & 12    & \multicolumn{4}{l|}{
        \cube{12}\, \cube{16}\, \cube{24}\, \cube{32}\, \cube{40}\, \cube{48}}
        \\  \cline{2-6}
       & 20    & \multicolumn{4}{l|}{
        \cube{12}\, \cube{16}\, \cube{24}\, \cube{32}\, \cube{40}\, \cube{48}}
        \\ \hline
\end{tabular}
\caption[0]{Lattice sizes used for the simulations at the transition
temperature for each ($m_H^*$,$\beta_G$)-pair.  In most of the cases,
several $\beta_H$-values were used around the transition point.
Multicanonical simulations are marked by subscript
($\mc$).\la{table:lattices}}
\end{table}

For each of the three values of $m_H^*$ we use
the gauge couplings $\beta_G=8$,
12 and 20, and for $m_H^*=60$\,GeV also $\beta_G = 5$ (remember that
$\beta_G$ is directly related to the lattice spacing through $\beta_G
= 4/(g_3^2 a)$).  In table \ref{table:lattices} we list the lattice
sizes for each $(m_H^*,\beta_G)$ -pair.  Each lattice has
several runs with different values for $\beta_H$ in order to
accurately locate the transition; typically smaller lattices have 5--20
and larger ones 1--3 values of $\beta_H$.
Separate runs are then joined
together with the Ferrenberg-Swendsen multihistogram method
\cite{Ferrenberg}.  When the interface tension causes noticeable
supercritical slowing down, the multicanonical algorithm is used.
All in all, the total number of separate `runs' --- different
combinations of lattice sizes and coupling constants --- described in
this paper is 289; this includes also runs away from the transition
region.

The large number of lattice volumes with several lattice spacings
makes it possible to accurately extract continuum values of the
physical observables:
\begin{itemize}\itsep
\item[(1)] For fixed lattice spacing $a$ (fixed $\beta_G$), we
extrapolate the lattice measurements to the thermodynamical limit
$V\rightarrow\infty$.

\item[(2)] Each of the $V=\infty$ values are in turn extrapolated to
the continuum limit $a\rightarrow 0$ ($\beta_G\rightarrow\infty$).
\end{itemize}

Note that in the 3d continuum theory the lowest-dimensional
gauge-invariant operator $\phi^{\dagger}\phi$ has the dimensionality
GeV. This means that the scaling violations in physical quantities
start from the first power of the lattice spacing $a$ (in the 4d
theory, scaling violations are proportional to $a^2$).

The transition becomes weaker --- the latent heat and the interface
tension become smaller --- when $m_H^*$ increases.  Measured in
dimensionless lattice units, the transition also becomes weaker when
the lattice spacing $a$ decreases for fixed $m_H^*$.  This can be
observed from the probability distributions of the average Higgs field
squared: $R^2 = \fr{1}{V}\sum_x R(\bfx)^2$.  In \fig\ref{m60-all-Rhg}
we show the distributions for some of the largest volumes for
$m_H^*=60$~GeV, $\beta_G=5$, 8, 12 and 20.  When $\beta_G$ increases,
the separation between the peaks becomes smaller and the minimum
between the peaks becomes shallower.  As we will explain in
Secs.~\ref{sec:latent} and \ref{sec:tension}, these features are
directly related to the latent heat and the interface tension,
respectively.

\begin{figure}[tb]
\vspace*{-1cm}
\centerline{\hspace{-3.3mm}
\epsfxsize=10cm\epsfbox{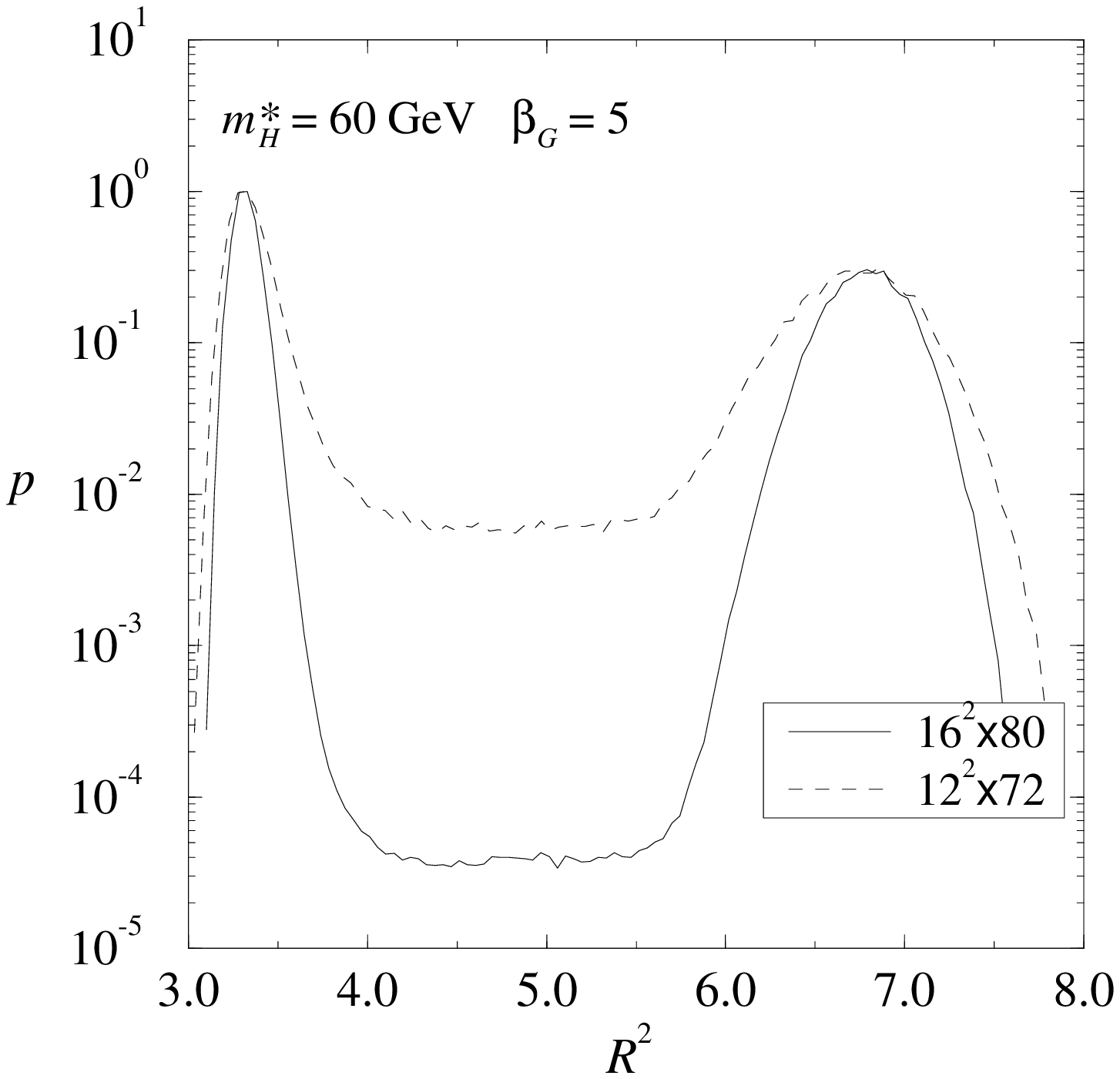}
\hspace{-2cm}
\epsfxsize=10cm\epsfbox{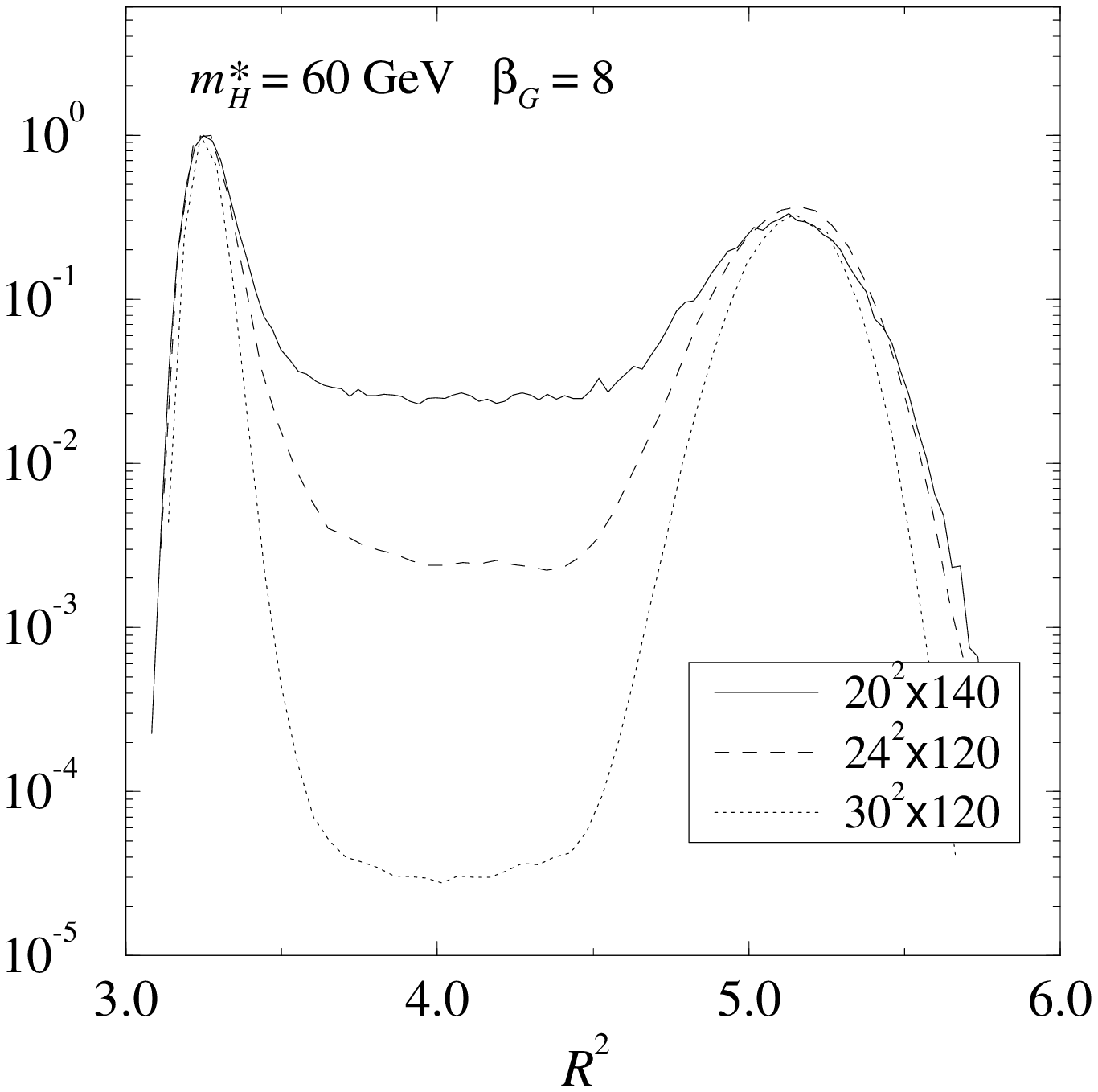}}
\vspace{-5.4cm}
\centerline{\hspace{-3.3mm}
\epsfxsize=10cm\epsfbox{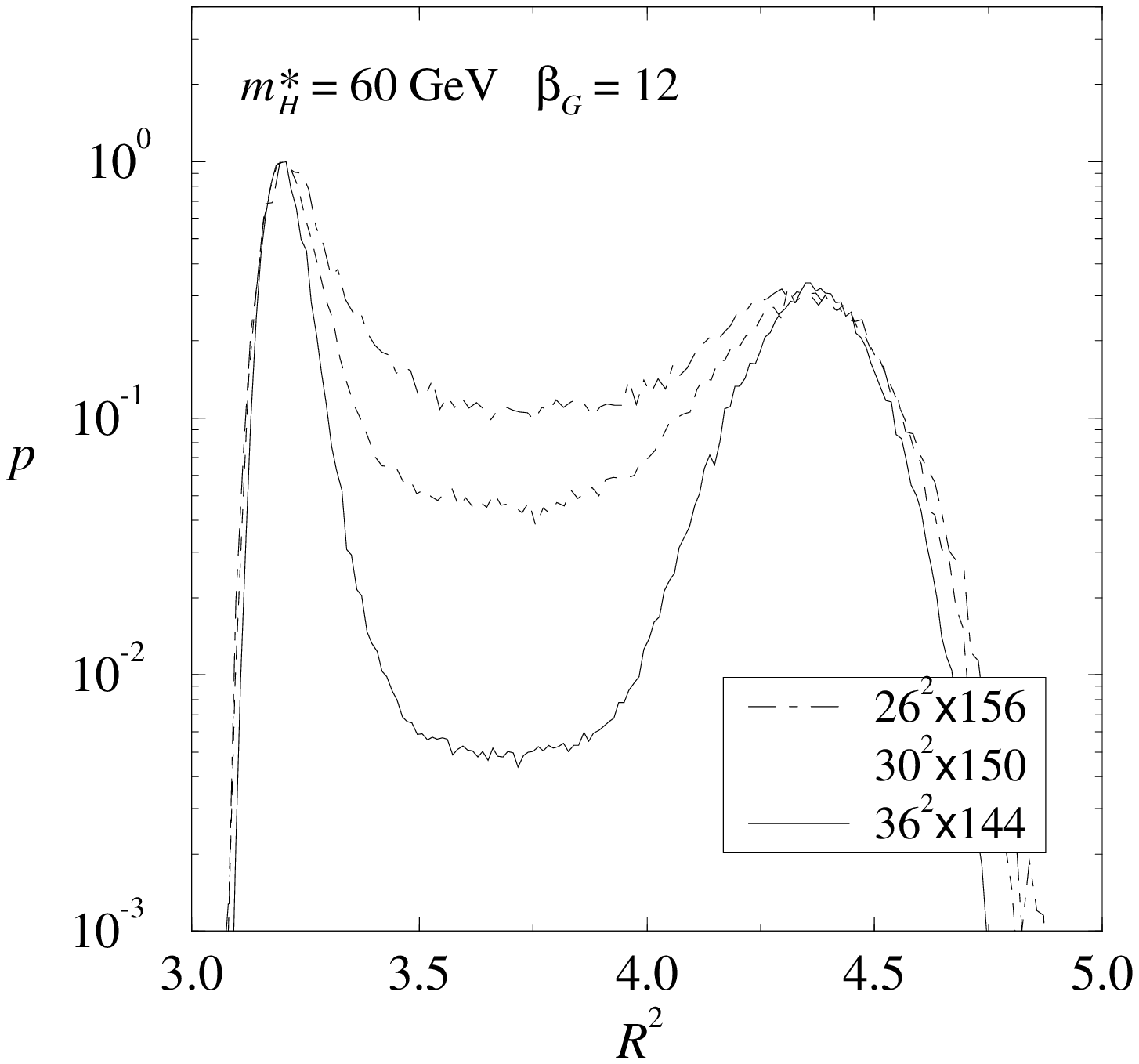}
\hspace{-2cm}
\epsfxsize=10cm\epsfbox{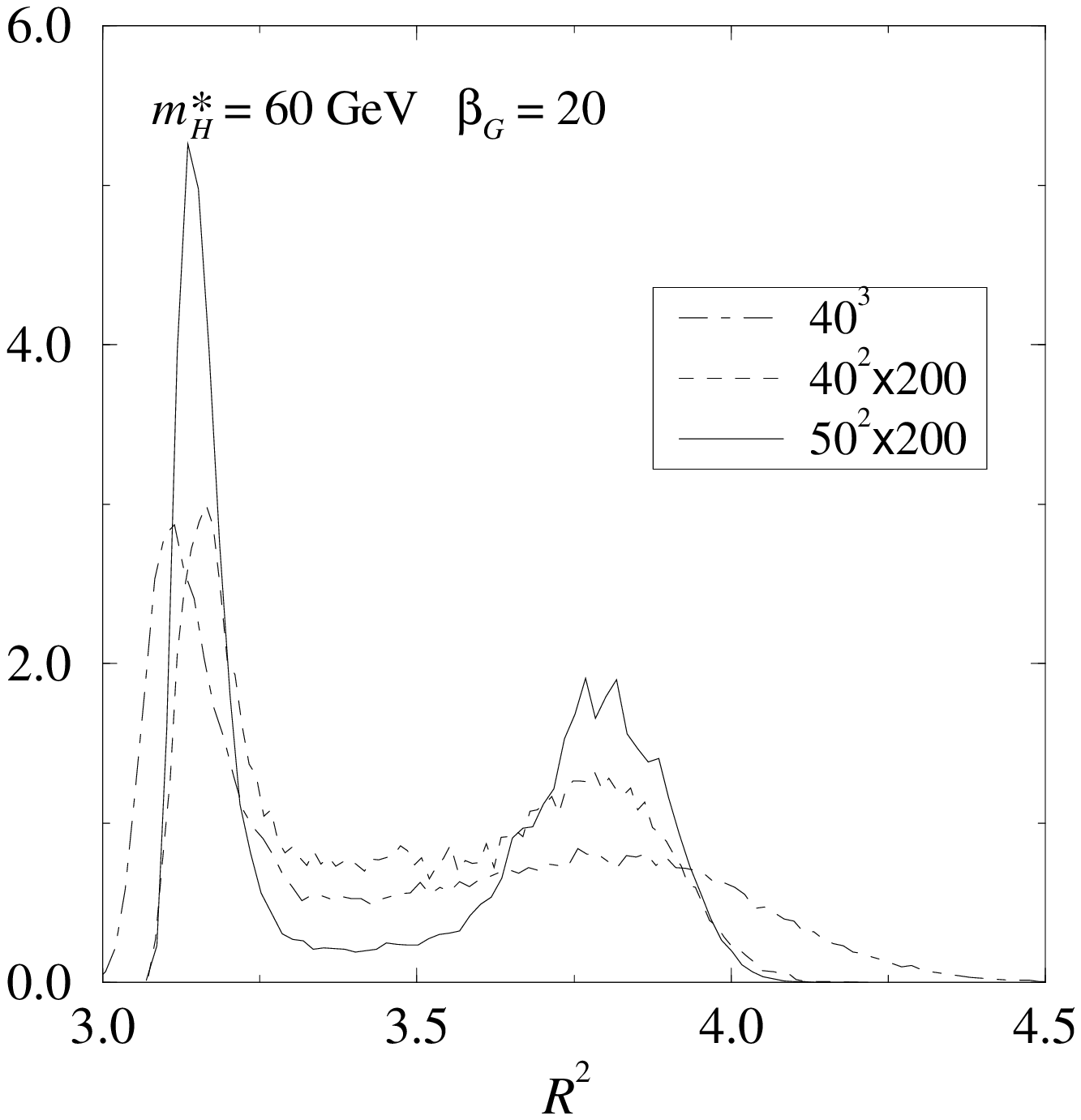}}
\vspace*{-5cm}
\caption[a]{The probability distribution of the average Higgs
length squared $R^2$ for $m_H^*=60$ and $\beta_G=5$, 8, 12 and 20.}
\la{m60-all-Rhg}
\end{figure}

\begin{figure}[tb]
\vspace*{-1cm}
\centerline{\hspace{-3.3mm}
\epsfxsize=10cm\epsfbox{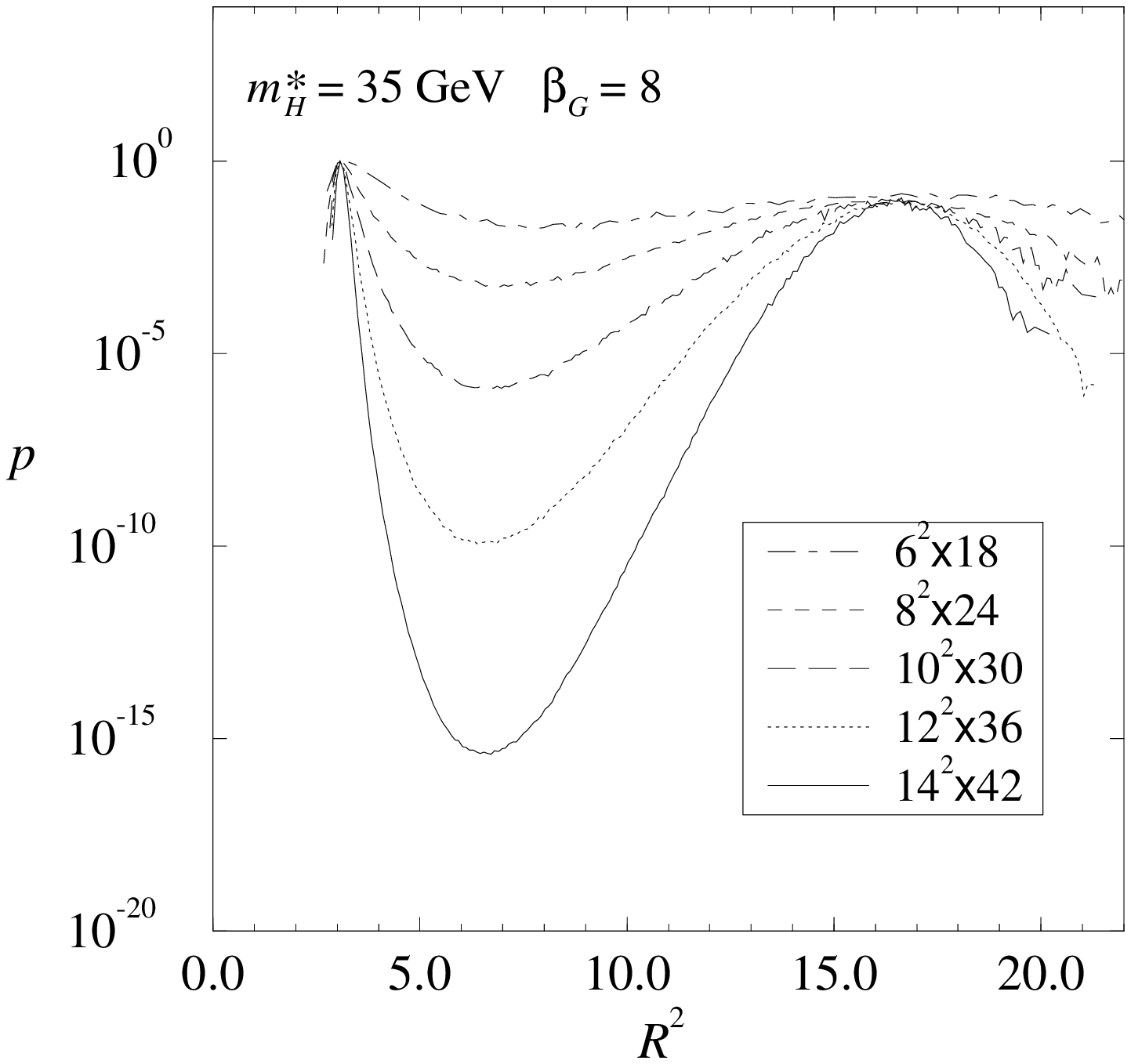}
\hspace{-2cm}
\epsfxsize=10cm\epsfbox{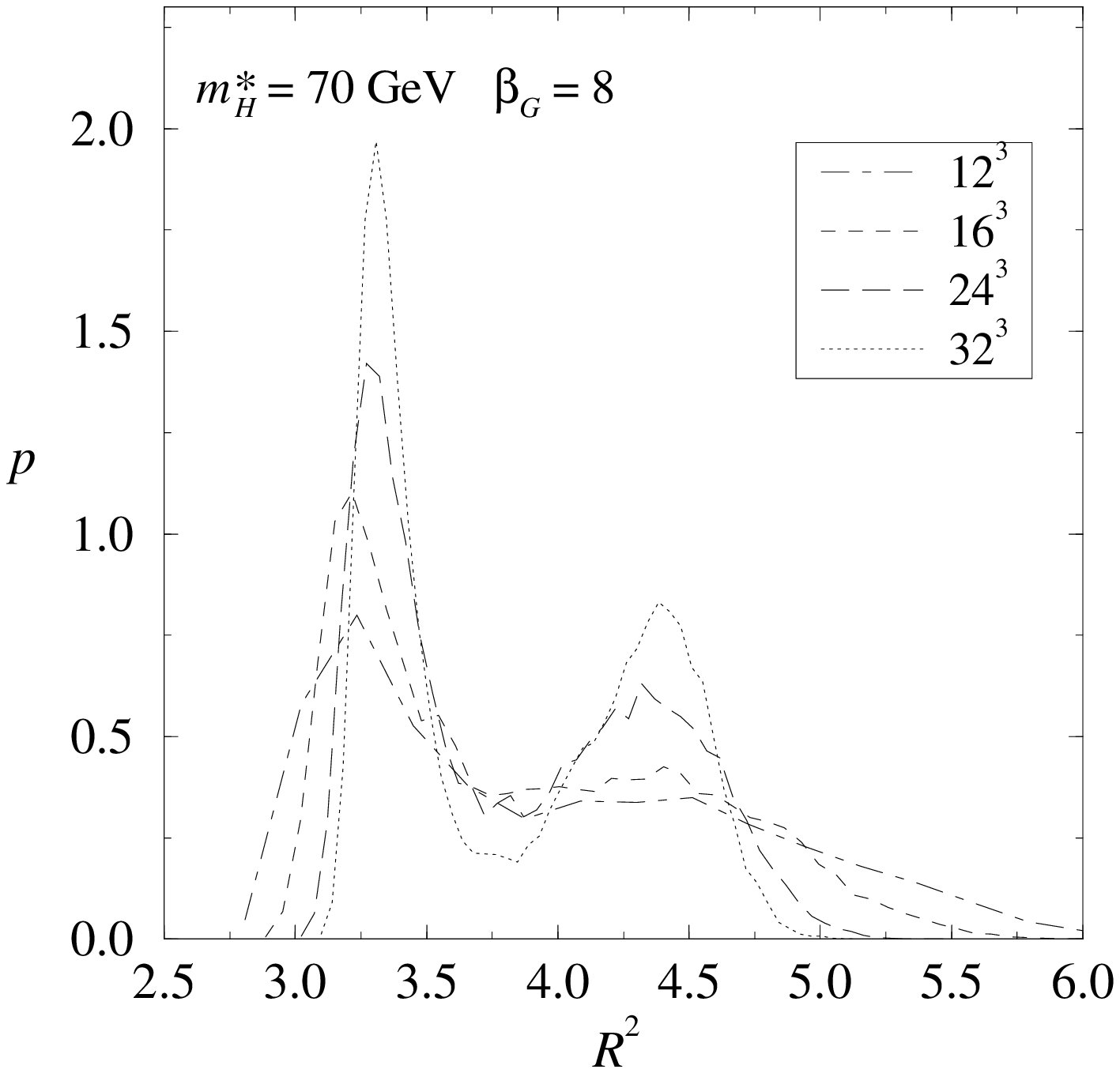}}
\vspace*{-5cm}
\caption[a]{$R^2$ distributions for $\beta_G=8$, $m_H^*=35$ and
70~GeV\@.}
\la{m35-m70-b8-Rhg}
\end{figure}

In \fig\ref{m35-m70-b8-Rhg} the corresponding $\beta_G=8$ histograms
are shown for $m_H^*=35$ and 70~GeV\@.  The dramatic effect of the Higgs
mass to the strength of the phase transition is clearly evident.  Note
that all the distributions in figs.~\ref{m60-all-Rhg} and
\ref{m35-m70-b8-Rhg} correspond to different values of $\beta_H$; each
of the histograms has been reweighted to the ``equal weight''
$\beta_H$-value (see Sec.~\ref{sec:critical}).

\subsection{The critical temperature}\la{sec:critical}

The critical temperature can be determined extremely accurately from
the Monte Carlo data.  In the SU(2)+Higgs model there are no known
local order parameters, which would acquire a non-zero value only in
one of the two phases of the model.  Instead, we use order parameter
like quantities which display a discontinuity at the transition point
(when $V\rightarrow\infty)$.  The quantities we use are $R^2$
and the hopping term
\be
  L=\fr1{3V}\sum_{x,i} \fr12 \tr V^\dagger(\bfx)U_i(\bfx)V(\bfx+i)
  \la{defineL}
\ee
where $V(\bfx)$ is the SU(2) direction of the Higgs variable
$\Phi(\bfx)=R(\bfx)V(\bfx)$.
The behaviour of $\langle L\rangle$ as a function of
$\beta_H$ is shown in \fig\ref{m60-b8-L} for $m_H^*=60$~GeV,
$\beta_G=8$ case for lattice sizes up to $32^3$.  The development of
the discontinuity is clearly visible.  The continuous lines are a
result of the multihistogram method calculation.

\begin{figure}[tb]
\figtopspace \hspace{1cm}
\epsfysize=\figysize
\centerline{\epsffile{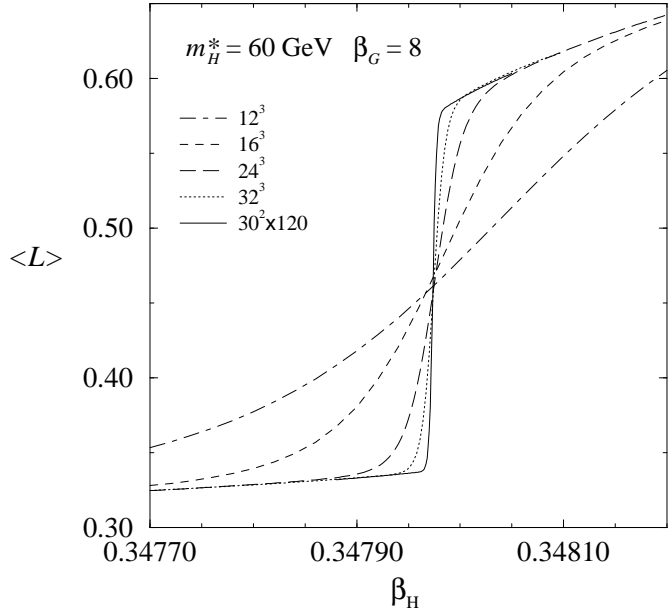}}
\figbottomspace
\caption[a]{The hopping term $\langle L\rangle$
as a function of $\beta_H$ for
different lattice sizes, for $m_H^*=60$~GeV, $\beta_G=8$.}
\la{m60-b8-L}
\end{figure}

For each individual lattice volume and ($m_H^*,\beta_G$)-pair, we
locate the {\em pseudocritical coupling\,} $\beta_{H,c}$ with several
different methods (see, for example, \cite{Herrmann92} and references
therein):

\vspace{3mm}
(1) maximum of the $L$-susceptibility $C(L) = \langle (L -
\langle L \rangle)^2\rangle$

(2) maximum of $C(R^2) = \langle (R^2 - \langle R^2
\rangle)^2\rangle$

(3) minimum of the 4th order Binder cumulant of $L$:
$B(L)= 1-\langle L^4\rangle/(3\langle L^2\rangle^2)$

(4) ``equal weight'' $\beta_H$ -value of the distribution $p(R^2)$

(5) ``equal height'' $\beta_H$ -value of the distribution $p(L)$
\vspace{3mm}

The locations of the extrema of the observables $C(L)$, $C(R^2)$ and
$B(L)$ are computed by reweighting the original measurements; the
error analysis is performed with the jackknife method, using
independent reweighting for each of the jackknife blocks.  As an
example, we show $C(R^2)$ and $B(L)$ for $m_H^*=60$~GeV, $\beta_G=8$
lattices in \fig\ref{m60-b8-R2-LB} as functions of $\beta_H$.  For
clarity, the errors are omitted from the figures.  Again, we would
like to point out the unambiguous first order scaling displayed by the
data in these figures.


\begin{figure}[tb]
\vspace*{-1cm}
\centerline{\hspace*{-3mm}
\epsfxsize=10cm\epsfbox{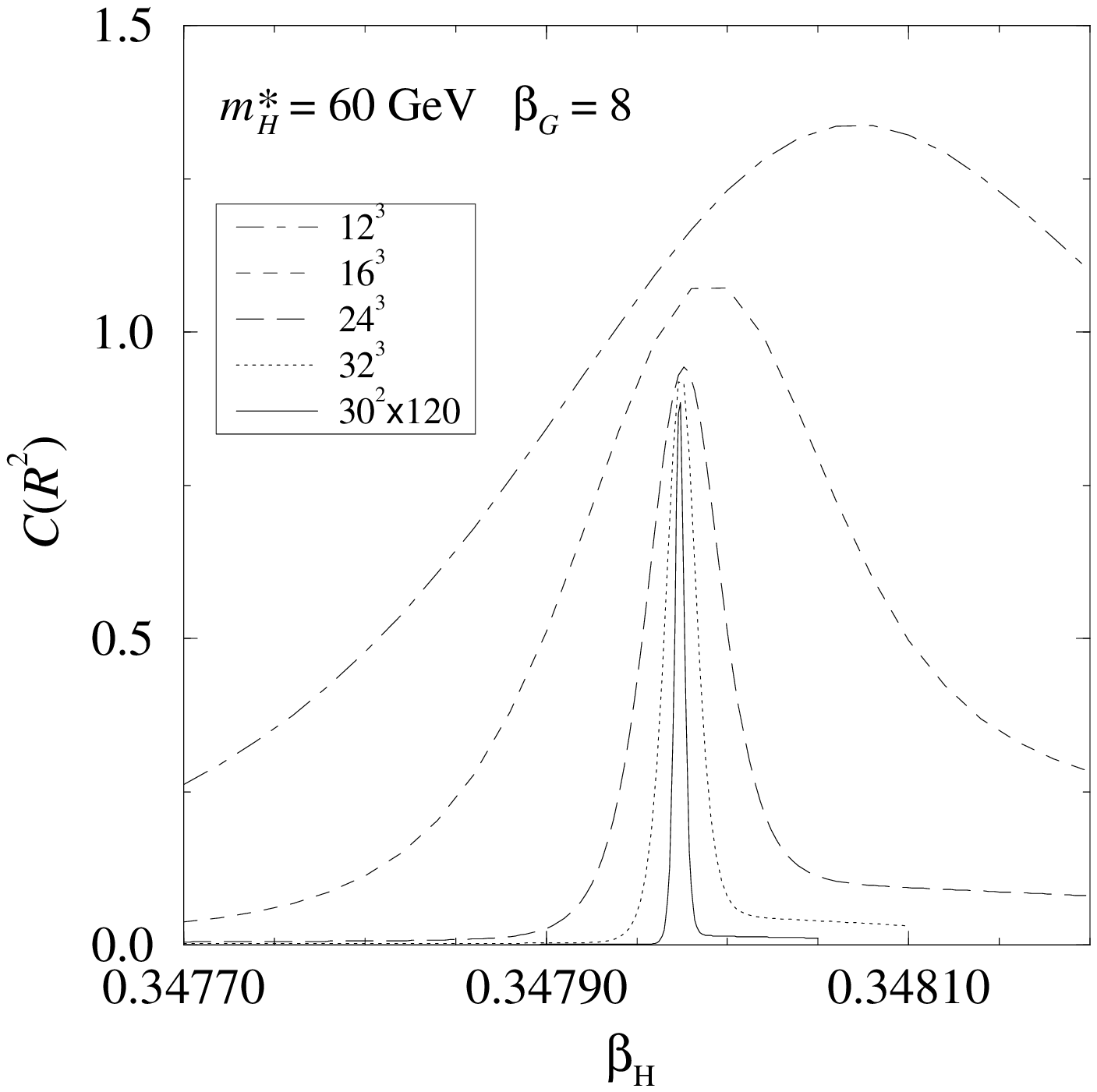}
\hspace*{-2cm}
\epsfxsize=10cm\epsfbox{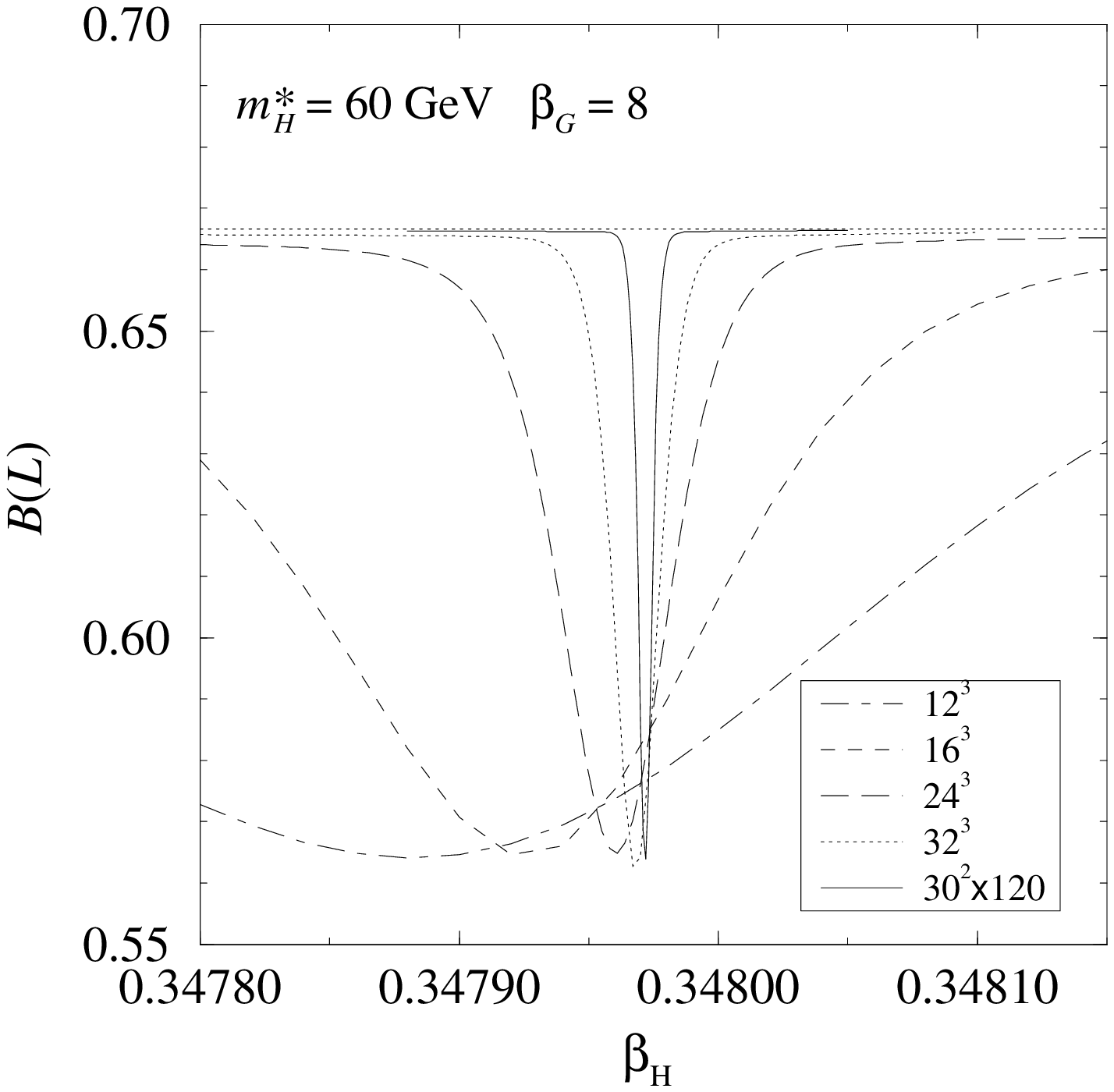}}
\vspace{-4.5cm}
\caption[a]{$C(R^2)$, the susceptibility of the order parameter $R^2$
(top), and the Binder cumulant $B(L)$ of the order parameter
$L$ (bottom), plotted as a function of $\beta_H$ for
$m_H^*=60$~GeV and $\beta_G=8$ runs.}
\la{m60-b8-R2-LB}
\end{figure}


The $\beta_H$-values for the ``equal weight'' and the ``equal height''
distributions are also found by reweighting the histograms
independently for each jackknife block.  The histograms in
\figs\ref{m60-all-Rhg} and \ref{m35-m70-b8-Rhg} are all equal weight
$R^2$ histograms; that is, the areas in the symmetric phase and broken
phase peaks are equal.  The calculation of the peak area requires an
arbitrary selection of the value of $R^2$ which is used to separate the
peaks; we used a fixed value for all the lattices in each
($m_H^*,\beta_G$)-set, defined by the minimum of the distribution of
the largest volume.

\paragraph{The infinite volume limit:}
The values of $\beta_{H,c}$ determined with the methods (1)--(5) above
differ for each individual lattice, but the $V\rightarrow\infty$
extrapolations are very well compatible within the statistical errors.
It should be noted that the different methods for determining
$\beta_{H,c}$ do {\em not\,} give statistically independent results,
and it is not justified to combine the values given by different
methods together.  However, they serve the purpose of checking the
consistency of the infinite volume and continuum limits.

\begin{figure}[tb]
\figtopspace
\epsfysize=\figysize
\centerline{\epsffile{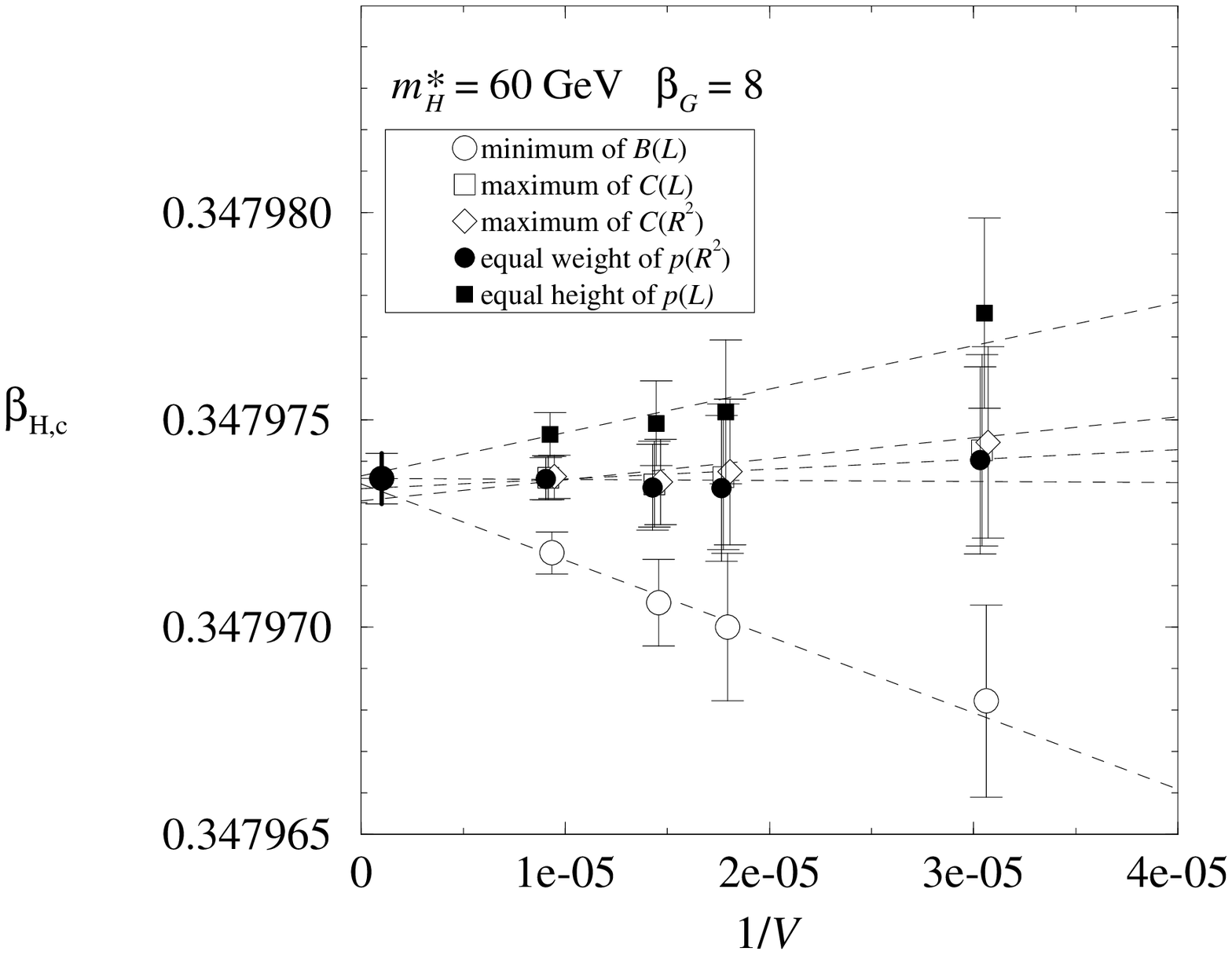}}
\figbottomspace
\caption[a]{The $V=\infty$ limit of the pseudocritical couplings
$\beta_{H,c}$ for $m_H^*=60$~GeV, $\beta_G=8$, calculated with five
different methods.  All the methods give compatible $V=\infty$
limits; the point near $1/V=0$ is the result from linear extrapolation
of the ``equal weight of $p(R^2)$'' -values.  Only the largest volumes are
shown.}
\la{m60-b8-bc}
\end{figure}
\begin{figure}[tb]
\vspace*{-1cm}
\centerline{\hspace{-3.3mm}
\epsfxsize=10cm\epsfbox{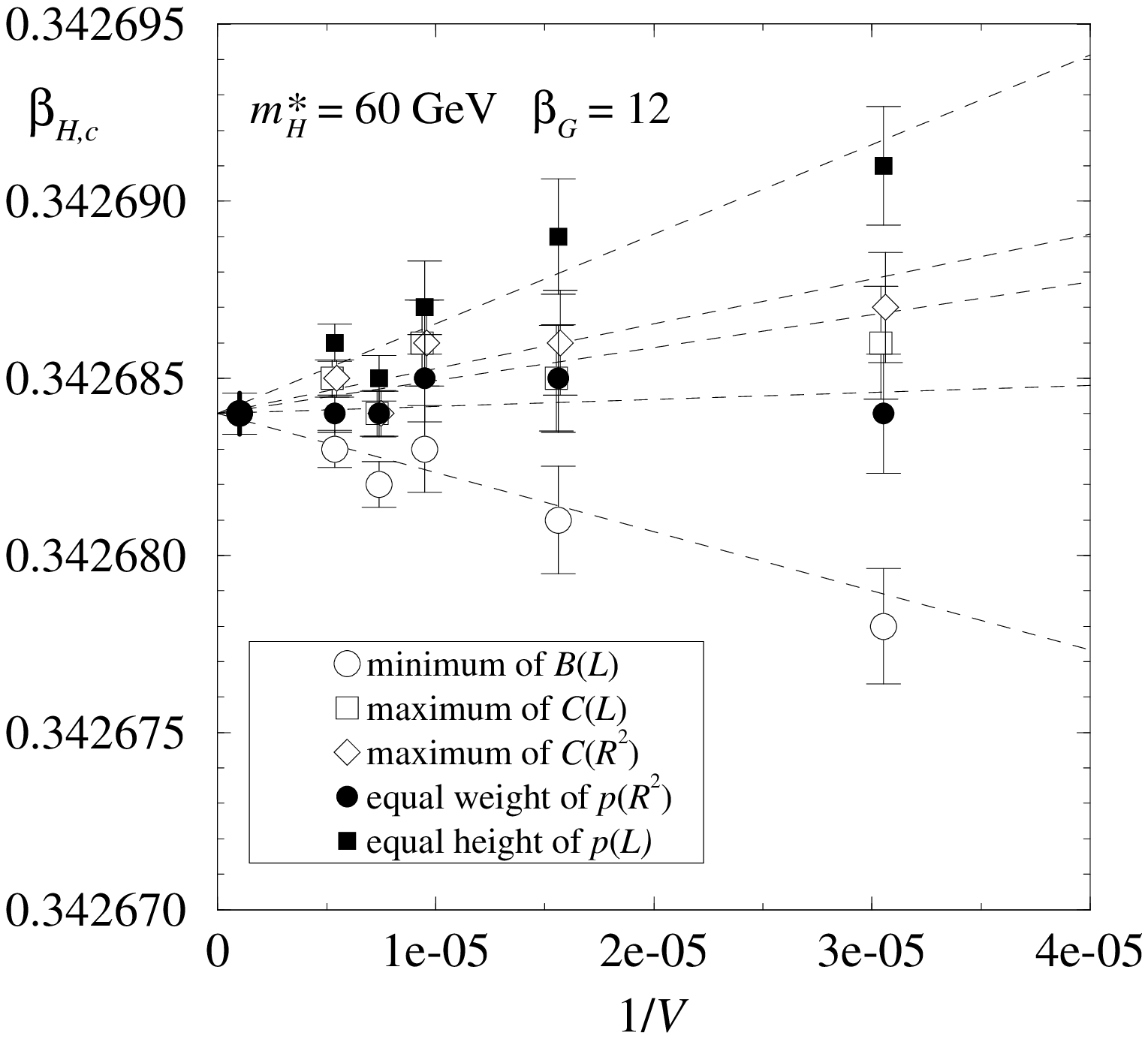}
\hspace{-2cm}
\epsfxsize=10cm\epsfbox{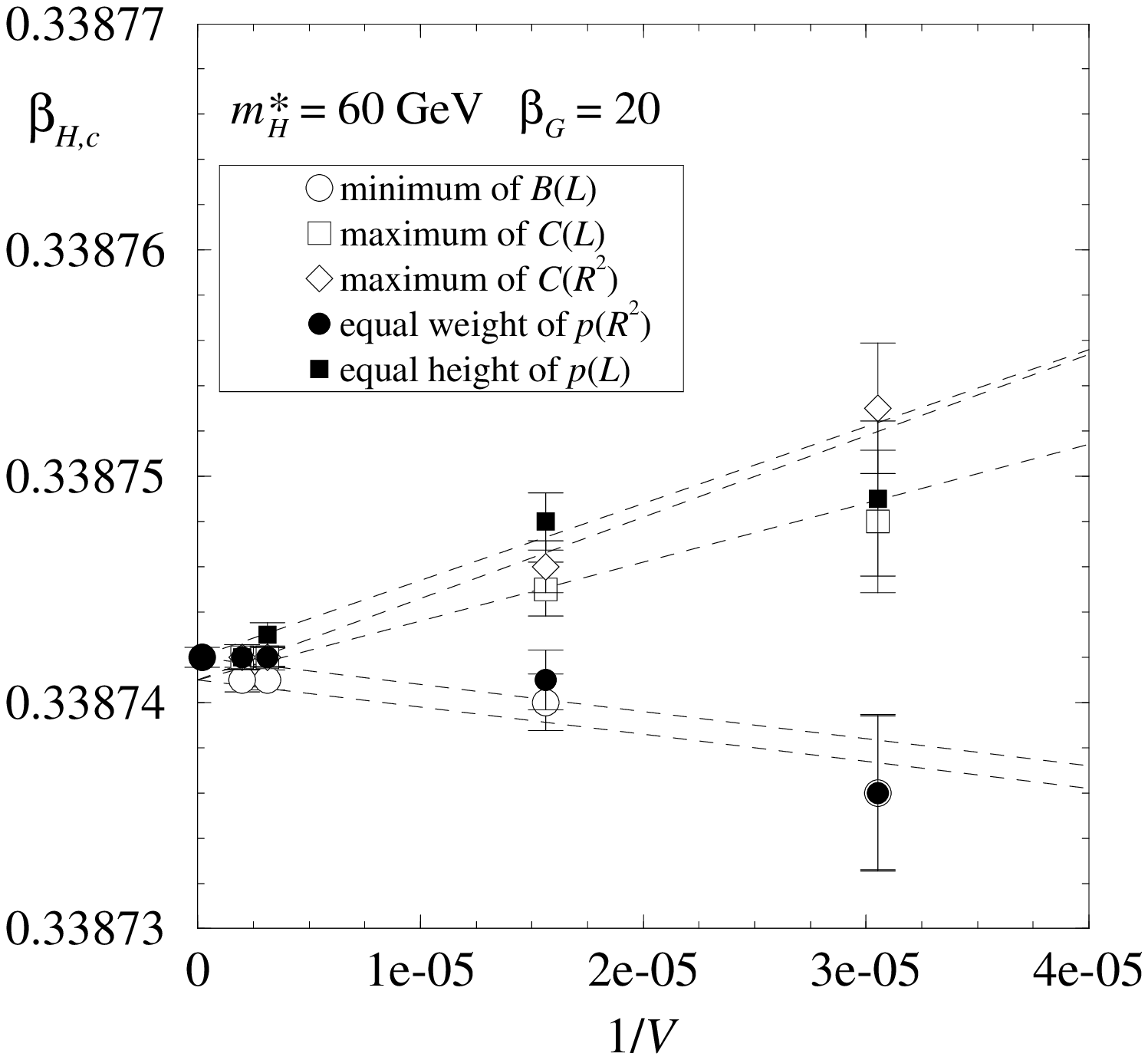}}
\vspace*{-5cm}
\caption[a]{The same as \fig\ref{m60-b8-bc} for $\beta_G=12$
and $\beta_G=20$.}
\la{m60-b12-bc}
\end{figure}


In \figs\ref{m60-b8-bc}--\ref{m60-b12-bc} we show the infinite
volume limits for $m_H^*=60$~GeV and $\beta_G=8$, 12 and 20.  As can
be observed, different methods converge extremely well (the intercepts
of the dashed lines at $1/V=0$ are nearly equal).  The same holds true
for Higgs masses $m_H^*=35$ and 70~GeV, which are not shown here.  In
table \ref{table:betac} we show the $V\rightarrow\infty$
extrapolations of $\beta_{H,c}$, using the data obtained with the
equal weight of $p(R^2)$ -method.  The corresponding values of the
critical temperature $T^*_c$, calculated with
eqs.~\nr{y}, ~\nr{yeff}, are also shown.

\begin{table}[ht]
\center
\begin{tabular}{|c|r|l|l|}
\hline
 $m_H^*$/GeV & $\beta_G$ & \cen{$\beta_{H,c}$} & \cen{$T^*_c$/GeV} \\
\hline
 35    &        8  & 0.3450806(17) & 94.181(15)  \\
       &       12  & 0.3411047(11) & 93.666(23)  \\
       &       20  & 0.3379421(28) & 93.27(16)   \\
\hline
 60    &        5  & 0.358495(5)   & 137.534(17) \\
       &        8  & 0.3479735(6)  & 137.669(5)  \\
       &       12  & 0.3426840(6)  & 137.842(12) \\
       &       20  & 0.3387418(4)  & 138.019(27) \\
\hline
 70    &        8  & 0.3491523(39) & 153.620(37) \\
       &       12  & 0.3433841(12) & 153.930(26) \\
       &       20  & 0.3391279(26) & 154.03(16)  \\
\hline
\end{tabular}
\caption[a]{The infinite volume critical couplings
$\beta_{H,c}$, and the critical temperatures $T^*_c$.
The values of $\beta_{H,c}$ are
calculated from the ``equal weight of $p(R^2)$'' data.}\la{table:betac}
\end{table}

\begin{figure}[tb]
\figtopspace \vspace{1cm}
\epsfysize=\figysize
\centerline{\epsffile{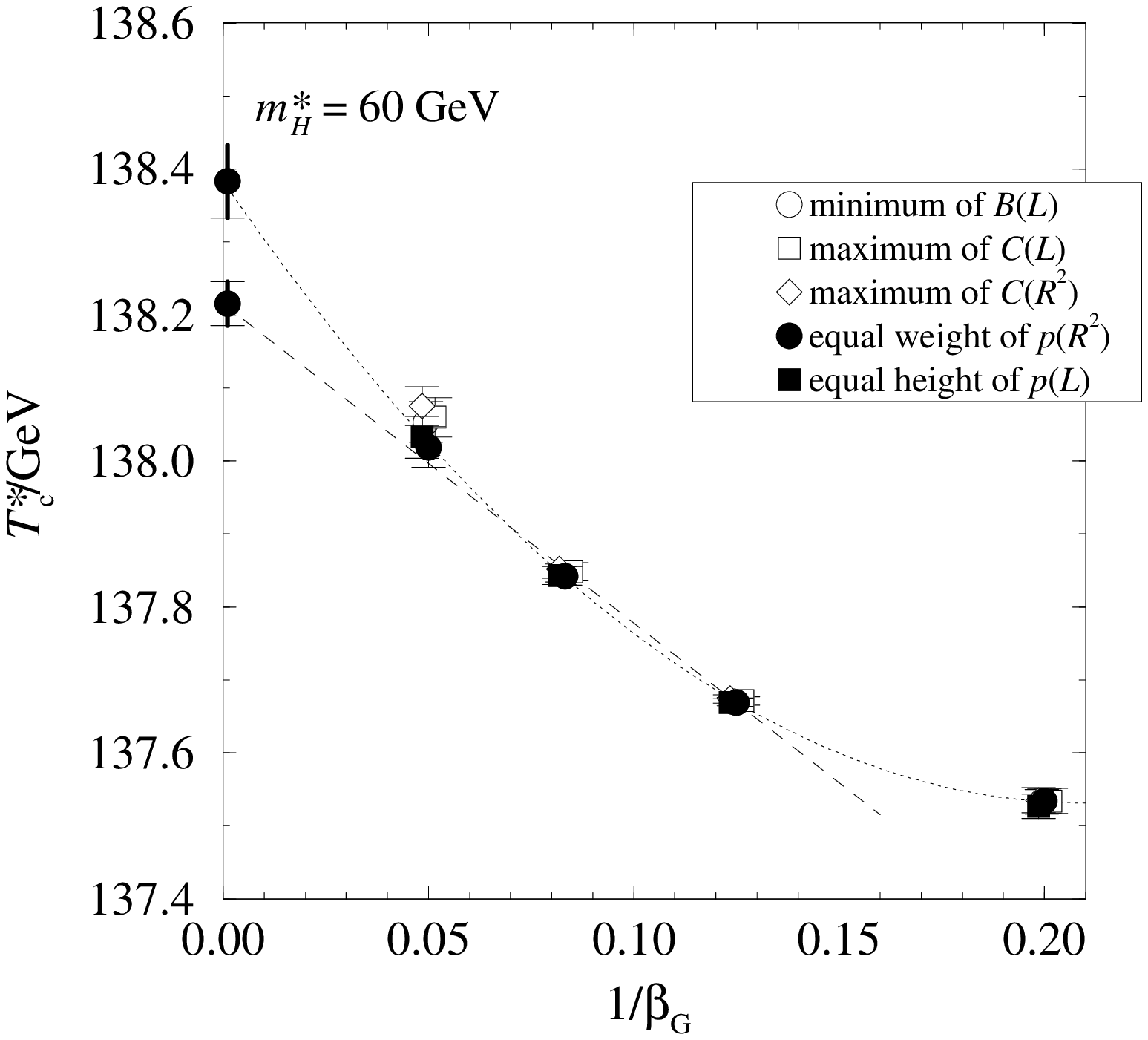}}
\figbottomspace
\caption[a]{The continuum limit ($\beta_G\rightarrow\infty$) of the
critical temperature for $m_H^*=60$~GeV.
A quadratic fit (dotted line) in $1/\beta_G$
to all four $\beta_G$-values gives good $\chi^2$/d.o.f.\@, whereas
a linear fit (dashed line)
using only $\beta_G\ge 8$ does not have acceptable $\chi^2$.
The extrapolations shown are for ``equal weight of $p(R^2)$'' data.}
\la{m60-tc}
\end{figure}

\paragraph{The continuum limit:}
In \fig\ref{m60-tc} the $V=\infty$ values of $T^*_c$ are extrapolated
to the continuum limit $a\rightarrow 0$ for $m_H^*=60$~GeV\@.  We expect
the leading deviation from the continuum limit value to be of order
${\cal O}(a)$; in this case, there are 4 values for $\beta_G$, and the
accuracy of the data is high enough that quadratic fits are needed in
order to have good $\chi^2$/d.o.f.\@ values for the fits.  The quality
of the fits is very good for the $T^*_c$ values calculated with any of
the five criteria, and the final extrapolations are statistically
compatible.  For concreteness, we use the equal weight of $p(R^2)$
-results for our final numbers.
\begin{figure}[tb]
\vspace*{-1cm}
\centerline{\hspace*{-2.3mm}
\epsfxsize=9.5cm\epsfbox{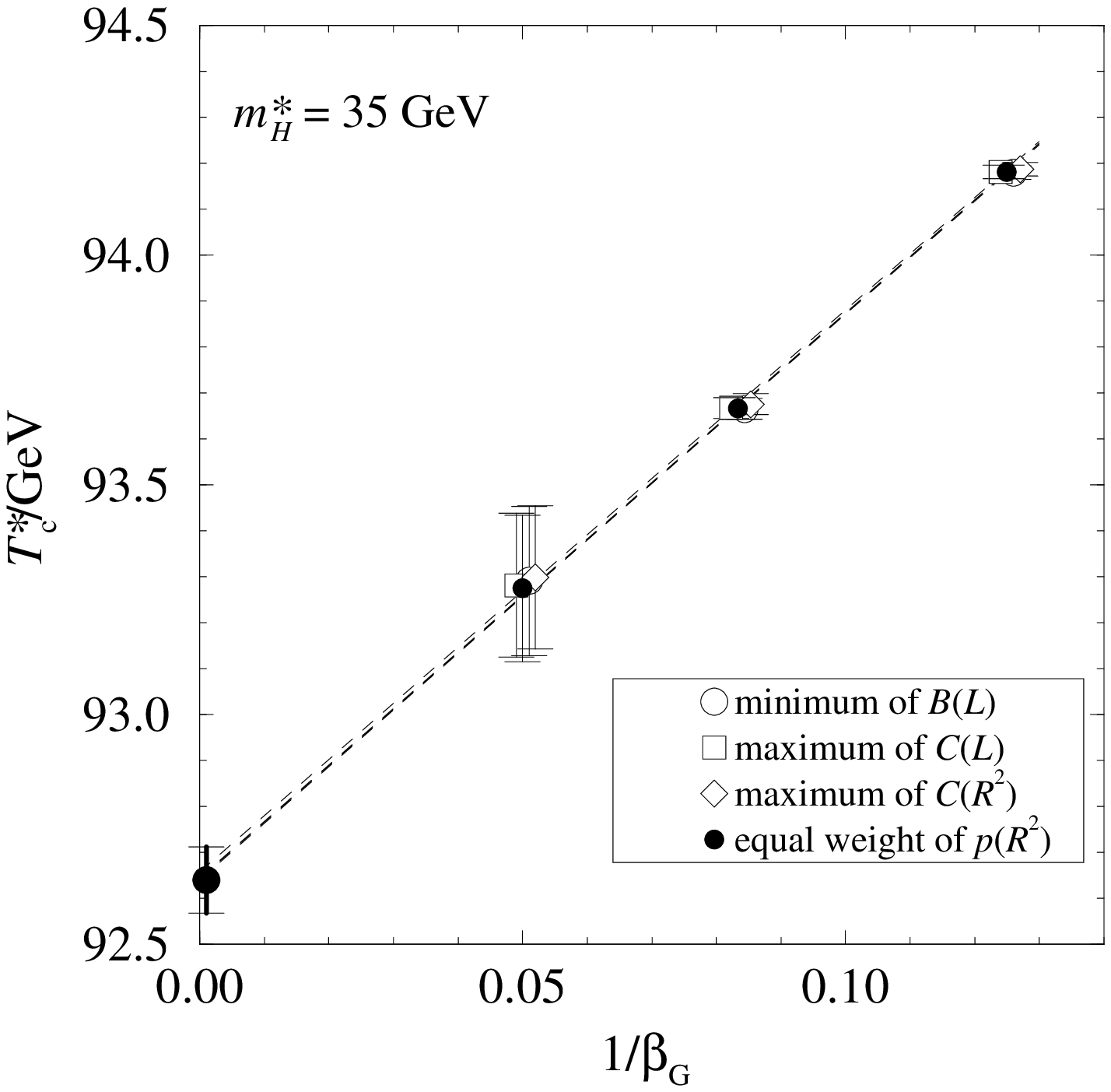}
\hspace*{-1.8cm}
\epsfxsize=9.5cm\epsfbox{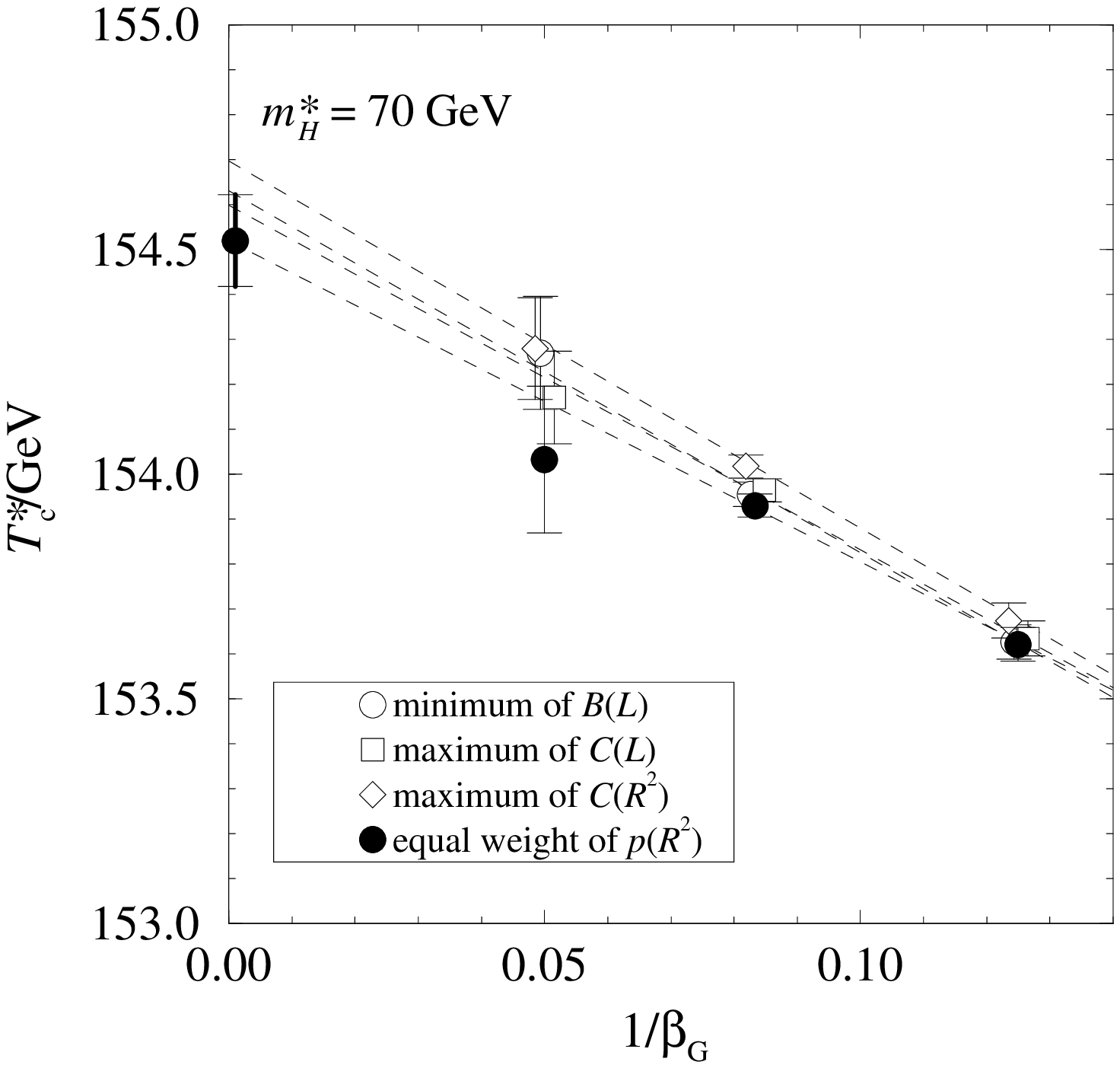}}
\vspace*{-4.5cm}
\caption[a]{The same as \fig\ref{m60-tc} for $m_H^*=35$~GeV and 70~GeV,
using linear fits.}
\la{m35-tc}
\end{figure}

In \fig\ref{m35-tc} we show the corresponding
extrapolations for $m_H^*=35$ and 70~GeV.  In these cases
we use linear fits.  The final results are summarized in table
\ref{table:critical}, together with the perturbative values of $T^*_c$.

\begin{table}[ht]
\centerline{
\begin{tabular}{|c|c|c|}
\hline
 $m_H^*$/GeV  & $T^*_c$/GeV & $T^{*\,\rm pert}_c$/GeV\\
\hline
 35           & \n92.64(7)\n &\n93.3 \\
 60           &  138.38(5)\n & 140.3  \\
 70           &  154.52(10)  & 157.2  \\
\hline
\end{tabular}}
\caption[a]{The continuum limit extrapolations of the critical
temperatures.  The $m_H^*=60$~GeV point has been calculated with a
quadratic fit, others with linear fits.}
\la{table:critical}
\end{table}

There is a systematic difference between the perturbative and the
lattice results; the perturbative $T^*_c$ is considerably larger.  The
values are closest to each other when $m_H^*=35$~GeV, but for
$m_H^*=60$ and 70~GeV the difference is more than 20 standard
deviations.  These results agree qualitatively with the results from
the simulations with
the $A_0$ field~\cite{krs,fkrs2}, but the errors here
are almost an order of magnitude smaller.
They also agree qualitatively with 4d
simulations~\cite{desylattice1,desylattice2} and recent 3d simulations
by Ilgenfritz \etal \cite{ikps}. Note, however, than in 4d case the
errors are considerably larger, and in \cite{ikps} the extrapolation
to continuum limit is not taken.

\subsection{The latent heat and $v(T^*_c)$}\la{sec:latent}

The latent heat $L$ --- the energy released in the transition --- can
be calculated from
\be
  \fr{L}{T} = \fr{d\Delta p}{dT} =
  \fr{T}{V}\fr{d}{dT}\Delta\log Z = \fr{T}{V} \fr{d}{dT} \Delta P\,,
  \la{latent}
\ee
where the derivatives are evaluated at the critical temperature,
$\Delta p$ is the difference of the pressures of the symmetric
and  broken phases, and $\Delta P$ is the difference of the
probabilities of the phases in volume $V$.  In \eq\nr{latent}, $T$ is
the physical (4d) temperature; for simplicity, in the following we
substitute $T\rightarrow T^*$.  In Sec.~\ref{sec:phys} we discuss
how the correct physical result can be obtained. The quantity $\Delta P$ is
directly proportional to the difference of the areas of the two peaks
in the order parameter distributions near $T^*_c$, and $d(\Delta
P)/dT^*$ is readily calculable by reweighting.

An alternative method is to evaluate $d(\Delta p)/d T^*$ directly from
the action in eq.~\nr{lagr}:
\be
  \fr{L}{T_c^{*4}} = \frac{m_H^{*2}}{T_c^{*3}}\Delta\langle
        \phi^\dagger\phi\rangle =
        \fr{1}{8}\fr{m_H^{*2}}{T_c^{*3}}
        g_3^2 \beta_H\beta_G \Delta \langle R^2 \rangle \,.
\la{Lwithpdp}
\ee
Both methods give compatible results in the
$V\rightarrow\infty$ limit.  The results shown here have been
calculated with \eq\nr{latent}.

\begin{figure}[tb]
\vspace*{-1cm}
\centerline{\hspace{-3mm}
\epsfxsize=9.3cm\epsfbox{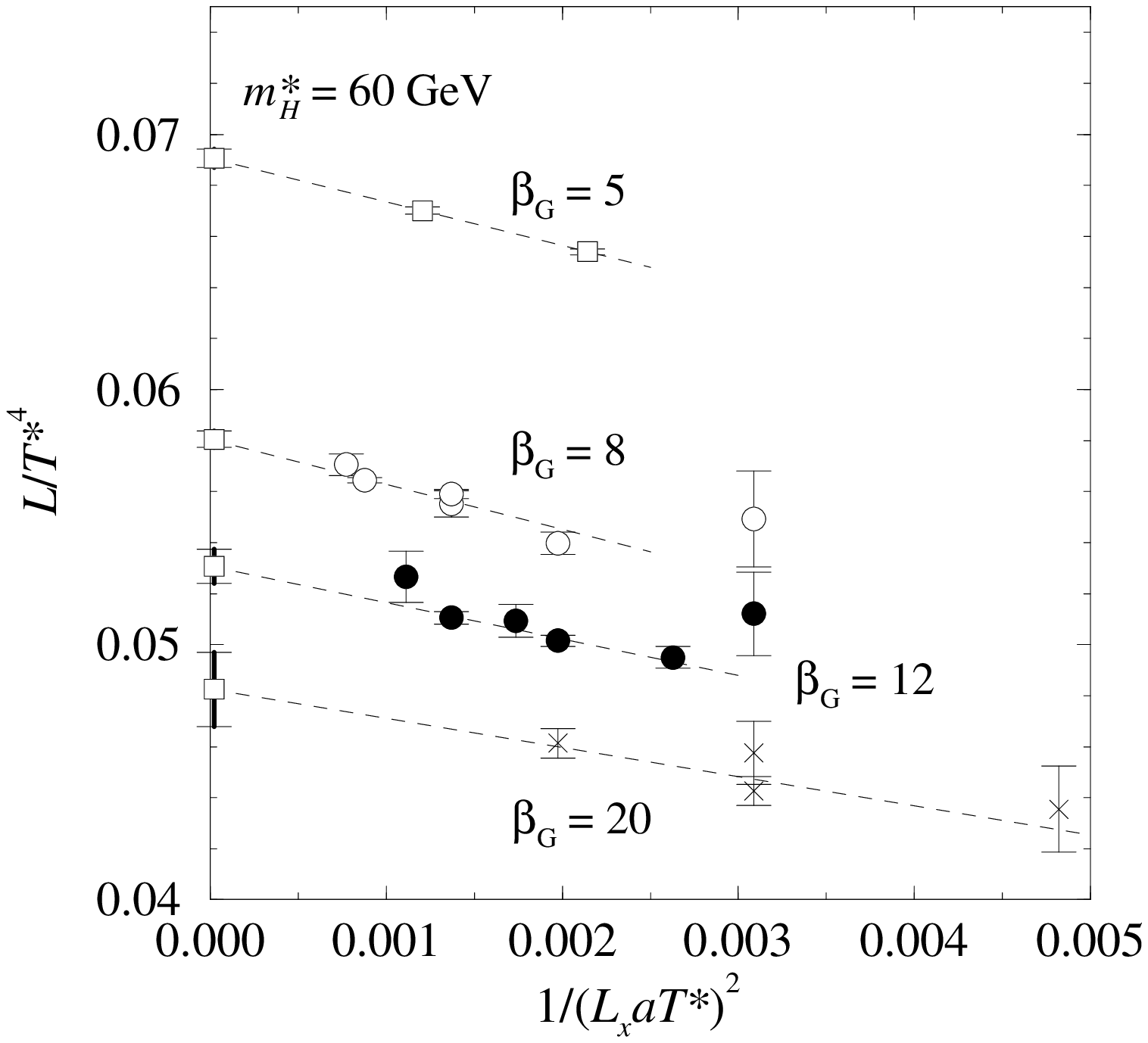}
\hspace{-1.5cm}
\epsfxsize=9.3cm\epsfbox{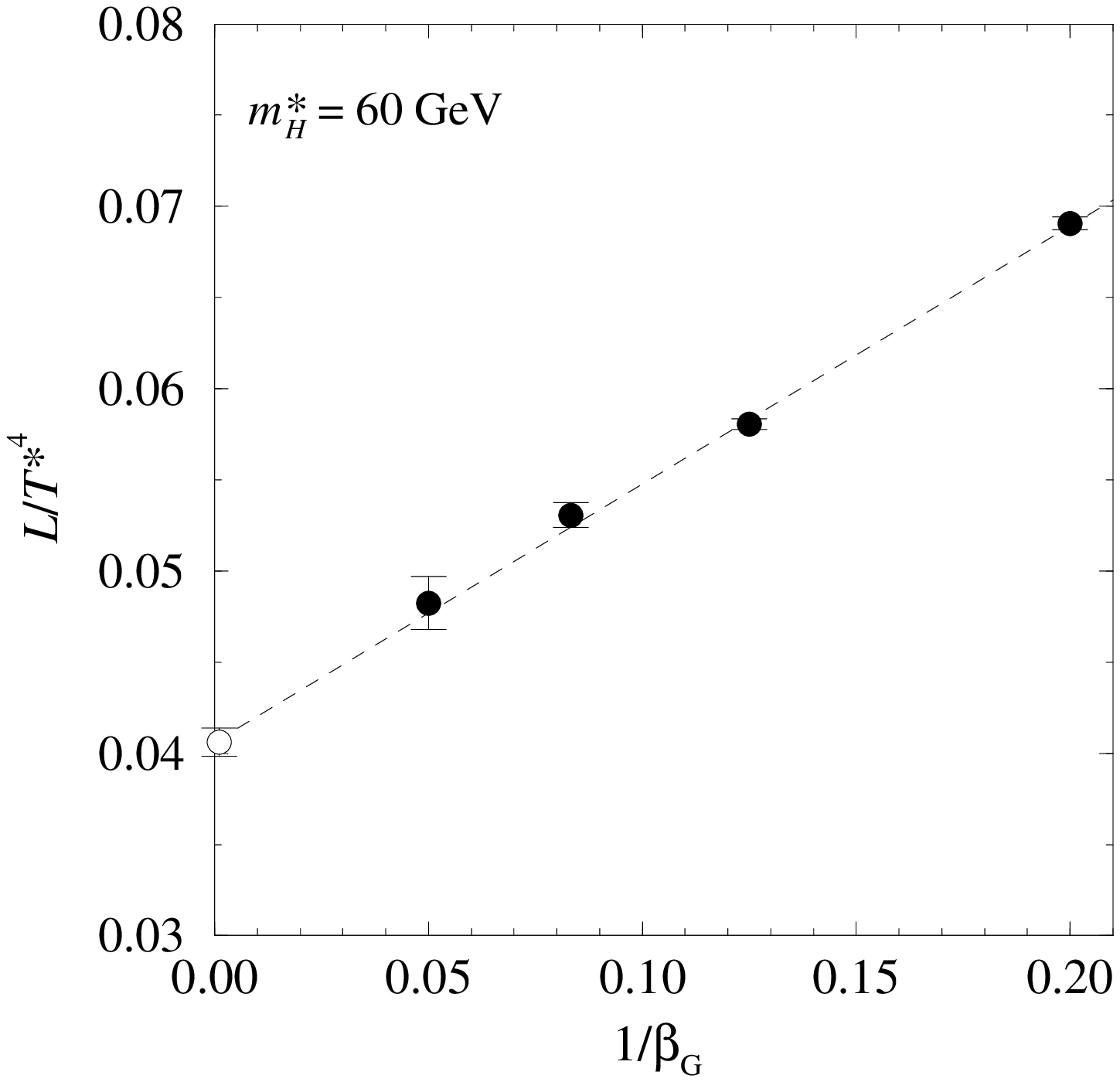}}
\vspace*{-4.5cm}
\caption[a]{Extrapolation of the latent heat to $V\rightarrow\infty$ limit
(left figure) and to the continuum limit $a\rightarrow 0$ (right figure)
for $m_H^*=60$~GeV lattices.}
\la{m60-latent}
\end{figure}

Fig.~\ref{m60-latent} displays the limits $V\rightarrow\infty$ and
$a\rightarrow 0$ for the latent heat for $m_H^*=60$~GeV systems.  The
infinite volume limit is taken by extrapolating linearly with respect
to the inverse area $1/(AT^{*2}) = (L_x a T^*)^{-2}$
of the system,
where $L_x a$ is the linear length of the lattice (one of the short
dimensions for the cylindrical volumes).  The extrapolation with
respect to the inverse volume would fail to accommodate the cubical and
cylindrical lattices simultaneously with the same scaling ansatz.  The
inverse area -type behaviour of the latent heat is known to occur for
the Potts models in 2 dimensions \cite{Borgs90,potts} (the area in this case
being the linear dimension of the lattice).

The Higgs field expectation value in the broken phase $v(T^*)$,
defined here by
\be
\frac{v^2(T^*)}{T^{*2}} \equiv
2 \frac{\langle\phi^\dagger\phi(T^*)\rangle}{T^*},
\label{defvoT}
\ee
can be calculated from $\langle R^2 \rangle$ using
\eq\nr{rl2}.  Because of the close relation between the equations
\nr{latent} and \nr{rl2}, the limits are very similar in both
cases.  In \fig\ref{m60-v2} we show the approach of $v^2(T^*_c)$ to
the continuum limit.
\begin{figure}[tb]
\figtopspace
\epsfysize=\figysize
\centerline{\epsffile{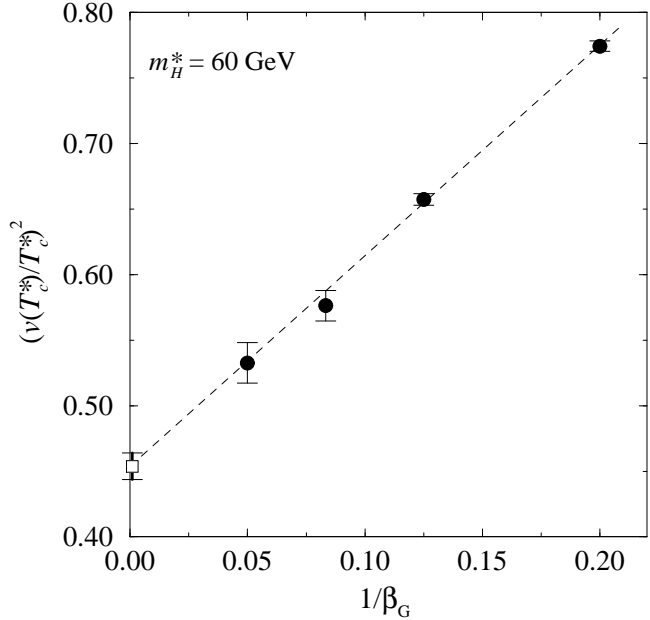}}
\figbottomspace
\caption[a]{The continuum limit of the square of the Higgs field
vev $v^2(T^*_c)$.}
\la{m60-v2}
\end{figure}

Higgs masses 35 and 70~GeV are analyzed in a similar way; the results
for $v(T^*_c)$ and the latent heat are shown in table \ref{table:latent}.

\begin{table}[ht]
\centerline{
\begin{tabular}{|c|c|c|c|c|c|}
\hline
$m_H^*$/GeV & $L/(T_c^*)^4$ & $L^{\rm p}/(T_c^{*\rm p})^4$ &
        $v(T_c^*)/T_c^*$ & $v^{\rm p}(T_c^{*\rm p})/T_c^{*\rm p}$ &
        $ v^{\rm p}(T_c^*)/T_c^* $  \\
\hline
  35      & 0.256(8)\n\n& 0.22\n & 1.86(3)\n & 1.75 & 1.87  \\
  60      & 0.0406(7)\n & 0.040  & 0.674(8)  & 0.68 & 0.82  \\
  70      & 0.0273(16)  & 0.027  & 0.57(2)\n & 0.55 & 0.70  \\
\hline
\end{tabular}}
\caption[a]{The latent heat $L$ and the Higgs field expectation value
$v(T^*_c)$ in the broken phase.  Here $T_c^{*\,\rm p}$ is the
perturbative critical temperature (see
table~\ref{table:critical}). The comparison between
lattice results and perturbation theory is
discussed in Sec.~\ref{sec:comparison}.\la{table:latent}}
\end{table}

Note that the values for both $v$ and $L$ are quite close to the
perturbative values, {\em evaluated at the perturbative critical
temperature $T_c^{*\rm p}$}.  If we use the lattice critical
temperature $v^{\rm p}$ becomes larger, as can be seen
from the last column in table~\ref{table:latent}. This shows the
presence of higher-loop perturbative corrections in the broken phase,
discussed in detail in the previous section.


\subsection{The interface tension\la{sec:tension}}

The interface tension is one of the primary quantities which
characterize the strength of the phase transition.  We measure it with
the {\em histogram method\,} \cite{Binder82}: at the pseudocritical
temperature, a system in a finite volume predominantly resides in
either the broken or the symmetric phase, but it can also exist in a
mixed state consisting of domains of the two states.  The probability
of the mixed state is suppressed by the extra free energy associated
with the interfaces between the phases.  This causes the typical
2-peak structure of the probability distribution of the order
parameter at the critical temperature (see figs.~\ref{m60-all-Rhg}
and \ref{m35-m70-b8-Rhg}): the midpoint between the peaks corresponds
to a state which consists of equal volumes of the symmetric and broken
phases.  Because of the associated extra free energy, the area of the
interfaces tends to minimize.  Assuming a lattice with periodic
boundary conditions and geometry $L_x^2\times L_z$, where $L_x \le
L_z$, the minimum area is $2\times A = 2(L_x a)^2$ --- the number 2
appears because there are two separate interfaces.  The interface
tension $\sigma$ can be extracted from the limit
\be
  \fr{\sigma}{T} = \lim_{V\rightarrow\infty} \fr{1}{2A}
        \log \fr {P_{\rm max}}{P_{\rm min}}\,,
  \la{sigma}
\ee
where $P_{\rm max}$ and $P_{\rm min}$ are the probability distribution
maximum and the minimum between the peaks.

In practice, the infinite volume value of $\sigma$ is reached in such
large volumes that careful finite size analysis of \eq\nr{sigma} is
necessary.  Numerous studies exist in the literature
\cite{Bunk92,Caselle93,Iwasaki94}; here we follow \cite{Iwasaki94}:
\be
  \sigma \fr{a^2}{T} = \fr{1}{2L_x^2}\log\fr{P_{\rm max}}{P_{\rm min}}
        + \fr{1}{L_x^2}\left[\fr34 \log L_z - \fr12 \log L_x
        + \fr12 G + const.\right].
  \la{fs-sigma}
\ee
The function $G$ interpolates between lattice geometries; the limiting
values are $G = \log 3$ for cubical volumes ($L_z=L_x$) and $G=0$ for
long cylinders ($L_z \gg L_x$).

The finite size scaling ansatz \nr{fs-sigma} assumes that the two
interfaces are far enough apart from each other that their mutual
interaction is negligible.  In practice, this is very difficult to
achieve in cubical volumes and usually requires the use of long
cylindrical lattices.  The order parameter histograms develop a flat
minimum when this condition is fulfilled: the flatness signals a
constant free energy when the volume fractions of the two phases
slightly change, and the interfaces move with respect to each other.

\begin{figure}[tb]
\vspace*{-1cm}
\centerline{\hspace{-3mm}
\epsfxsize=9.3cm\epsfbox{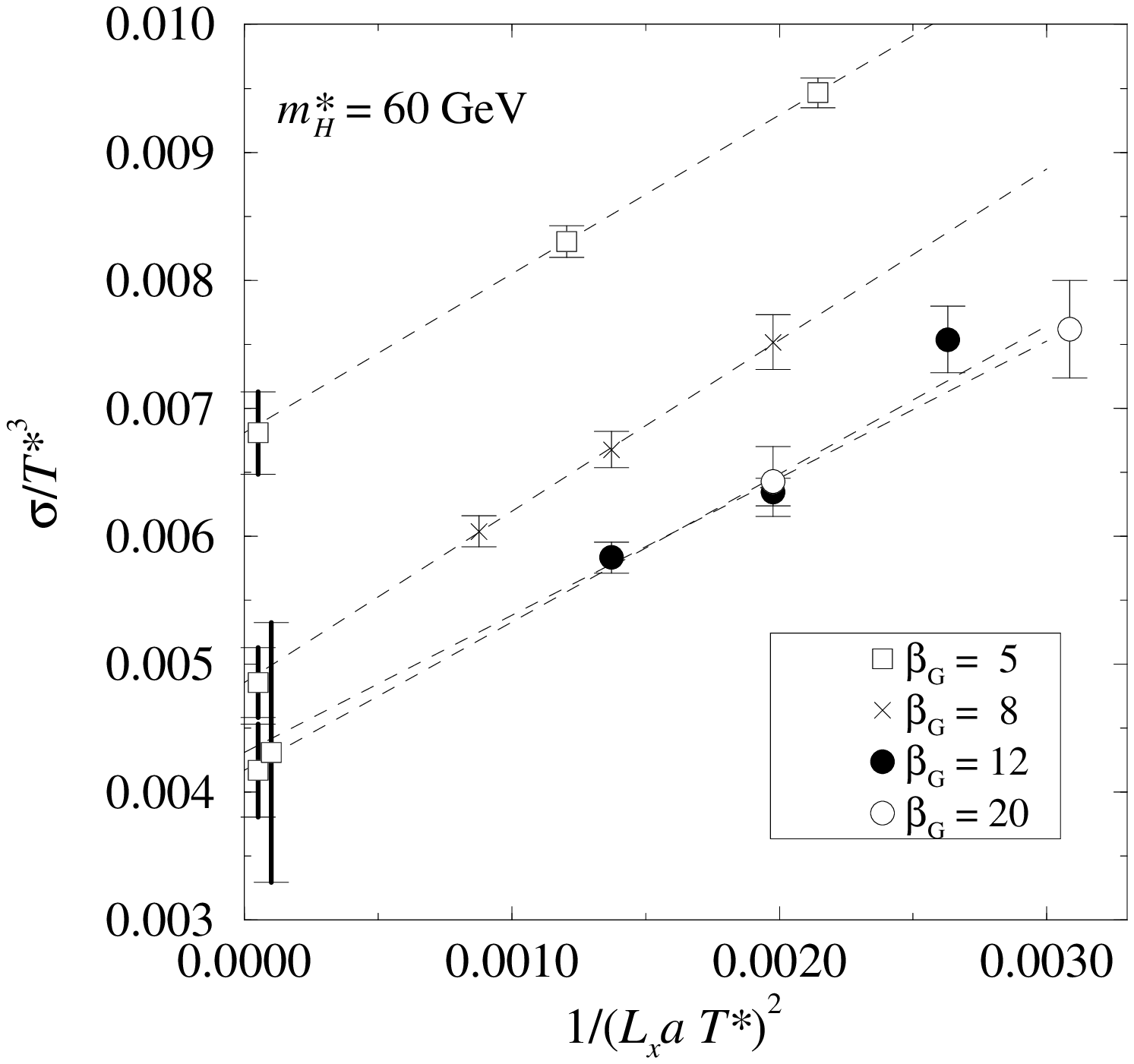}
\hspace{-1.5cm}
\epsfxsize=9.3cm\epsfbox{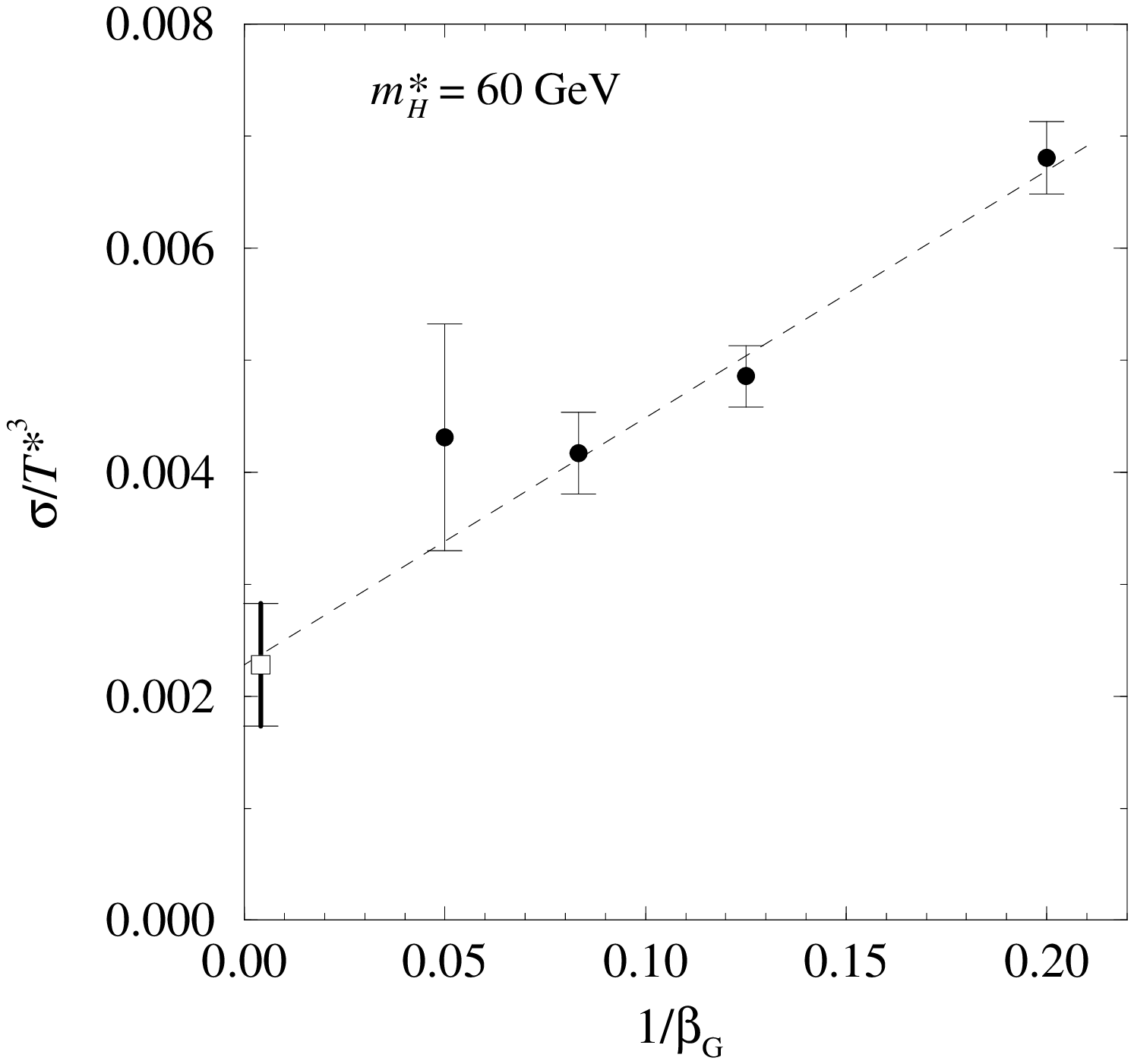}}
\vspace*{-4.5cm}
\caption[a]{The interface tension for $m_H^*=60$~GeV extrapolated to
$V\rightarrow\infty$ (left) and $a\rightarrow 0$ (right).\la{m60-sigma}}
\end{figure}

In order to find the flat part of the histograms we use large
cylindrical volumes for all $\beta_G$ values for $m_H^*=60$~GeV\@.  The
volumes included for the analysis are: $\beta_G=5$: $12^2\times 72$
and $16^2\times 80$; $\beta_G=8$: $20^2\times 140$, $24^2\times 120$,
and $30^2\times 120$; $\beta_G=12$: $26^2\times 156$, $30^2\times
150$, and $36^2\times 144$; $\beta_G=20$: $40^2\times 200$ and
$50^2\times 200$.  In \fig\ref{m60-all-Rhg} we show the equal
weight histograms of $p(R^2)$ for these lattices.

In the interface tension analysis we use {\em equal height histograms of
$p(L)$\,}.  Equal height histograms were chosen instead of equal
weight because the determination of $P_{\rm max}$ in equation
\ref{fs-sigma} then becomes unambiguous; and $p(L)$ instead of
$p(R^2)$ since the former histograms have more symmetric shapes than
the latter, and the equal height and equal weight $\beta_{H,c}$ values
are quite close to each other.  Nevertheless, we check the
measurements with equal weight $p(R^2)$ histograms (using $P_{\rm
max}$ which is a linear interpolation of the two peak heights to the
$R^2$ value of the minimum); the results are very well
compatible within statistical errors.

A comment about extracting the extrema from the histograms is in order:
we find the maximum values by fitting a parabola close around the peaks
of the histograms, and the value of the flat minimum by fitting a
constant.  This method gives much smaller errors and more reliable
results than simply using the absolute extrema values, which are
very prone to statistical noise.  The same method was used also to
locate the equal height $\beta_{H,c}$-values.

In the left part of \fig\ref{m60-sigma} we show the interface tension
measurements from each lattice.  The values shown here include the
finite size scaling correction $1/L_x^2 (\fr32 \log L_z - \fr12 \log
L_x)$ from \eq\nr{fs-sigma}, so that only a factor $const./L_x^2$
remains.  For each $\beta_G$, the behaviour of the data is linear in
$1/L_x^2$.  On the right part of \fig\ref{m60-sigma} we extrapolate
$\sigma$ to the continuum limit.  The results
are shown in table~\ref{table:sigma}.

\begin{table}[ht]
\centerline{
\begin{tabular}{|c|c|c|}
\hline
$m_H^*$/GeV & $\sigma/(T_c^*)^3$ & $\sigma^{\rm p}/(T_c^{*\rm p})^3$ \\
\hline
 35       &  0.0917(25)  & 0.066\n \\
 60       &  0.0023(5)\n & 0.0078 \\
 70       &  ---         & 0.0049 \\
\hline
\end{tabular}}
\caption[a]{The interface tension $\sigma$.
Only the $m_H^*=60$~GeV result
is an extrapolation to the continuum limit;
the $m_H^*=35$~GeV value is
only from $\beta_G=8$ simulations.\la{table:sigma}}
\end{table}

The $m_H^*=60$ GeV continuum limit result in table~\ref{table:sigma} is
obtained by a linear extrapolation in $1/\beta_G$.  However, on closer
inspection the data from $\beta_G=12$ and 20 lattices on the left part
of \fig\ref{m60-sigma} seem to indicate that the interface tension is
already scaling when $\beta_G\ge 12$, even in finite volumes.  If we
use the finite volume $\beta_G=8$ and 12 data and extrapolate to
infinite volume, we obtain the result $\sigma/(T^*_c)^3 = 0.0042(3)$,
which is not compatible with the result in table~\ref{table:sigma}.
Nevertheless, since the data for the latent heat or $v(T^*)$ do not
display similar scaling behaviour, we use the linear extrapolation in
$1/\beta_G$ in all cases.

For $m_H^*=35$ GeV, only $\beta_G=8$ lattices are cylindrical enough so that
we can estimate the $V\rightarrow\infty$ limit.  However, the
continuum value can not be extrapolated.  For $m_H^*=70$ GeV we did not
observe good enough flat parts in the order parameter histograms, and
we do not attempt to estimate the interface tension.

\section{The correlation lengths for $m_H^*=60$~GeV\la{sec:correlation}}

The measurement of the Higgs and $W$ masses around the transition
temperature is somewhat orthogonal to the measurements of the
quantities directly associated with the transition itself: instead of
attempting to enhance the tunnelling of the system from one phase to
another, in this case it is imperative that the system stays in one
homogenous phase throughout the measurement.  This is due to the
spurious signal caused by the {\em tunnelling correlations\,}: $m_{\rm
tunnel} \propto \exp(-\sigma A/T)$ \cite{Herrmann92}.  Even an
incomplete tunnelling can make the extraction of the physical mass
very complicated.  Since the tunnellings are suppressed by the
exponential factor $\exp(-\sigma A/T)$, we perform a separate set of
simulations around the critical temperature using large volumes and
monitor the simulation time history of order parameters in order to
ensure that the system stays in a single phase throughout the
measurement.

\begin{figure}[p]
\vspace*{-1cm}
\centerline{\hspace{-3.3mm}
\epsfxsize=10cm\epsfbox{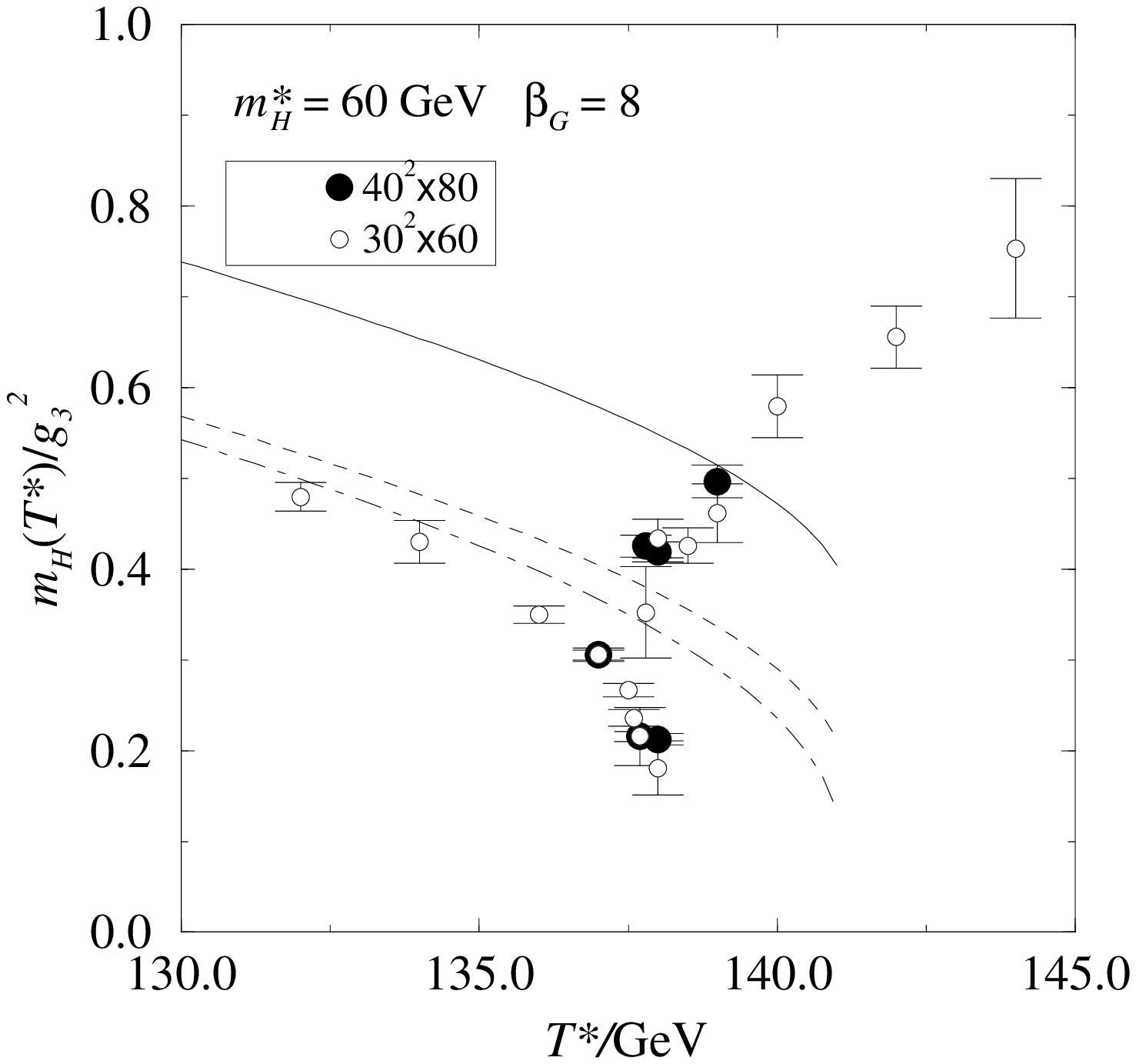}
\hspace{-2cm}
\epsfxsize=10cm\epsfbox{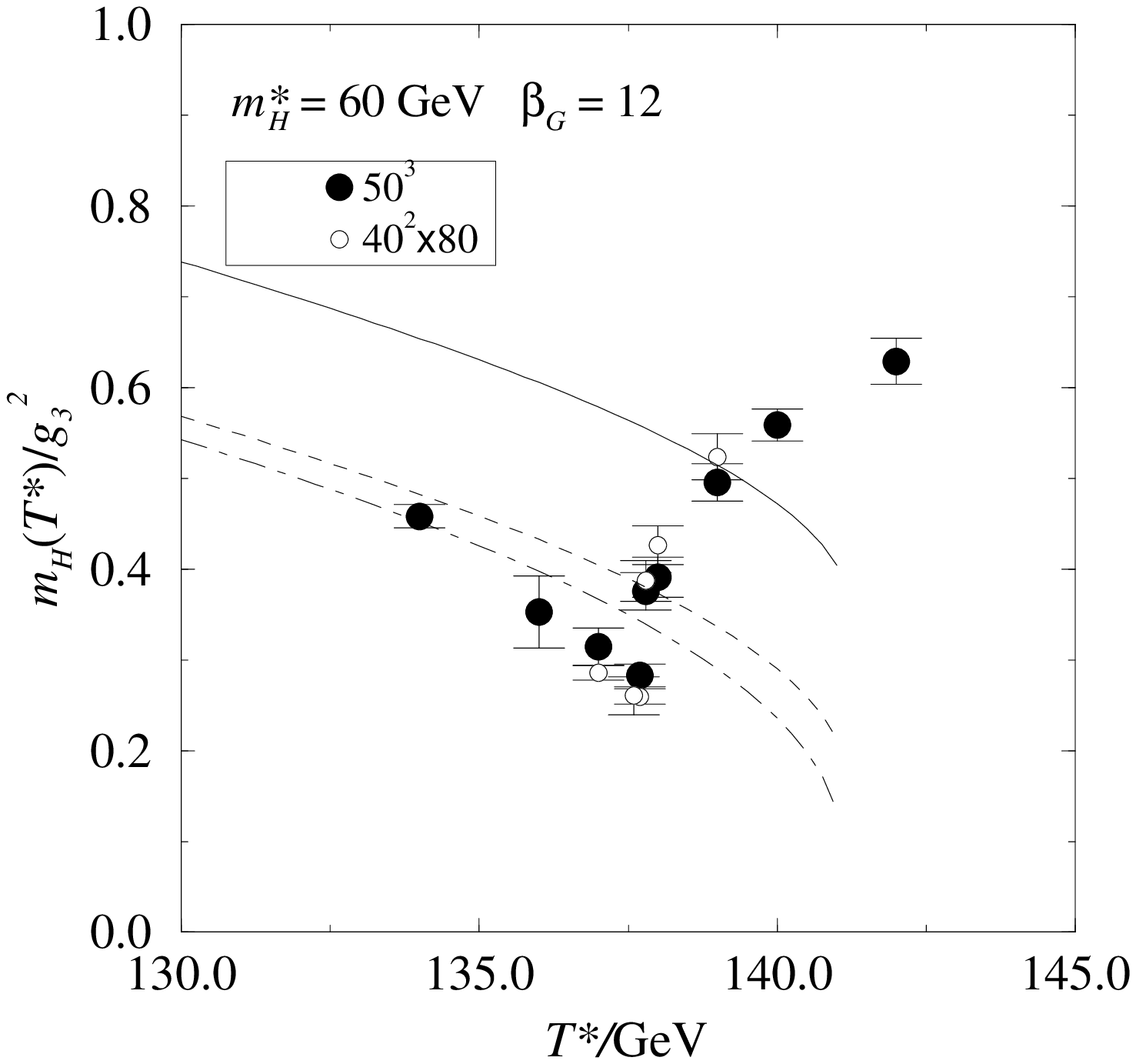}}
\vspace*{-5cm}
\centerline{\hspace{-3.3mm}
\epsfxsize=10cm\epsfbox{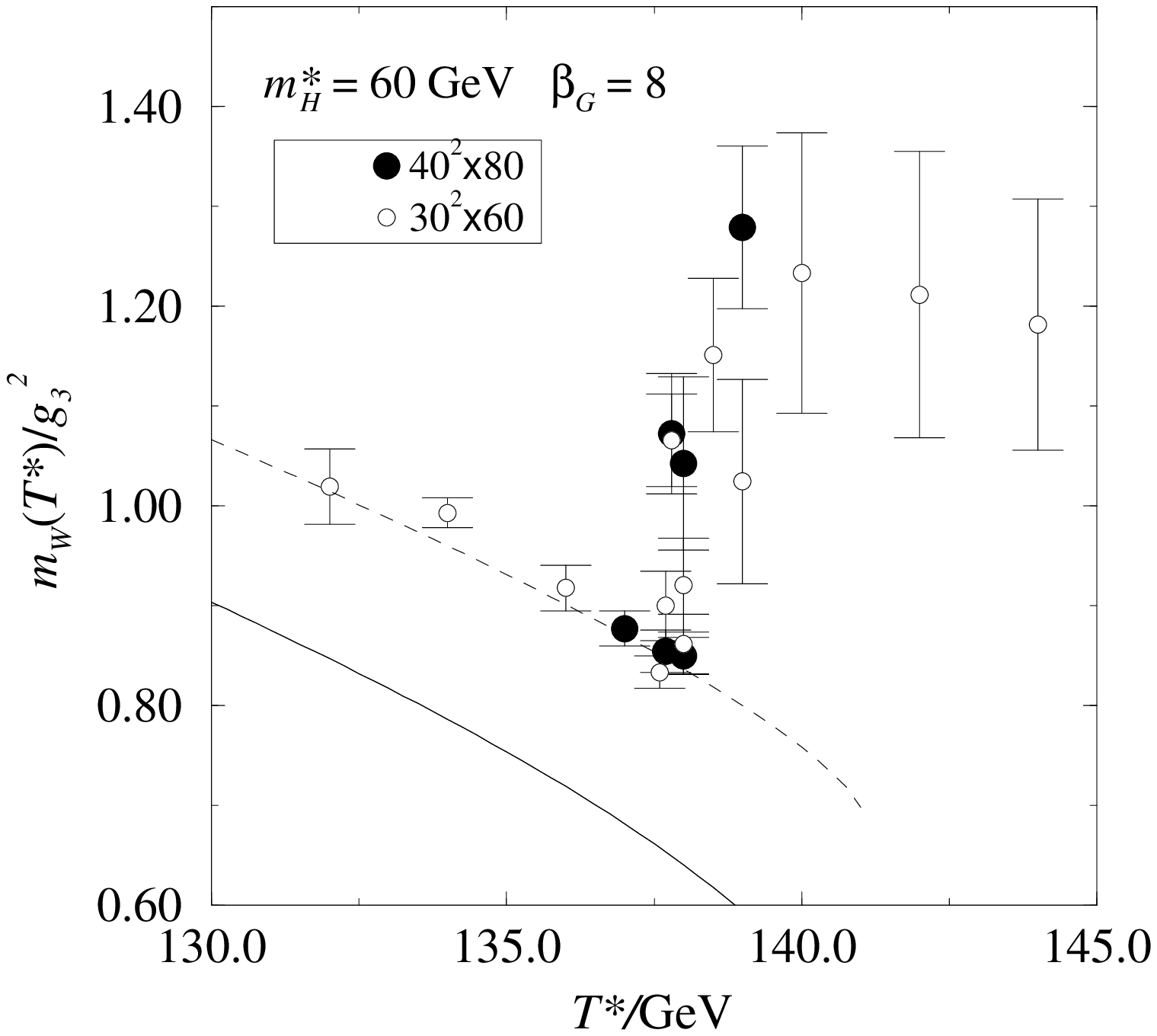}
\hspace{-2cm}
\epsfxsize=10cm\epsfbox{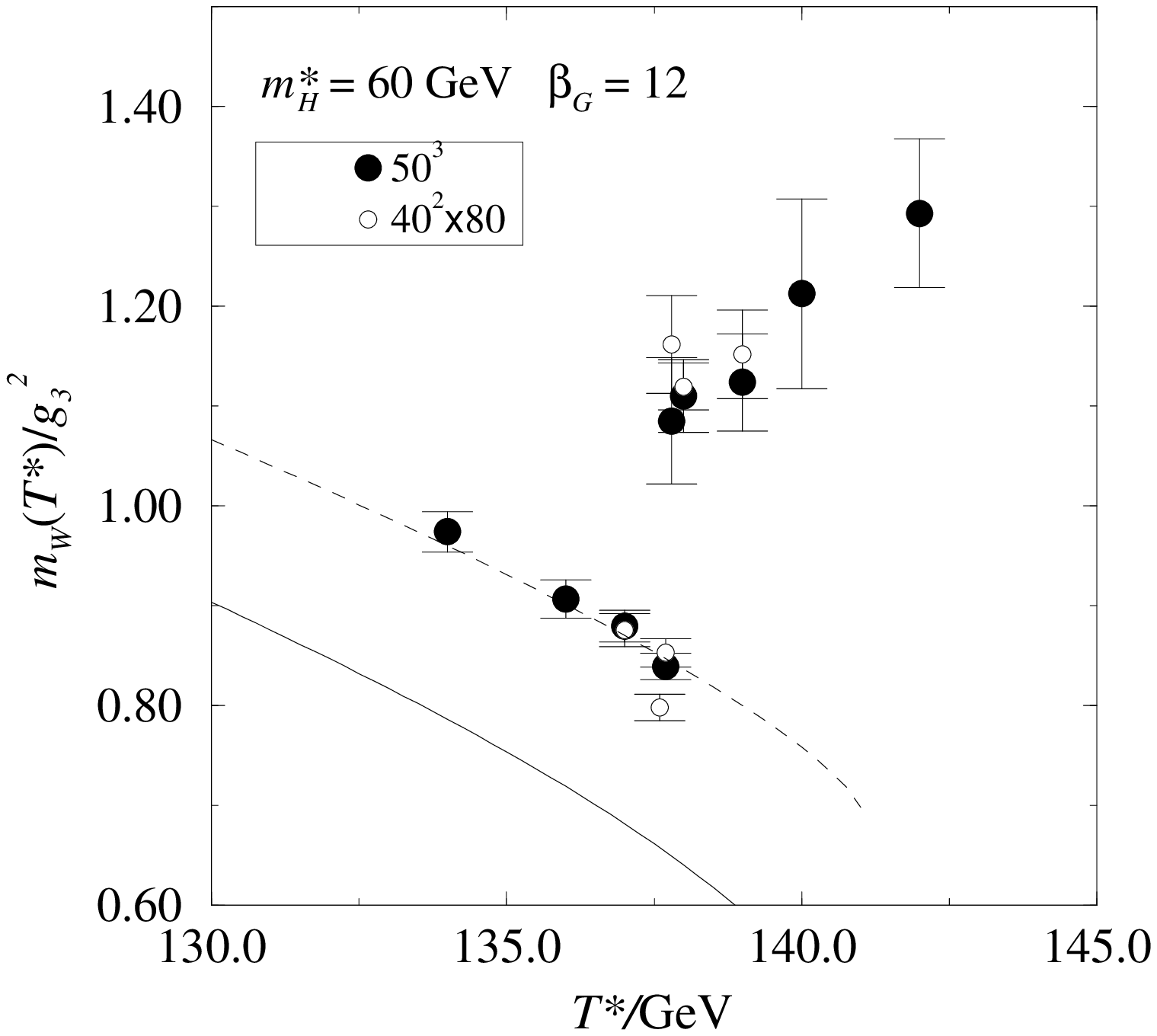}}
\vspace*{-5cm}
\caption[a]{The Higgs masses (top) and the $W$ masses (bottom) for
$m_H^*=60$~GeV, $\beta_G=8$ and 12 systems as functions of the
temperature.  The perturbative results are shown for $m_H$ with solid
(tree-level value), dashed (1-loop pole)
and dot-dashed (some 2-loop corrections,
see Sec.~\ref{cwpt}) lines; for $m_W$
with solid (tree-level) and dashed (1-loop) lines.
Note that the ratio $m_W/m_H$ is much larger than
indicated by the tree-level curves, since radiative
corrections make $m_W$ larger and $m_H$ smaller.
\la{m60-masses}}
\end{figure}

We perform the measurement only for $m_H^*=60$~GeV systems, using
lattice sizes $30^2\times 60$ and $40^2\times 80$ for $\beta_G=8$ and
$40^2\times 80$ and $50^3$ for $\beta_G=12$.  Let us define operators
\be
  c_d^a(z)  =
        \sum_{i=1,2} \sum_{\bfx}\delta_{x_3,z}
        \tr [ \tau^a\, \Phi^\dagger(\bfx)\, U_i(\bfx) \ldots
        U_i(\bfx+(d-1){\bf e}_i)\,\Phi(\bfx+d{\bf e}_i) ]
 \la{corr-operator}
\ee
where $a=0,\ldots 3$; $\tau^a$, $a=1,2,3$, are the Pauli matrices and
$\tau^0 = {\bf 1}$; and $d = 0,\ldots, 4$ is the length of the `chain'
of link matrices between $\Phi^\dagger$ and $\Phi$.  Using
\eq\nr{corr-operator} we can define correlation functions sensitive to
the Higgs and $W$ channels:
\ba
  h_d(l) &=& \fr{1}{V}  \sum_z c_d^0(z) c_d^0(z+l) \\
  w_d(l) &=& \fr{1}{3V} \sum_z \sum_{a=1}^3 c_d^a(z) c_d^a(z+l)\,.
\ea
The masses
$m_H(T^*)$ and $m_W(T^*)$ are found from the exponential fall-off of $h_d$
and $w_d$.  The results are independent of the parameter $d$, and it
is chosen to minimize the statistical errors.  We obtain best results
with $d=0$ for $h_d$ ($c_d^0 = \Phi^\dagger\Phi$) and $d=4$ for~$w_d$.

The measured values of $m_H(T^*)$ and $m_W(T^*)$ are shown in
\fig\ref{m60-masses}.  The scaling between $\beta_G=8$ and 12 is very
good, and both $m_H$ and $m_W$ display a discontinuity at the
transition.  At the transition temperature we are able to maintain the
systems in either the broken or the symmetric phase throughout the
measurement, so that immediately around $T_c$ we have two values for
the masses.  Both $m_H(T^*)$ and $m_W(T^*)$ are {\em higher\,} in the
symmetric phase than in the broken phase.

\section{The metastability ranges and scaling properties of the
symmetric phase}
\la{sec:metastab}

An important characteristic of the phase transition is the range of
metastability of the broken and symmetric phases.  The symmetric phase
is metastable below the critical temperature down to $T = T_{-}$, and
the broken phase is metastable above the critical temperature up to
$T=T_{+}$.  The aim of the present section is to estimate the
metastability range, and to study the dynamics of the composite field
$\pdp$ by determining its effective action.
We shall concentrate on the case $m_H^* = 60$ GeV (x=0.06444) and
$T^*$ near $T_c^*$ ($|y-y_c|<0.06$).

\subsection{The metastability range from the correlation lengths}

Consider the behaviour of the scalar mass near the phase transition,
\fig\ref{m60-masses}.  When the temperature decreases, $m_H$ in the
symmetric phase rapidly decreases. This behaviour suggests that it
reaches zero at some point, and the symmetric correlation length
diverges. This is the lower spinoidal decomposition point and
corresponds to the temperature at which the symmetric phase ceases to
be metastable,~$T_{-}$.  Similar behaviour takes place in the broken
phase, when the temperature increases towards~$T_{+}$.  This observation
allows us to estimate the metastability range, i.e.,\@ the upper and lower
spinoidal decomposition temperatures. In general, one would expect the
following dependence of the Higgs correlation lengths on the
temperature:
\be
  \xi_H^i = c_i |T^*/T_i - 1|^{-\gamma_i}\,,
  \la{pow-correlation}
\ee
where $i=b,s$ labels the broken and symmetric phases,
and $T_b\equiv T_{+}$, $T_s\equiv T_{-}$.  The data we
have does not allow one to determine the critical exponents and the
temperatures $T_\pm$ simultaneously with good accuracy, so that
we have chosen $\gamma_i = 1/2$, following the guidance from mean
field theory. The results for the metastability region are given in
table \ref{tab:metastable}, using the $m^*_H=60$~GeV, $\beta_G=8$ and
12 results from \fig\ref{m60-masses}.
\begin{table}[hbt]
\center
\begin{tabular}{|c|cc|cc|}
\hline
 $\beta_G$ & $T_{-}$/GeV & $c_s g_3^2$ & $T_{+}$/GeV & $c_b g_3^2$ \\
\hline
 \n8 &       135.0(4) & 0.35(2) & 140.1(4) & 0.49(2) \\
  12 &       135.5(3) & 0.33(2) & 139.6(4) & 0.45(2) \\
\hline
\end{tabular}
\caption[1]{The endpoints of the metastability temperature ranges for
the symmetric ($T_{-}$) and broken ($T_{+}$) phases, determined from
$m^*_H=60$~GeV Higgs correlations with the ansatz
\nr{pow-correlation}.\la{tab:metastable}}
\end{table}

\subsection{Metastability from reweighting\la{sec:met-reweight}}

A direct method to measure the metastability range is to use the order
parameter distributions and reweighting.  Let us define an effective
variable $\pi \equiv \phi^\dagger\phi = \fr{\beta_H}{2a} R^2 +
\mbox{\it const.}$, where the last equality follows from \eq\nr{rl2}.
Around the pure phase peaks, the probability distribution $p(R^2)$ is
related to an effective potential $U(\pi)$:
\be
  p(R^2) \propto e^{-V U(\pi)}\,,
  \la{prob}
\ee
where $V$ is the volume of the system.  Eq.~\nr{prob} has
pre-exponential corrections; however, to the accuracy we are working
here the above formula is sufficient \cite{tsypin}.  By reweighting
the distribution $p(R^2)$ we obtain the temperature dependence of the
potential.  It should be noted that \eq\nr{prob} is only valid in the
immediate neighbourhood of the pure phases; it does not correctly
describe the mixed state between the pure phases.

\begin{figure}[t,b]
\vspace*{-0cm}
\centerline{\hspace{2mm}
\epsfxsize=10.4cm\epsfbox{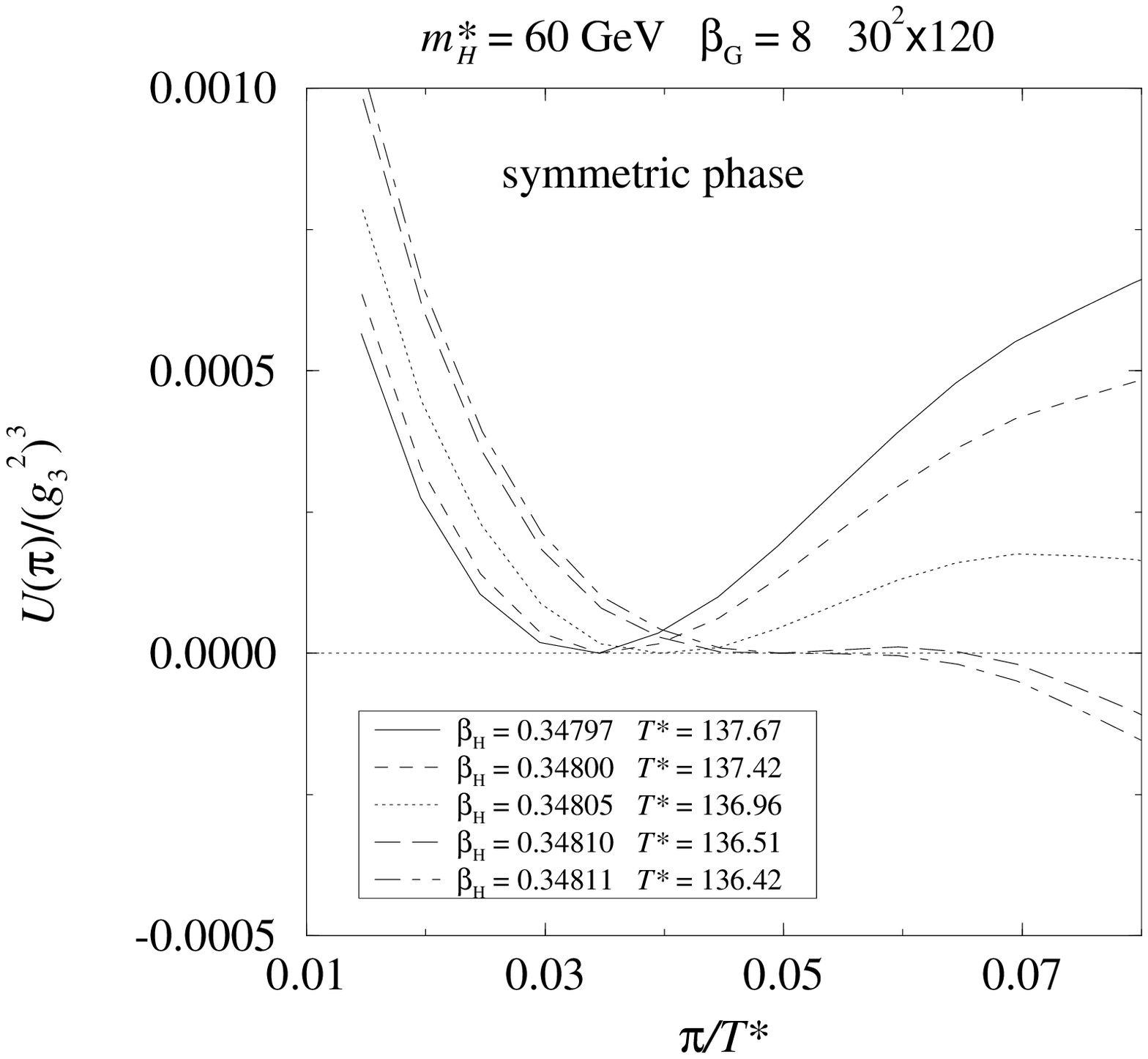}
\hspace{-3.7cm}
\epsfxsize=10.4cm\epsfbox{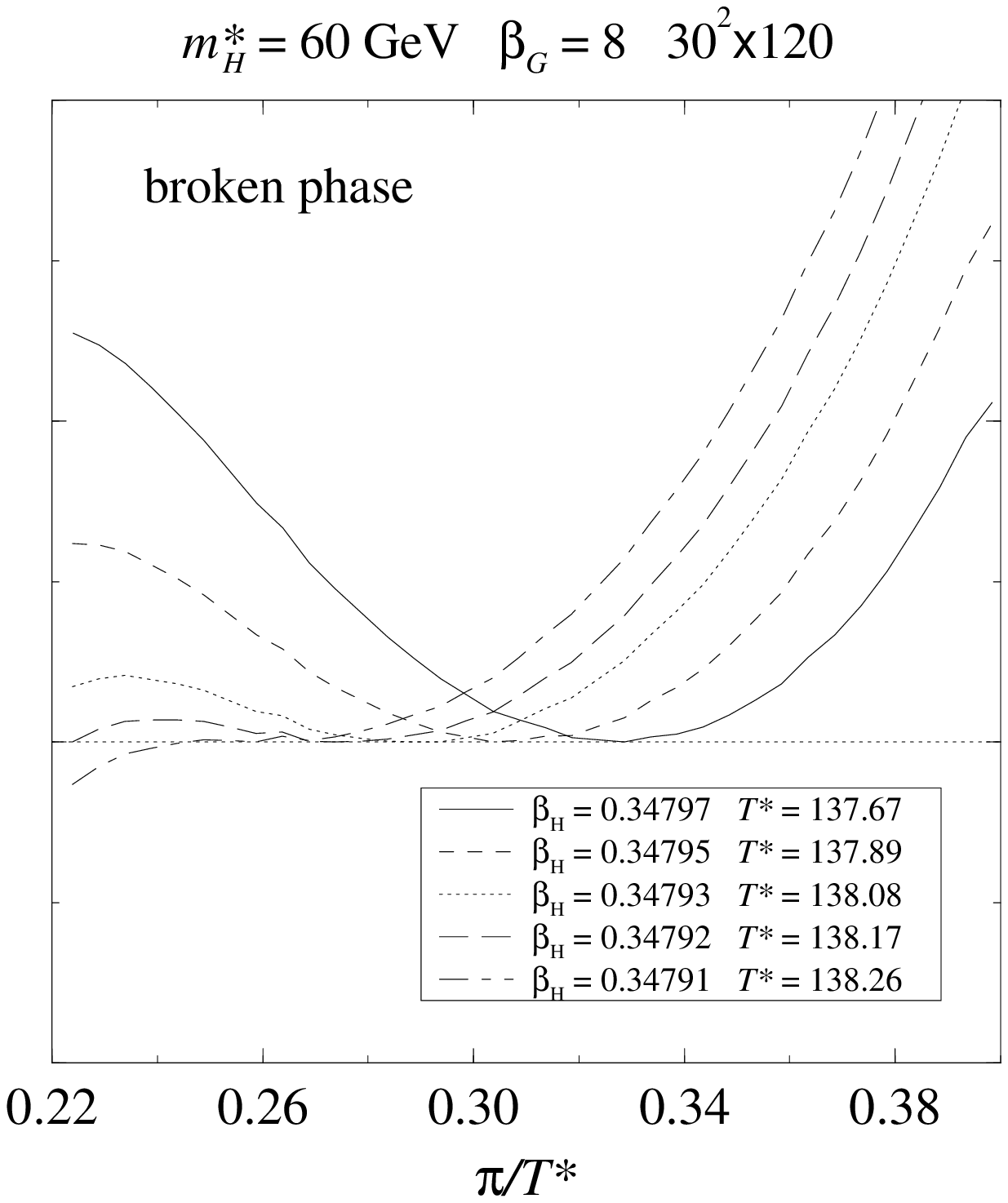}}
\vspace*{-5cm}
\caption[a]{Evolution of the potential $U(\pi)$ for the effective
field $\pi=\phi^\dagger\phi$,
related to the lattice variable $R^2$ through relation
(\ref{rl2}), in the vicinity of the phase transition for $m_H^*=60$
GeV and $\beta_G=8$.  Different $\beta_H$ values are reached by
reweighting a single (multicanonical) simulation.\la{Vphase}}
\end{figure}
\begin{figure}[t,b]
\vspace*{-1cm}
\centerline{\hspace{-3.3mm}
\epsfxsize=10cm\epsfbox{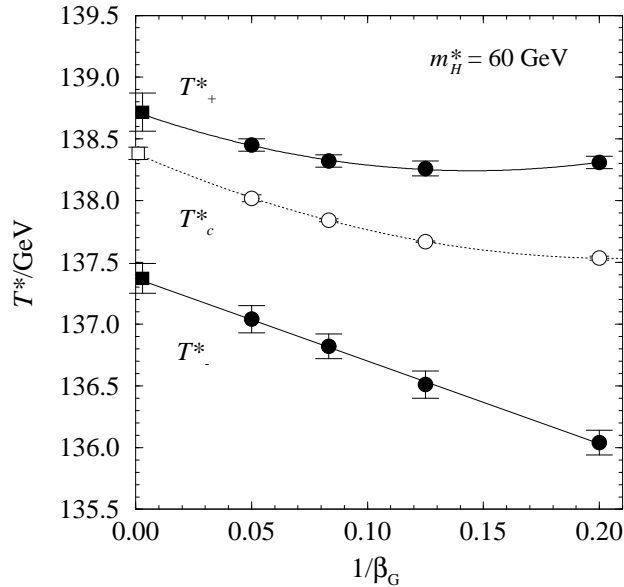}}
\vspace*{-5cm}
\caption[a]{The metastability range for $m_H^*=60$~GeV Higgs systems,
calculated with the reweighting analysis and extrapolated
to $a\rightarrow 0$.\la{m60-spinoidal}}
\end{figure}

The endpoints of the metastability branches can be found by locating
the temperature at which the barrier against the tunnelling vanishes
--- the minimum of the potential in \eq\nr{prob} turns into an
inflection point.  We present the evolution of the potential in
\fig\ref{Vphase} for the symmetric and the Higgs phases for an
$m_H^*=60$~GeV, $\beta_G=8$ system.  These potentials are reweighted
from the $30^2\times 120$ lattice histograms shown in
\fig\ref{m60-all-Rhg}.  In \fig\ref{m60-spinoidal} we extrapolate
the temperatures $T_{-}$ and $T_{+}$ to the continuum limit, with the
result
\be
  T^*_{-} = 137.37(12) \mbox{~GeV} \h\h T^*_{+} = 138.72(15) \mbox{~GeV}.
  \la{spinoidal-values}
\ee
The metastability range here is smaller than the range determined by
fitting the correlation lengths (table \ref{tab:metastable}).
We believe that the values in \eq\nr{spinoidal-values} are more
reliable, because the power law extrapolation of the correlation
lengths is very sensitive to statistical errors and finite volume
effects.
Note also that the range of $\beta_H$ over which reweighting
is carried out is very small.

\subsection{The effective theory description}

More information on the dynamics of the composite field $\pi$ is
contained in its effective action. We shall write it separately for
each phase in the form
\be
S_{\rm eff}=\int d^3x\,\biggl[\frac{1}{Z^2}\fr12 (\partial_i \pi)^2 +
U(\pi)\biggr],
\label{effective}
\ee
where near the minima corresponding to the two phases $s,b$:
\be
U(\pi)=\fr12 v_2^2(\pi-\pi_\rmi{min})^2+v_3(\pi-\pi_\rmi{min})^3+
v_4(\pi-\pi_\rmi{min})^4+\ldots .
\ee
The quantities
\be
{\pi\over g_3^2},\quad {Z^2\over g_3^2},\quad {v_2^2\over g_3^2},
\quad v_3,\quad {v_4g_3^2}
\ee
are dimensionless functions of $x,y$ and are different in the broken and
symmetric phases.

Including only the quadratic terms the correlator is
\be
\langle\pi(\bfx)\pi(0)\rangle={Z^2\over4\pi|\bfx|}\exp(-Zv_2|\bfx|).
\ee
However, it is more convenient to define a plane-averaged field
\be
\pi(z)=\int{dx\,dy\over{\rm Area}}\pi(x,y,z)
\ee
satisfying
\be
\langle\pi(z)\pi(0)\rangle={1\over{\rm Area}}{Z\over2v_2}\exp
(-Zv_2z). \label{planecorr}
\ee
Determining the exponential decay and the magnitude of
plane-averaged correlator thus gives the parameters $Z$ and $v_2$
of the effective action.

By scaling $\pi=(\beta_H/2a)R^2+\mbox{\it const.}$,
these formulas can be directly
rewritten for~$R^2$. For a lattice with the
geometry $N^2 \times L$ the plane Higgs variable is defined as
$R^2_p(z)=[\sum_{x,y}R^2(x,y,z)]/N^2$.  At large distances the
plane-plane correlator has the asymptotic form
\be
  \langle R_p^2(z)R_p^2(0) \rangle =
  A N^{-2} (e^{-z m_H}+ e^{-(La-z)m_H})
  + \mbox{\it const.},
\ee
where $m_H=1/\xi_H$. Using eq.~(\ref{planecorr}) this determines
the quadratic parameters of $S_\rmi{eff}(\pi)$ as
\be
 m_H=Zv_2,\qquad  Z^2= \frac{1}{2} A \beta_H^2 m_H.
\la{z2}
\ee
For the potential we write
\ba
  U(\pi) a^3 & = & U({\beta_H\over2a}R^2+\mbox{\it const.})a^3 \nonumber \\
& = & V_2(R^2-R^2_{\rm min})^2 +
V_3(R^2-R^2_{\rm min})^3
        +V_4(R^2-R^2_{\rm min})^4+ \ldots \,,
  \label{effR}
\ea
where the couplings are related to the previous ones by
\ba
V_2&=&\fr12{v_2^2\over g_3^2}\biggl(\fr12\beta_H\biggr)^2 \times ag_3^2,
\label{V2}\\
V_3&=&v_3\biggl(\fr12\beta_H\biggr)^3,
\nonumber\\
V_4&=&v_4g_3^2\biggl(\fr12\beta_H\biggr)^4\times{1\over ag_3^2}.
\ea

{\bf The wave function normalization.~~}
The lattice data for the quantity $Z^2/g_3^2$ is shown in
\fig\ref{wave-norm} for $m_H^*=60$~GeV and for $T$ near $T_c^*$.
One can see that $Z$ depends considerably on $T$ in the
broken phase, but is very accurately constant in the symmetric phase.
The value of $Z^2$ in the symmetric phase is
\be
  Z_s^2  =  0.320(5)\,g_3^2,
\label{zdef}
\ee
where the result is the average of the constant fits to the
$\beta_G=8$ and 12 data.  In the broken phase the behaviour of $Z_b$
can be easily understood via the tree level relation
\be
  \frac{Z_b^2}{g_3^2} = \frac{v^2(T^*)}{g_3^2 T^{*}}\,.
  \label{zbrok}
\ee
The quantity $v^2/(g_3^2T^{*})$ is also plotted in \fig\ref{wave-norm},
calculated from \eq\nr{rl2}.  We can observe that the relation
\nr{zbrok} is valid to good accuracy.

\begin{figure}[t,b]
\vspace*{-1cm}
\centerline{\hspace{-3.3mm}
\epsfxsize=10cm\epsfbox{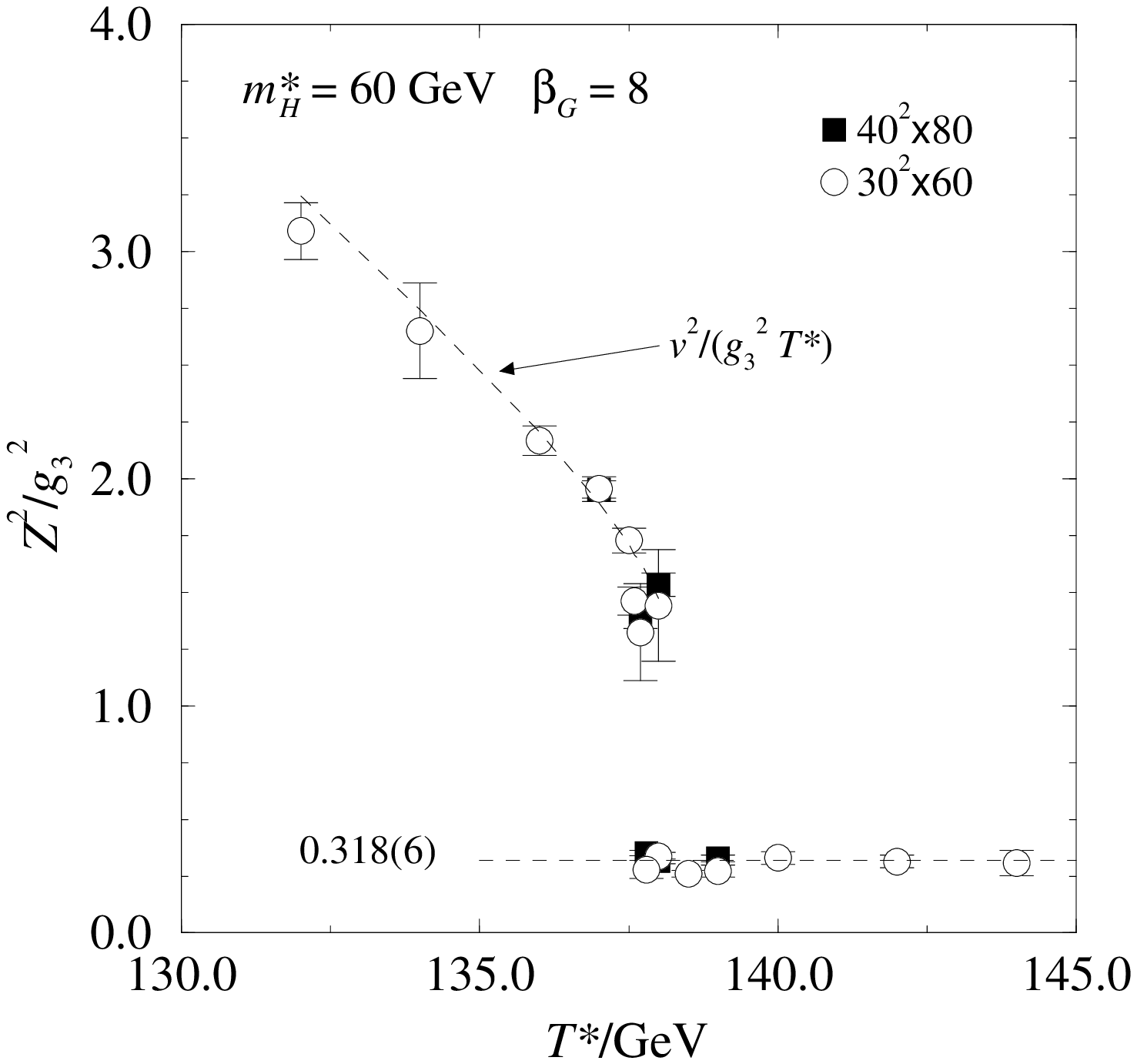}
\hspace{-2cm}
\epsfxsize=10cm\epsfbox{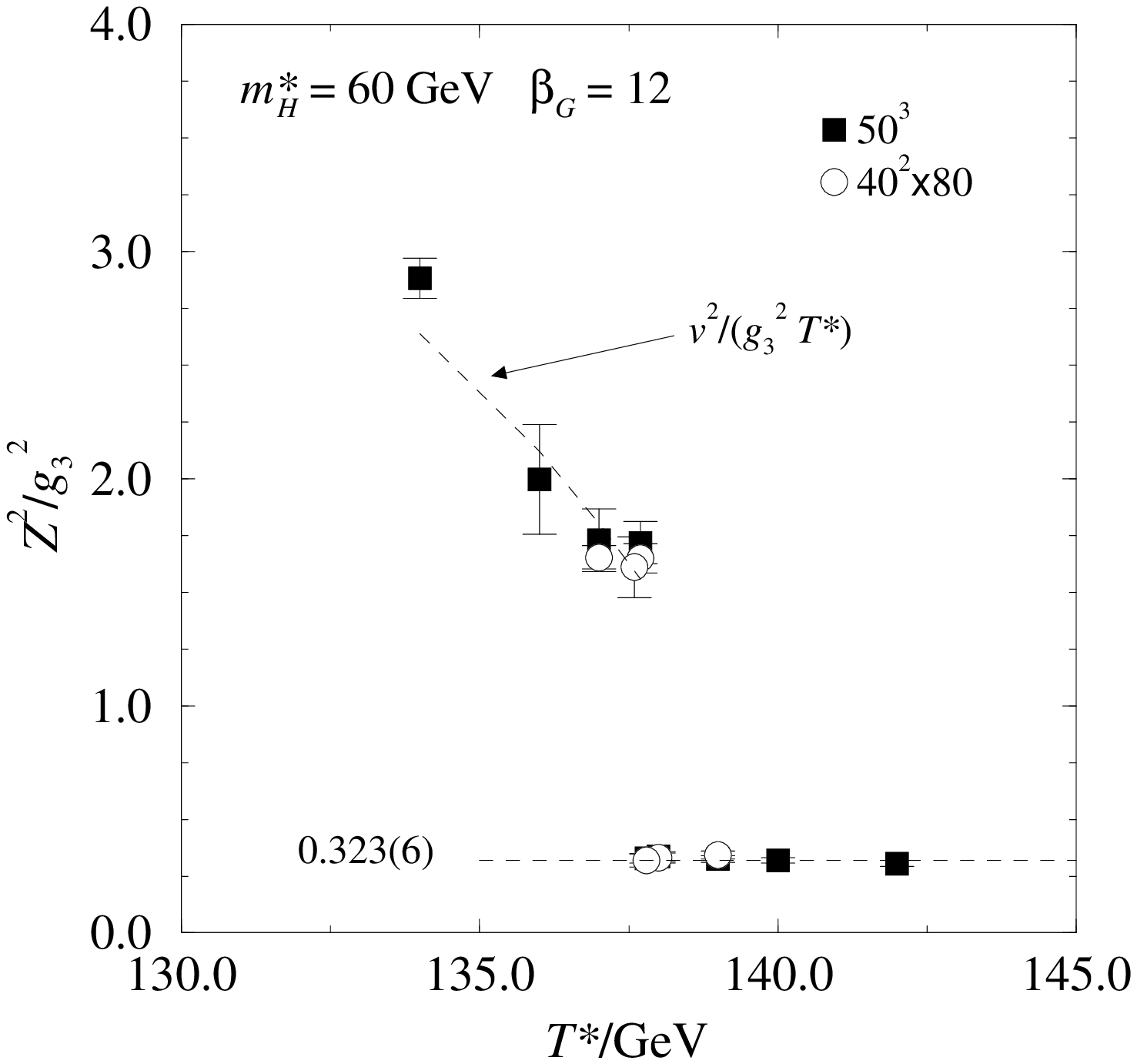}}
\vspace*{-5cm}
\caption[a]{The wave function normalization of the scalar field near
the phase transition for $m_H^*=60$ GeV and $\beta_G=8$, $12$.  The
horizontal lines are constant fits to the symmetric phase data.  The
dashed curves labeled $v^2/g_3^2T^*$ connect the MC datapoints for that
quantity.\la{wave-norm}}
\end{figure}

{\bf The potential.~~} The potential part $U(\pi)$
of the effective theory can
be derived from the probability distribution $p(R^2)$ using
the relation \nr{prob} and by fitting the parameters
 $R^2_{\rm min}$ and $V_i$ in \nr{effR}\footnote{This method
is known as the constrained effective
potential \cite{tsypin}.}.
Using again the $m_H^*=60$~GeV, $\beta_G=8$ data we are able to determine
the three parameters $R^2_{\rm min}$, $V_2$ and $V_3$; the statistical
accuracy
does not allow one to fix $V_4$ with reasonable precision.  The results
of the fits are shown in \fig\ref{Vcoeff}.

\begin{figure}[t,b]
\vspace*{-1cm}
\centerline{\hspace{-3.3mm}
\epsfxsize=8.6cm\epsfbox{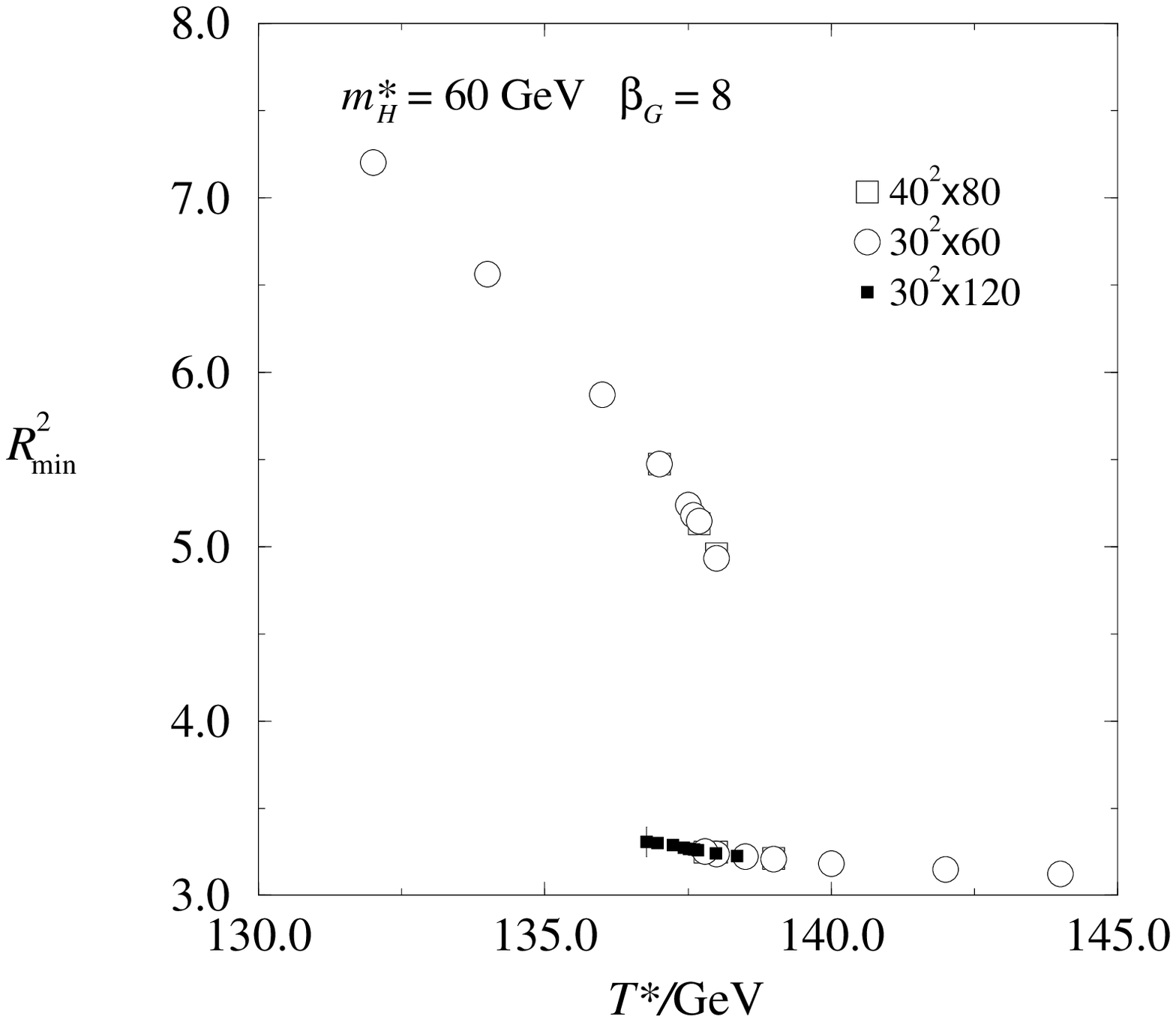}
\hspace{-1cm}
\epsfxsize=8.6cm\epsfbox{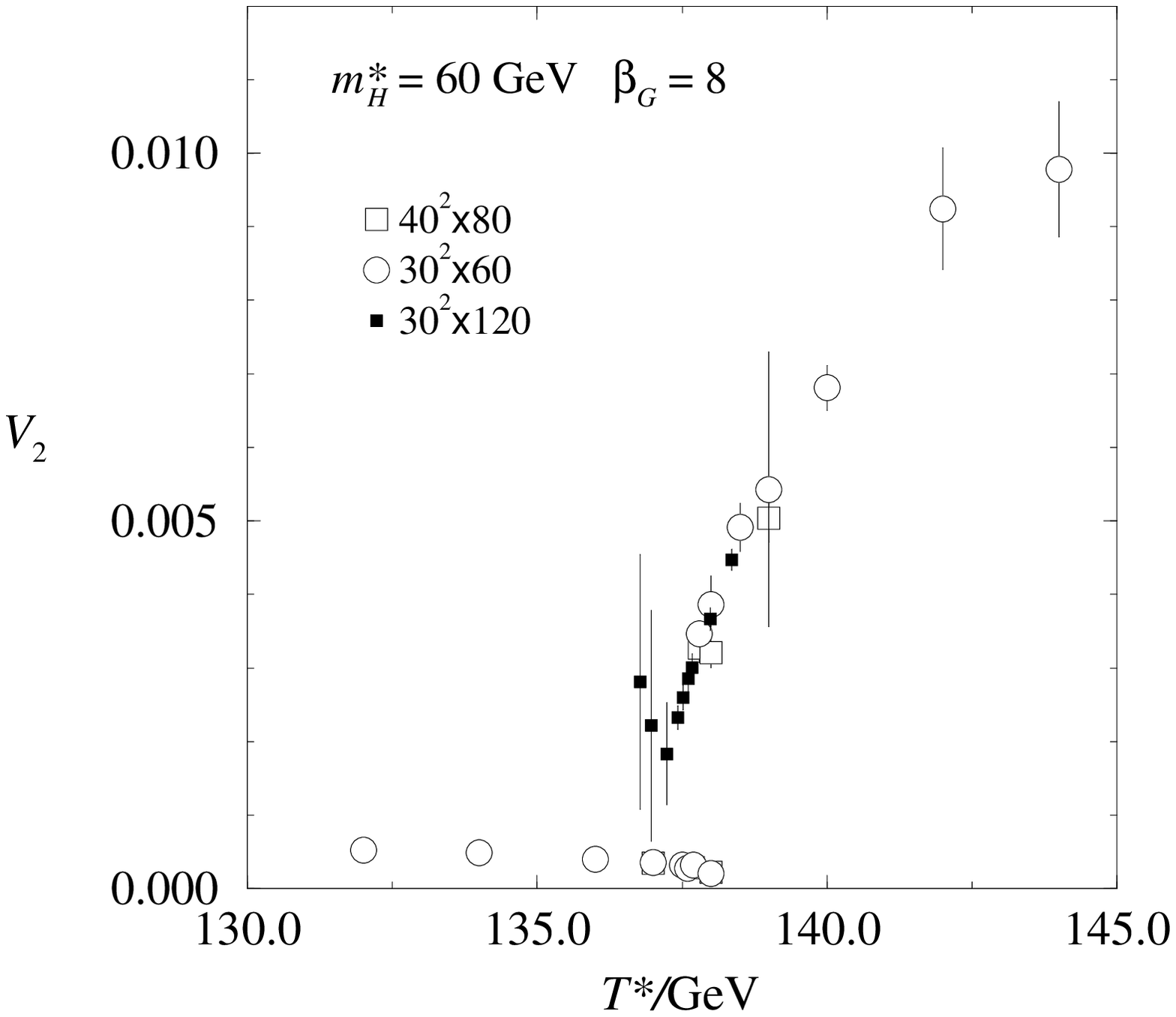}}
\vspace*{-5cm}
\centerline{\hspace{-3.3mm}
\epsfxsize=8.6cm\epsfbox{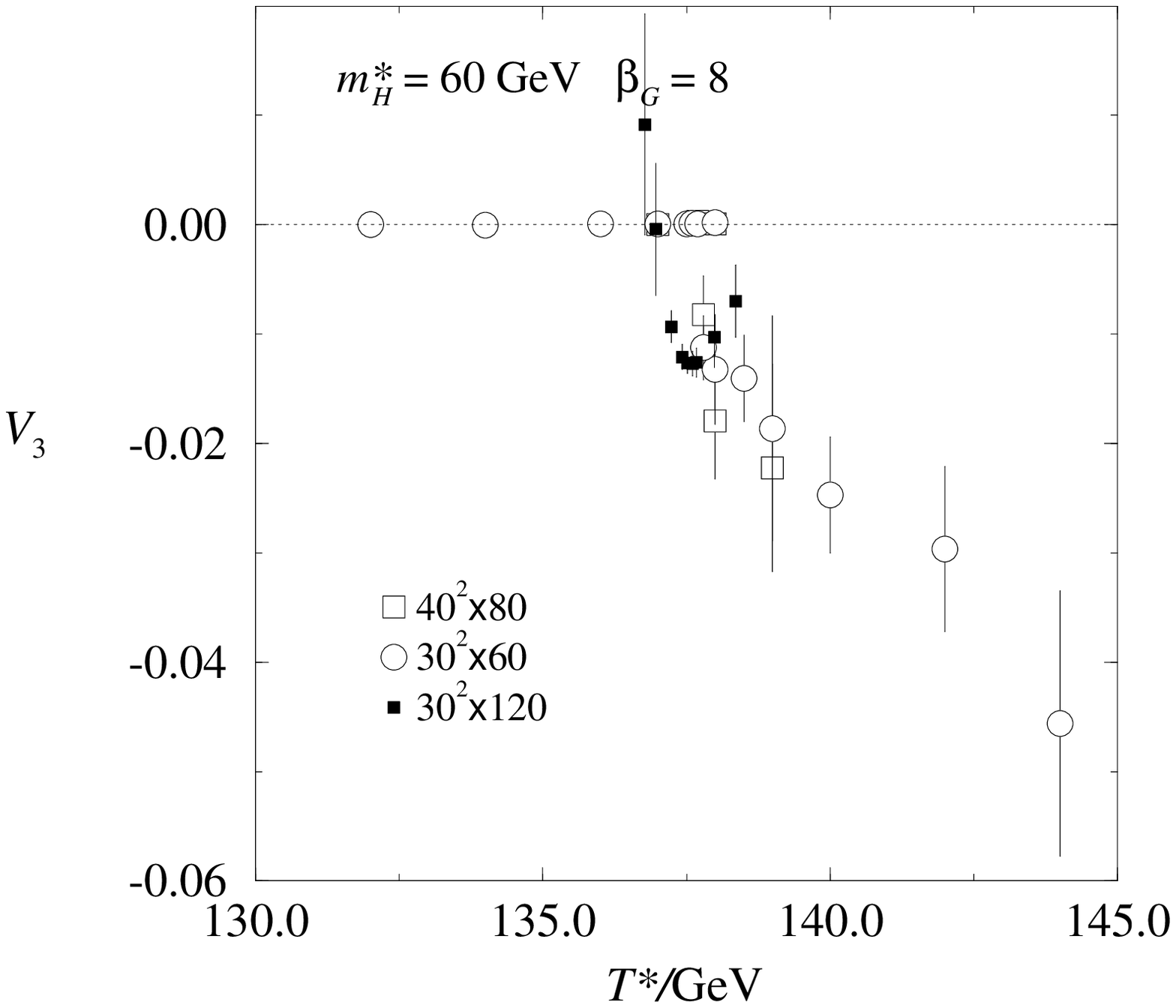}
\hspace{-1cm}
\epsfxsize=8.6cm\epsfbox{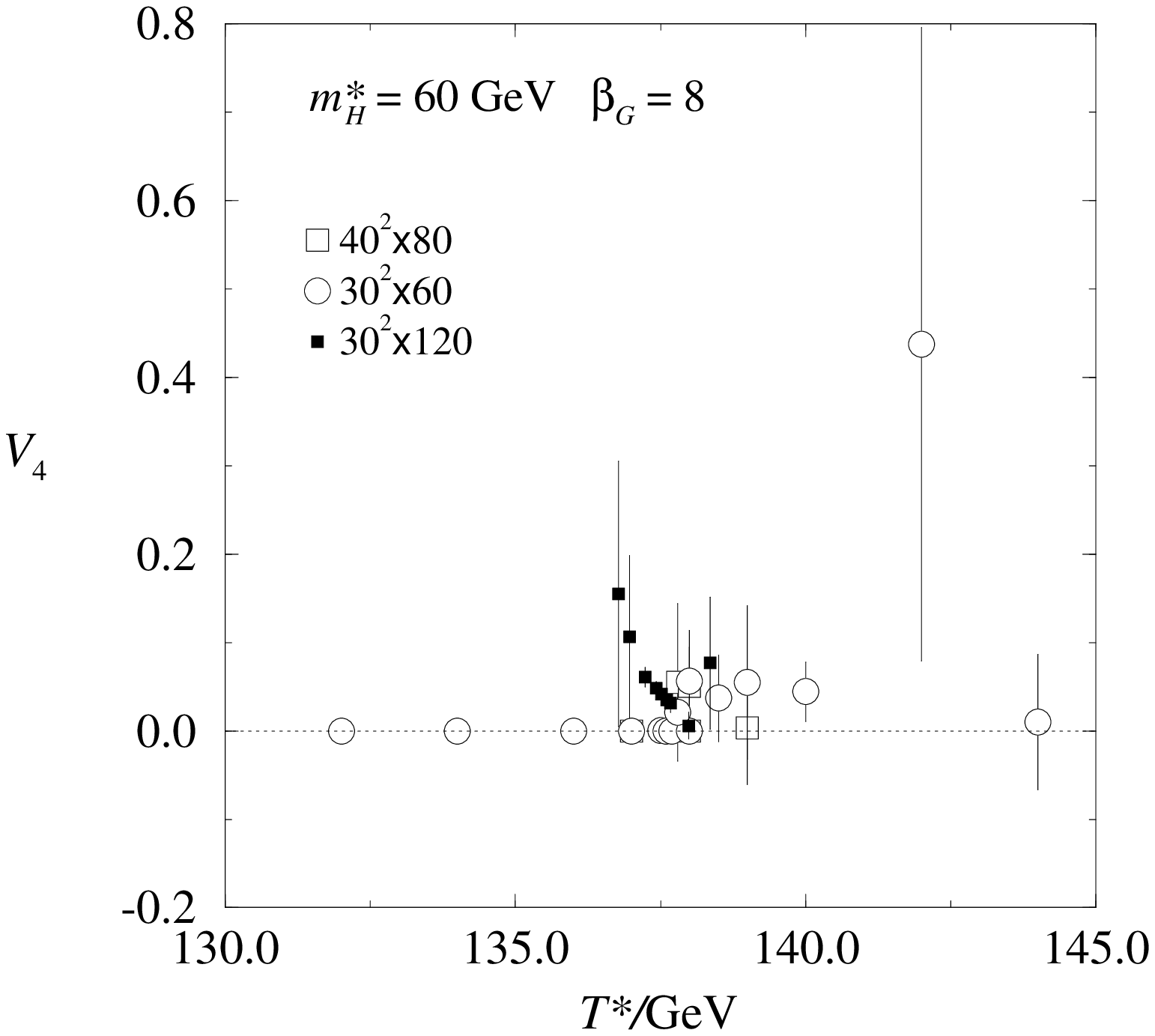}}
\vspace*{-4.2cm}
\caption[a]{The lattice results for the effective potential
coefficients, determined from $m_H^*=60$~GeV, $\beta_G=8$
systems.\la{Vcoeff}}
\end{figure}

There is an important consistency check for the validity of the
effective theory description: using eqs.~(\ref{z2}) and (\ref{V2})
the mass of the scalar particle is
\be
  m_H^2 = \frac{8 V_2 Z^2}{a \beta_H^2}.
\ee
Substituting $Z^2$ from (\ref{z2}) this becomes
\be
   4 A (\xi_H/a) V_2 = 1\,.
\la{coch}
\ee
Here $\xi_H/a$ is the Higgs correlator in lattice units. Eq.~\nr{coch}
establishes a non-trivial connection between the different methods
of mass determination.  The first method is the direct measurement
via correlation functions and the second one is through the potential
curvature.  The lattice data for $4 A (\xi_H/a) V_2$ is presented in
\fig\ref{effac}.  It is seen that the prediction
of eq.~\nr{coch} is satisfied
reasonably well in both phases.
\begin{figure}[t,b]
\vspace*{-1cm}
\centerline{\hspace{-3.3mm}
\epsfxsize=10cm\epsfbox{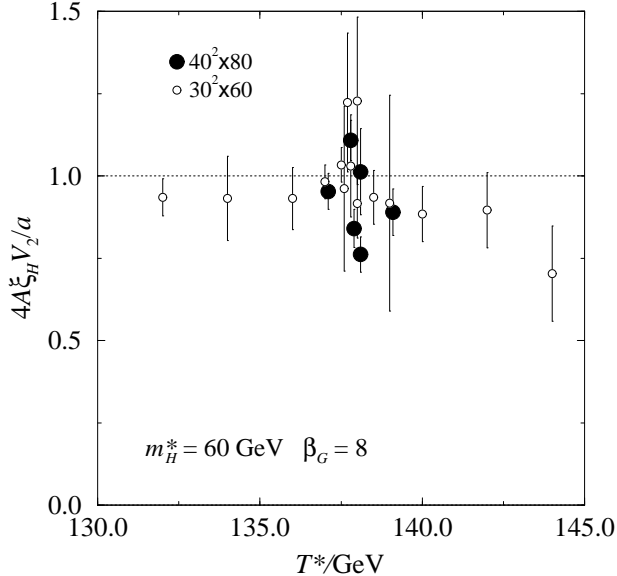}}
\vspace*{-5cm}
\caption[a]{The lattice results for the test of the effective theory
description.\la{effac}}
\end{figure}

Let us turn now to the discussion of the behaviour of the
coefficients $V_2$ and $V_3$. When the temperature decreases
(increases), $V_2$ in the symmetric (broken) phase rapidly
decreases, in fact repeating the behaviour of the scalar mass from
the correlation function measurements. This gives another
method for the metastability range determination.  We
write
\be
  V_2^i \propto |T-T_i|^{\gamma_i}
\ee
and fit $T_i$ and $\gamma_i$ separately for symmetric and broken phases.
In the symmetric phase, the results are
\be
\begin{array}{rcl}
\beta_G = \n 8\,: & T_{-} = 137.0(5)\,\mbox{GeV}\,. & \gamma_s = 0.51(14)  \\
\beta_G =  12 \,: & T_{-} = 137.3(5)\,\mbox{GeV}\,, & \gamma_s = 0.47(16).
\end{array}
\ee
These values are quite consistent with the numbers in
Sec.~\ref{sec:met-reweight}.
However, in the broken phase we could not
obtain acceptable fits.

\begin{figure}[t,b]
\vspace*{-1cm}
\centerline{\hspace{-3.3mm}
\epsfxsize=9cm\epsfbox{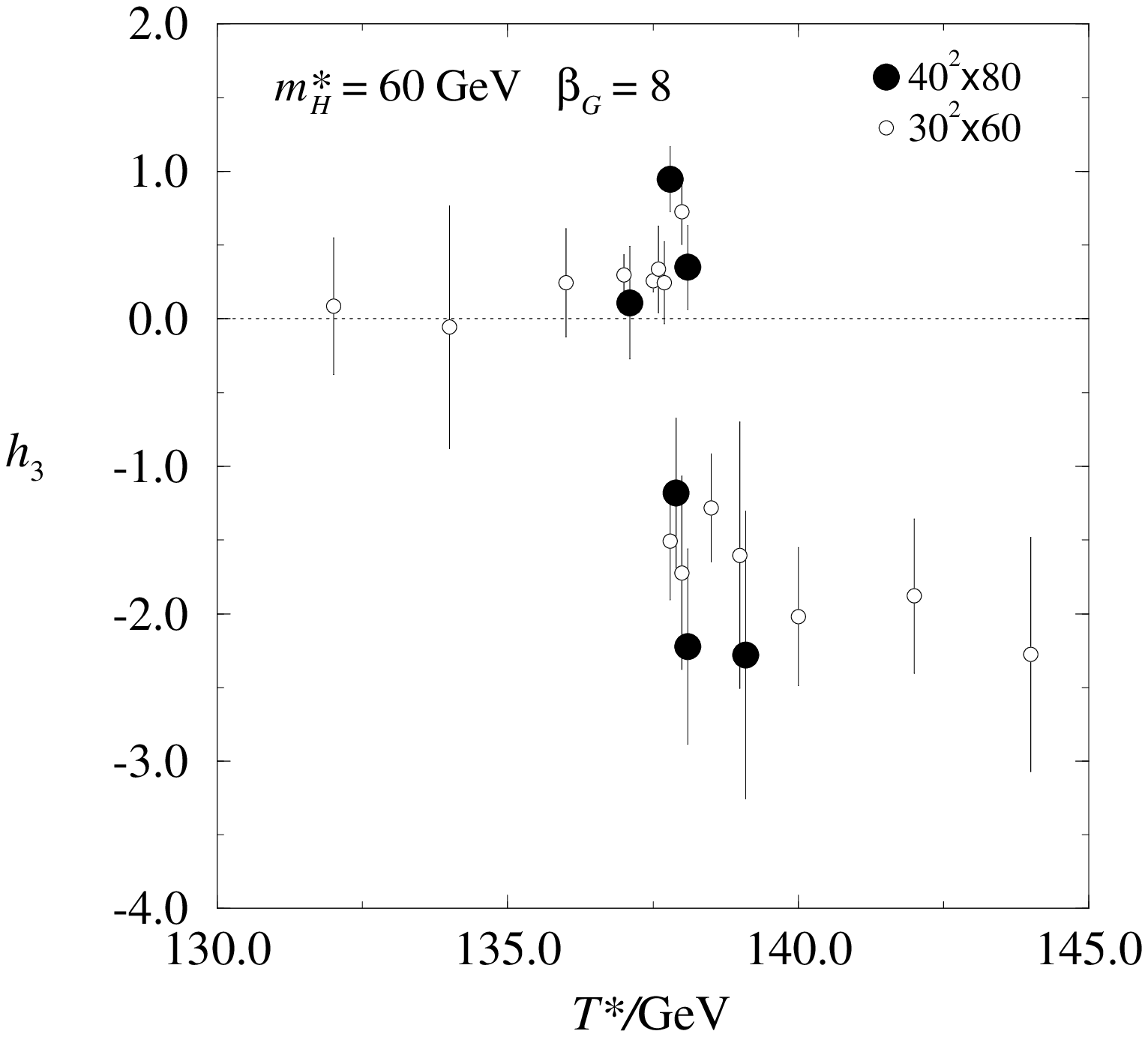}
\hspace{-1cm}
\epsfxsize=9cm\epsfbox{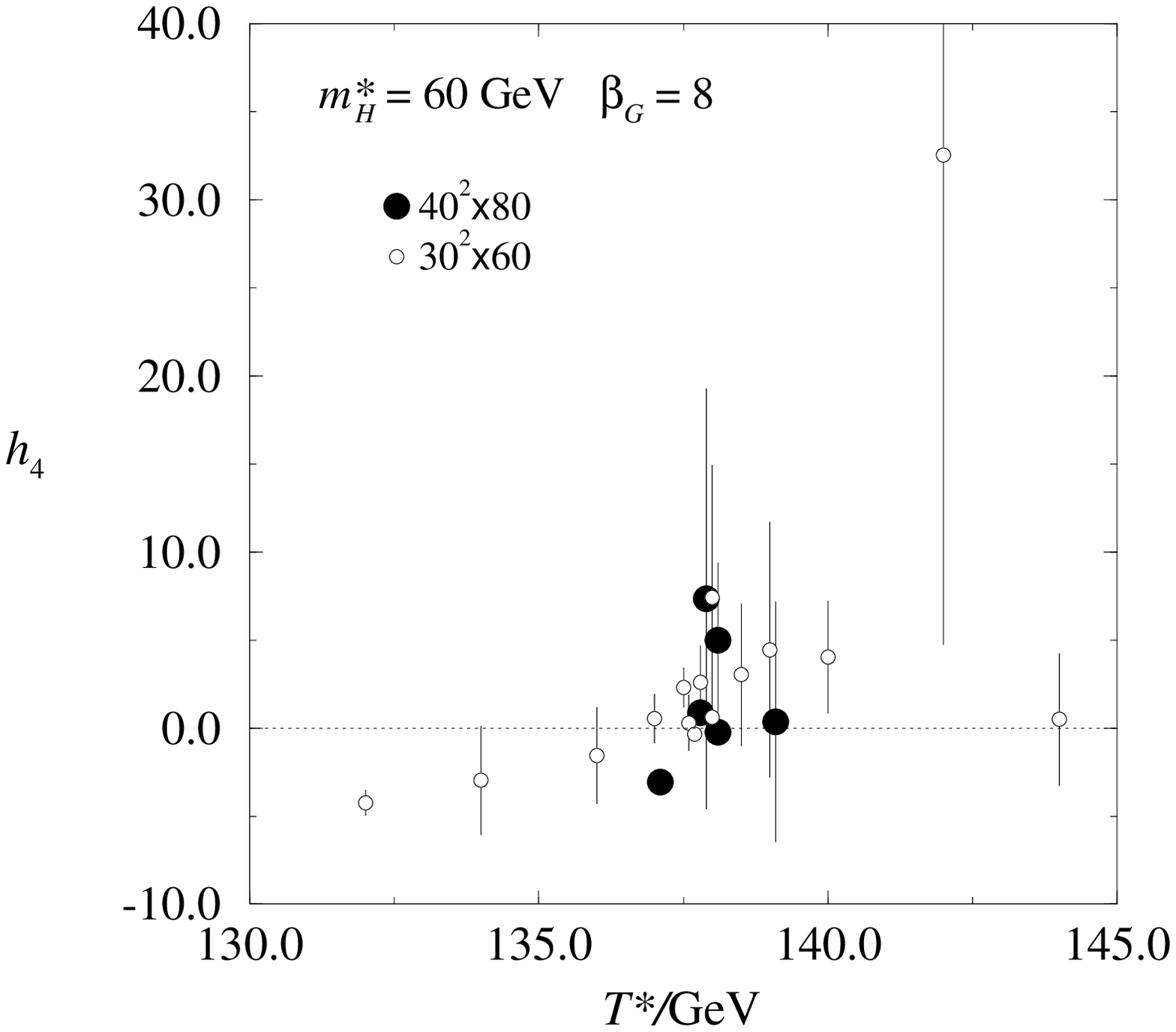}}
\vspace*{-4.3cm}
\caption[a]{The temperature dependence of the effective coupling
constants $h_3$ and $h_4$.\label{effcoupl}}
\end{figure}

It is interesting to note that the coefficient $V_3$ has a strong
dependence on temperature in the symmetric phase. This suggests
that the effective potential $U(\pi)$ has a scale-invariant form in
the vicinity of the lower spinoidal decomposition point. If we
introduce the dimensionless variables
\be
  \tilde\pi = (\pi-\pi_{\rm min})/(Z \sqrt{m_H}),\h \tilde x = x m_H,
\ee
then the action can be written in the form
\be
  S=\int d^3\tilde{x} \left[\fr12(\tilde{\partial}\tilde{\pi})^2 +
  \fr12 \tilde{\pi}^2 + h_3\tilde{\pi}^3 +
   h_4\tilde{\pi}^4+ \ldots \right] \,,
\ee
where the dimensionless couplings $h_3$ and $h_4$ are expressed via
lattice observables as
\be
  h_3=V_3 (2 A)^{3/2},\h h_4=16 V_2 V_4 A^3.
\ee
Now, if $h_3$ and $h_4$ do not depend on the temperature, then the
system has scale-invariant behaviour near the point of absolute
metastability. The lattice data is shown in
\fig\ref{effcoupl}.  Indeed, $h_3$ is reasonably temperature
independent, while the quality of the data does not allow to make a
definite conclusion on the higher interactions. The values of the
coupling constants are completely different in distinct phases; they
are larger in the symmetric phase, showing that interactions are
strong there.

\section{Simulations with $A_0$ field, $m_H^*=80$~GeV\la{sec:80}}

In \cite{fkrs2} Monte Carlo studies of the
$m_H^*=80$~GeV system, using a lattice action which includes the
adjoint Higgs field $A_0$, were reported. The $A_0$ field, a
remnant of the timelike component of the gauge field, is a heavy field
of mass $\sim gT$, and has been integrated out in
the theory of eq.~\nr{lagr}.
The lattice action is
\ba
S &=&
S[U,\Phi] + \fr12\beta_G \sum_{x,i} [ \tr
A_0(\bfx)U_i^\dagger(\bfx)A_0(\bfx+i)U_i(\bfx) - \tr A_0^2(\bfx) ]
\nonumber \\ &+& \beta_2^A \sum_x \fr12\tr A_0^2(\bfx) + \beta_4^A
\sum_x [\fr12\tr A_0^2(\bfx)]^2 \la{a0-action} \\ &-& \fr12 \beta_H
\sum_x [ \fr12\tr A_0^2(\bfx) \fr12\tr \Phi^\dagger(\bfx)\Phi(\bfx) ]
\nonumber
\ea
where $S[U,\Phi]$ is the action given in
\eq\nr{lagrangian}, and the lattice field $A_0$ is given in terms of
the continuum field $A_0^a$ as $A_0 = \fr{i}{2}g_3 a \tau_a A_0^a$.  The
parameters $\beta^A_2$ and $\beta^A_4$ can be written in terms of the
4d continuum variables (see \eqs(II.15--19) and (II.97--99)).

\begin{figure}[tb]
\figtopspace \hspace{1cm}
\epsfysize=\figysize
\centerline{\epsffile{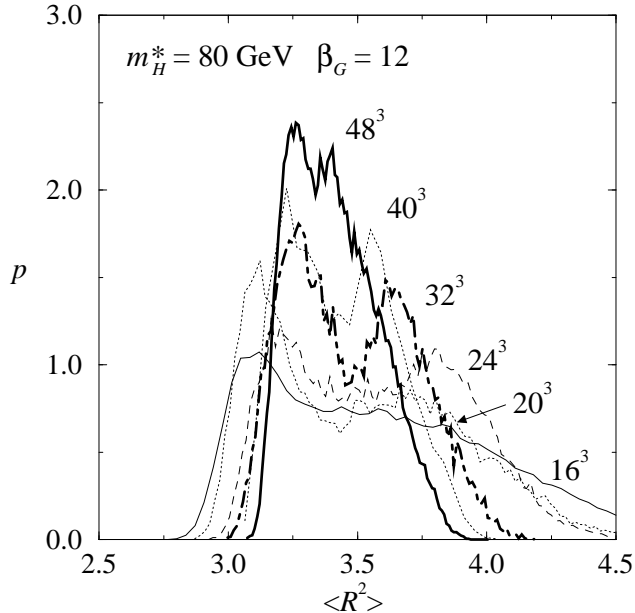}}
\figbottomspace
\caption[a]{The $p(R^2)$ distributions for $m_H^*=80$~GeV, $\beta_G=12$
lattices.  The histograms are evaluated at $\beta_H=0.34771$.  The
double-peak structure of the histograms becomes less pronounced when
the volume becomes larger than $32^3$.\la{m80-b12-Rhg}}
\end{figure}
\begin{figure}[t,b]
\vspace*{-1cm}
\centerline{\hspace{-3.3mm}
\epsfxsize=9.6cm\epsfbox{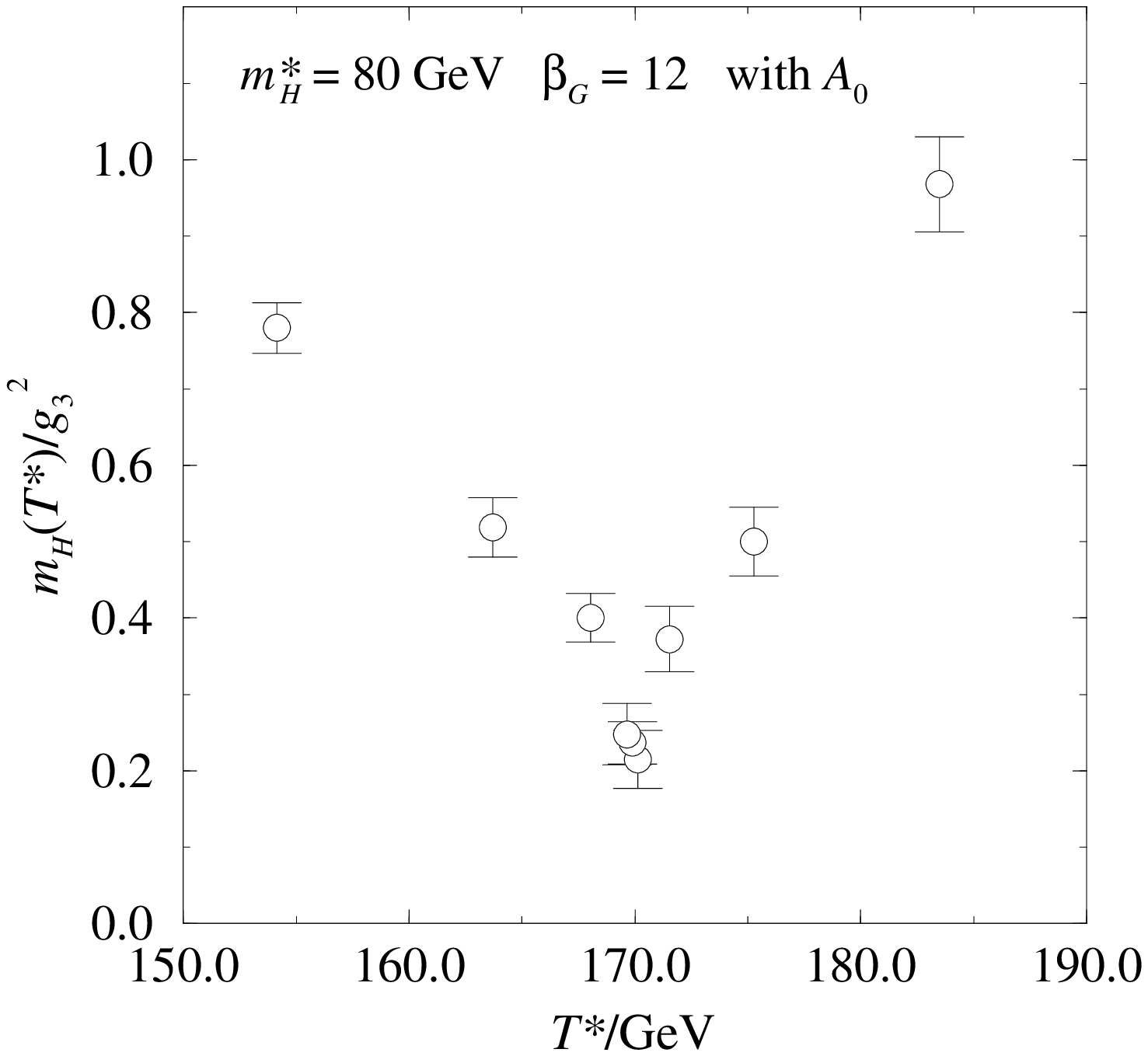}
\hspace{-2cm}
\epsfxsize=9.6cm\epsfbox{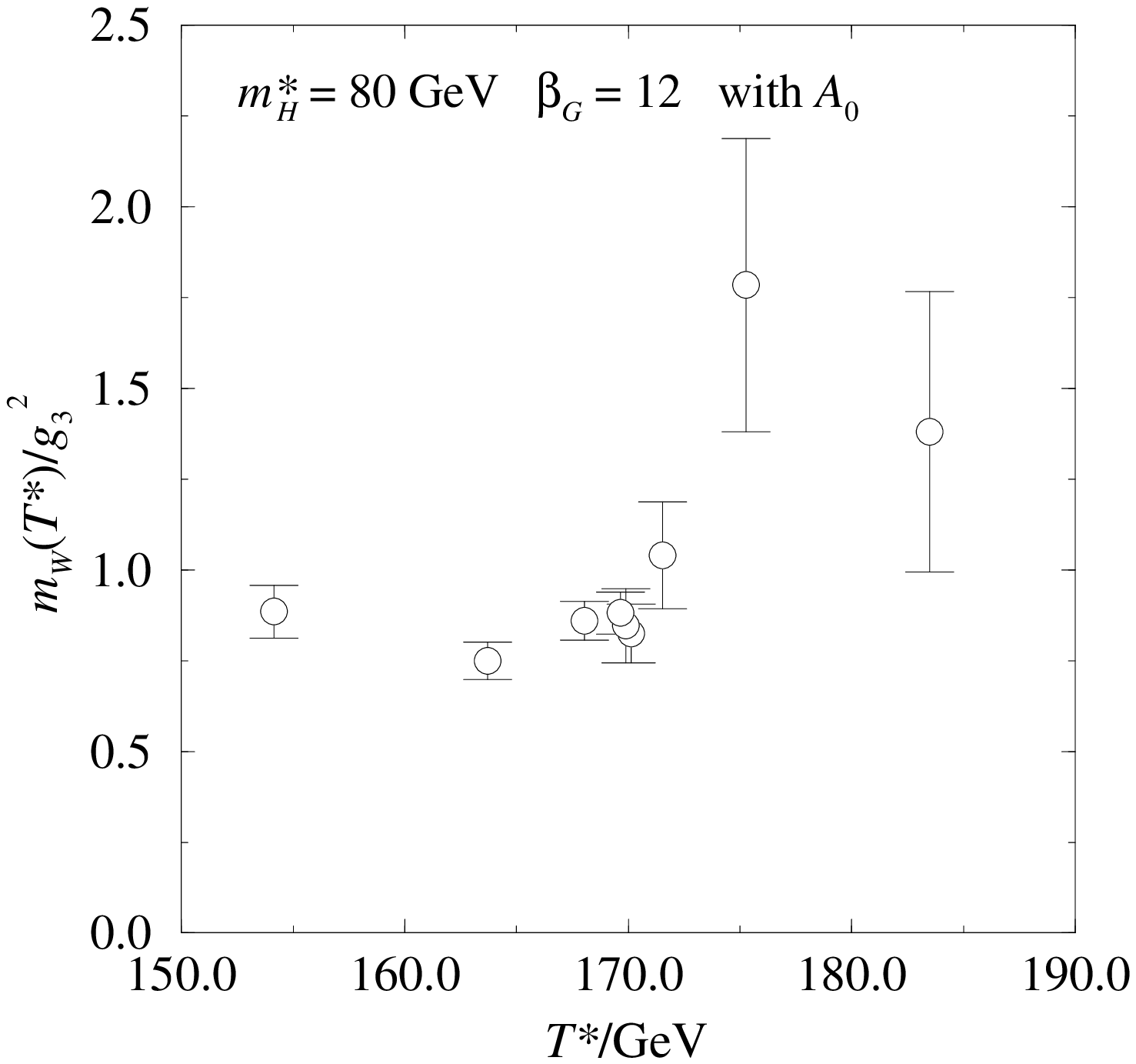}}
\vspace*{-4.6cm}
\caption[a]{The Higgs and $W$ masses of $m_H^*=80$~GeV, $\beta_G=12$,
and $V = 40^3$ lattices around the transition temperature.\la{mh80-mhmw}}
\end{figure}

In \cite{fkrs2} lattices of sizes up to $32^3$, with $\beta_G=12$ and
20, were studied.  Clear double-peak histograms were observed for both
values of $\beta_G$, signaling a first order transition.  However,
since the publication of \cite{fkrs2} we have performed simulations
using lattice volumes up to $48^3$.  Surprisingly, with increasing
volume the double-peak feature of the histograms becomes {\em less\,}
pronounced, in contrast to the expected behaviour in first order
transitions (see \fig\ref{m80-b12-Rhg}).  The most striking finite
volume effect is the shift of the peak positions towards each other
when the volume is increased.  Qualitatively similar finite volume
behaviour occurs also in several other systems, which can exhibit
either first or second order phase transitions in the infinite volume
limit (for example, the 2-dimensional 7-state and 4-state Potts
models, which have respectively first and second order transitions
\cite{Borgs90,potts}).  Thus we cannot yet make definite conclusions
about the order of the phase transition.

However, a note about the computer simulations is in order: the action
\nr{a0-action} used in these simulations includes the $A_0$ field, and
because of the $A_0$ hopping term in the action it is difficult to
write an efficient overrelaxation algorithm.  The algorithm we use
here uses heat bath and Metropolis updates, and it performs more than
an order of magnitude worse than the overrelaxation described in
Sec.~\ref{MC}.  Therefore, the statistical accuracy of the
$m_H^*=80$~GeV results is not nearly comparable to the $m_H^*\le
70$~GeV results, described in Secs.~\ref{pt}--\ref{sec:correlation}.
There are also more systematic uncertainties than
in the simulations in Secs.~\ref{pt}--\ref{sec:correlation},
since the relations of lattice and $\msbar$ schemes were not
fully known at the time of the $m_H^*=80$ GeV simulations.

In \fig\ref{mh80-mhmw} we show the Higgs and $W$ masses
around the transition temperature.  To the accuracy of our data,
we do not observe any discontinuities at the transition, as opposed
to the $m_H^* = 60$~GeV masses in \fig\ref{m60-masses}.  However,
the errors here are about a factor of $5$ larger.  In the immediate
vicinity of the transition, the ratio
of the W and Higgs masses is $m_W/m_H \approx 4$.

We measure the critical coupling $\beta_{H,c}$ for each individual
volume as described in Sec.~\ref{pt}.  The infinite volume
extrapolation is done using volumes larger than $20^3$, with the
results $\beta_{H,c} = 0.347715(8)$ for $\beta_G=12$, and $\beta_{H,c}
= 0.341700(10)$ for $\beta_G=20$.  In contrast to simulations with
$m_H^*\le 70$~GeV, in these simulations $m_H^*$ was not kept constant,
but instead the coupling constant $\beta_R$ was fixed to values
$\beta_R = 0.00124$ and $\beta_R = 0.000712$, respectively.  Using the
relation $m_H^{*2}/m_W^2 = 8\beta_R\beta_G/\beta_H^2$, these
correspond to $m_H^* = 79.98$ and 79.62~GeV\@.  The transition
temperatures and the linear extrapolations to the continuum limit are
given in table \ref{tab:m80results}. The numbers are somewhat
different from those presented in~\cite{fkrs2}. The discrepancy
is due to an error in the 2-loop term of the constant
physics curve in~\cite{fkrs2}.

\begin{table}[hbt]
\center
\begin{tabular}{|c|c|l|}
\hline
$\beta_G$ & $m^*_H$/GeV & \cen{$T^*_c$/GeV} \\
\hline
  12      & 79.976(1)       & 170.0(2)  \\
  20      & 79.614(2)       & 169.8(7)  \\
$\infty$  & 79.072(6)       & 169.4(17) \\
\hline
\end{tabular}
\caption[0]{Results from $m_H^*\approx 80$~GeV simulations, together
with the extrapolations to the continuum limit.
The perturbative critical temperature corresponding to
$m_H^*=79.072$ GeV is about 171.5 GeV.\la{tab:m80results}}
\end{table}

The continuum limit $T_c^*$ is somewhat below the perturbative
value; however, the statistical errors are too large for the
difference to be significant.  To resolve the situation at
$m_H^*=80$~GeV requires simulations performed without the $A_0$
field. Let us note that the
value of $x$ corresponding to $m_H^*=80$ GeV is  $x_c=0.1188$,
and the physical Higgs mass in the Standard Model corresponding
this $x_c$ is about 81~GeV.

\section{What happens at large Higgs masses}
\la{sec:largemH}
The phase transition in the 3d SU(2)+Higgs system becomes weaker and
weaker with increasing scalar self-coupling $x\sim m_H^{*2}$. An important
question is whether the transition for some value of $x$ terminates in a
2nd order transition or whether it continues as a 1st order one for
all $x$.

The main difficulty in resolving the order of the phase transition at
large scalar self-coupling is the presence of the two
distinct physical mass scales (inverse correlation lengths,
$m(T)=1/\xi$).
In fact, inspection of \fig\ref{m60-masses}
shows that for $m_H^*=60$ GeV  near the transition point,
\be
  \frac{m_H(T)}{m_W(T)}|_\rmi{symmetric}\approx
\frac{m_H(T)}{m_W(T)}|_\rmi{broken}\approx
  \frac{1}{3}.
\ee
Furthermore, \fig\ref{mh80-mhmw} indicates that this ratio is even
smaller for larger $m_H^*$, and data for smaller $m_H^*$ indicates
that it is larger for them.
Note also that vector masses do not seem to go to zero at the points
of absolute metastability defined in Sec.~\ref{sec:metastab}.

Given the observed hierarchy
$\xi_H>\xi_W$, the lattice spacing must be much smaller than the vector
correlation length, while the lattice size must be much larger than the
scalar correlation length. This puts stringent constraints on the
lattice size and makes definite numerical conclusions more and
more difficult with increasing $m_H$ (see the
discussion of the simulations with $m_H^*=80$ GeV in
Sec.~\ref{sec:80}).

However, if the ratio $m_H(T)/m_W(T)$ continues decreasing with
increasing $m_H$, a solution to the problem immediately suggests
itself: the vector degrees of freedom become relatively heavy, decouple,
and should be integrated out. From the effective 3d SU(2)+Higgs
theory one thus at large $m_H$ proceeds to an effective 3d scalar theory
with just one degree of freedom.
The transition to the scalar theory cannot be carried out by integrating
out $A_i$ perturbatively, and we first have to discuss the relevant degrees
of freedom in more detail.

The 3d Euclidean SU(2)+Higgs theory originating from dimensional
reduction corresponds to a (2+1)-dimensional
Minkowskian SU(2)+Higgs
theory. This theory has an analogy with QCD: we have doublets
of scalar ``quarks'' bound together by an SU(2) triplet of ``gluons''. We
have chosen a Minkowskian language  in order to
keep the analogy with QCD transparent.

Then, one would expect that the lowest lying particle states in the
symmetric phase are the spin zero scalar bound state and the three
degenerate vector bound states, with quantum numbers corresponding to the
composite operators
\be
\pi = \phi^{\dagger}\phi,
\ee
\be
W_j^{0}=i(\phi^{\dagger}D_j\phi-(D_j\phi)^{\dagger}\phi),\quad
W_j^{+}=i(\phi^{\dagger}D_j{\tilde\phi}-
(D_j{\tilde\phi})^{\dagger}\phi),\quad
W_j^{-}=(W_j^{+})^*,
\la{operators}
\ee
correspondingly. Here $\tilde{\phi}=i\tau_2\phi^*$.
Note that these operators provide also a
gauge-invariant description of the Higgs particle and intermediate
vector bosons in the broken phase. In other words, the particle
degrees of freedom are the same in both phases.
This fact alone suggests that there is no distinction between the two
phases, so that in some region in the parameter space \cite{banks,shenker}
there may be no phase transition at all. However, it does not exclude
the situation that going from one phase to another is only possible
through a phase transition.

The most general renormalizable local scalar field theory in 3d has
the action
\be
L=\frac{1}{2}(\partial_i\chi)^2 + P(\chi)
\ee
where $P(\chi)$ is a sixth order polynomial which contains five constants
(a linear term can always be removed by the shift of the scalar
field)\footnote{Note that there is no symmetry (discrete or
continuous) which forbids odd powers of $\chi \sim Z^{-1}\pi$.}:
\be
P(\chi)= \sum_{i=2}^{6}\frac{c_i}{i}\chi^i.
\ee
One of the constants (say, $c_4$) fixes the scale,
and the other four (say,
$c_2/c_4^2, c_3^{2}/c_4^3,c_5^2/c_4$ and $c_6$) are dimensionless
numbers completely fixing the dynamics of the theory. If the
assumption on the effective theory is correct, then all the four
parameters are some functions of the dimensionless ratios $x$ and $y$
characterizing the SU(2)+Higgs theory. In the vicinity of the phase
transition, the first of the ratios ($c_2/c_4^2$)
is fixed, and there is a mapping
of the three last ones to the single parameter $x$ of the gauge-Higgs
system at $T=T_c$.
Clearly, these relations cannot be computed in perturbation theory,
since at large $m_H^*$ the phase transition occurs in the strong
coupling regime. After they have been found by some nonperturbative
means, resolving the order of the phase
transition on the lattice at large $m_H$ becomes a much easier
task, because only one essential scale is present.

Finally, note that the action proposed here has a different number of
degrees of freedom, different symmetries, and a different range of validity
than the scalar action used in the simulations of~\cite{knp}.

\section{What did we learn from the lattice simulations?}
\la{sec:comparison}

The non-perturbative information we acquired by doing lattice
simulations can be confronted with different hypotheses on the nature
of the phase transition existing in literature. It can be used as
a test of validity of a number  non-perturbative approaches, such as
the $\epsilon$-expansion~\cite{ay},
the renormalization group approach~\cite{rw}, and
the Dyson-Schwinger equations~\cite{bph}.
In this section we
provide a summary of our findings and compare them with some
of the mentioned approaches.

\subsection{Comparison with perturbation theory}
\la{cwpt}

Perturbation theory can be used for the computation of different
quantities in the broken phase. Not many quantities are known to
high order in perturbation theory in the 3d SU(2)+Higgs theory, though.
The whole list consists of the 2-loop effective potential for the scalar
field in different gauges~\cite{fkrs1,mlaine,kls}
(and the values of different condensates
related to the effective
potential, see~\cite{fkrs2}), and the 1-loop pole masses of the gauge
and Higgs particles (see~\cite{bph} and below).

{\bf The effective potential.} If the temperature is fixed to some
value, the vev of the Higgs field,
defined by eq.~\nr{defvoT}, can be computed with the use of the
2-loop effective potential. At the same time, the ``exact'' vev of
the Higgs field can be found on the lattice (after appropriate
extrapolation to the continuum limit). The 2-loop effective
potential gives a reasonably good prediction for $v/T$. For example,
$\delta v/v \sim 0.3\%$ for the vev of the Higgs field when $v/T\simeq 2$,
and $\delta v/v \sim 10\%$ for $v/T\simeq 0.64$ (these numbers refer
to $m_H^*=80$ GeV).
Here $\delta v = v_{\rm 2-loop}-v_{\rm exact}$. The exact
vev is found to be {\em smaller} than that from 2-loop perturbation
theory. This proves that the 3-loop linear correction to the
effective potential comes with a positive sign. If the term
in eq.~\nr{3loop} is
added to the 2-loop effective potential with $\beta=50$,
then the agreement between the exact and
3-loop results is within $1\%$ even for $v/T$ as small as 0.6.
In other words, the temperature dependence of the vev of the Higgs field
can now be found {\em analytically} with 1\% accuracy
up to rather small values of $v$. This result is even
more non-trivial than a direct analytical 3-loop computation
of the effective potential, because it indicates
that higher loop corrections are indeed small in this region of vevs.

The determination of 3-loop corrections to the effective potential
allows one to get a better idea of the expansion parameter of the
SU(2)+Higgs theory.
As we have discussed in~\cite{fkrs1}
it is difficult to get a reliable estimate
of the expansion parameter for the effective potential
on the basis of 2-loop perturbative
computations. The expansion parameter (see Sec.~\ref{cwpt})
is proportional to $g_3^2/(\pi m_T)=2g_3/(\pi\phi)$, but how big
is the constant of proportionality?
Indeed, the loop expansion
may be reliable in the broken phase
in spite of the fact that the 1-loop correction is
comparable with the tree term, just because the scalar self-coupling
constant is small. Moreover, the magnitude of the 2-loop correction
may be changed by adjusting the scale $\mu$, so that
the 2-loop correction may even be tuned to zero by some choice of
$\mu$. Hence the determination of the 3-loop correction to the
effective potential is important.

Let us take for definiteness $m_H^*=82$ GeV, i.e., $\lambda_3/g_3^2=
1/8$ according to eq.~\nr{lameff}. In addition, consider the region near
$T_c$ (take $m_3^2=0$) and include only gauge field loops (neglect
$m_1$ and $m_2$).
Then the estimates of the different contributions to the
effective potential are
\[
V_0 = \frac{1}{32}g_3^2 \phi^4,
\]
\[
V_1 \simeq V_0\times \frac{g_3^2}{\pi m_T(\phi)},
\]
\be
V_2 \simeq V_0\times\frac{33}{32}\biggl[\frac{g_3^2}
{\pi m_T(\phi)}\biggr]^2
\log\frac{\mu}{\phi},
\ee
\[
V_3 \simeq V_0\times \frac{\beta}{32}\biggl[\frac{g_3^2}
{\pi m_T(\phi)}\biggr]^3,
\]
where $m_T(\phi)=g_3\phi/2$.
Inspection of these relations immediately shows that a reasonable
estimate of the expansion parameter is $\frac{g_3^2}{\pi m_T(\phi)}$.
This quantity is rather large ($\sim 0.8$) in the vicinity of
the phase transition for $m_H^*=70$ GeV, and the convergence of
perturbation theory is expected to be quite bad. The main reason why
the vev of the Higgs field can be found with a pretty high accuracy
is that the effective potential is known up to a high order
in perturbation theory.

{\bf Perturbative correlators in the broken phase.} The next two
quantities characterizing the broken phase are the vector and scalar
correlation lengths.

As in Sec.~\ref{sec:broken}, 
at a finite order in perturbation theory
there are different methods of calculating the correlators.
In principle, the most straightforward is a strict perturbative
calculation in powers of $\hbar$ in the loop expansion.
Such a calculation can be done by shifting the Higgs field
to the classical broken minimum where the Goldstone boson
mass $m_2$ vanishes, and by then calculating all the
1-loop diagrams, including the reducible tadpoles
(the calculation in~\cite{bph} is organized in this way).
This method gives an explicitly gauge-independent results for the
pole masses. Unfortunately, such a calculation gets increasingly
unreliable as one approaches the phase transition, since the
tree-level results $m_1$, $m_T$ for the Higgs and W correlators,
respectively, go to zero. The
same problem was met in Sec.~\ref{sec:broken}
in connection with the effective potential.
One should therefore solve the relevant equations numerically,
hoping that the higher-order corrections
included are the essential ones.
With this method, a small residual
gauge- and $\mu$-dependence remains,
and its magnitude can be used as an estimate of the
consistency of the approximation. We find that
the gauge-dependence of the final results for
the correlators is numerically small,
at most of the order $0.03g_3^2$.
Below we give results in the Landau gauge.
The $\mu$-dependence is also briefly discussed.

To calculate the correlators,
we first solve for the radiatively corrected location
of the broken minimum~$\phi_b$ from the RG-improved 2-loop
effective potential $V(\phi)$~\cite{fkrs1}. The requirement of
RG-improvement used here for the renormalization scale
$\mu(\phi)$ is that the 2-loop contribution
to the derivative of the effective potential,
$\partial_\phi V_2(\phi,\mu)$, vanishes. With $\phi_b$ fixed,
we calculate the inverse propagator, and solve numerically for
the real part of the location of the pole.
Since $\phi_b$ is an approximation to the
exact minimum of $V(\phi)$,
no reducible tadpole diagrams are needed
in the calculation of the propagator in contrast to
the $\hbar$-calculation.
The renormalization scale used in the calculation of
the propagators is chosen to be $\mu(\phi_b)$.
The dependence on $\mu$ is formally
of higher order than the accuracy of the 1-loop calculation,
but is numerically visible, see below.
The details of the calculation are in~\ref{oldC}.

The pole mass obtained from the correlator for
the W particle is shown in \fig\ref{m60-masses}.
In this figure we also show the tree-level value $m_T$.
It is seen that in the broken phase, 1-loop
perturbation theory gives excellent results. This is quite unexpected
since the expansion parameter is rather large at the transition point.
In \fig\ref{m60-masses}, we show also the pole mass of the Higgs
particle, together with the tree-level value. There is
also a third curve: from the known 2-loop
contribution $V_2(\phi)$ to
the effective potential, one
can derive the momentum-independent part
of the 2-loop self-energy: $\Pi^H_2(0)=-V_2''(\phi_b)$.
When this is added to the complete 1-loop result, one gets
the dotted curve. It is seen that the discrepancy between
lattice and perturbation theory is larger than for the W
correlator. The effect of $V_2''$ suggests
that a complete 2-loop calculation of the
Higgs self-energy might give a better estimate.
The importance of 2-loop corrections for the Higgs
particle is in accordance with experience from
the effective potential, see~\cite{fkrs1}.
Finally, let us point out that the tree- and 1-loop $W$
masses, as well as the the Higgs mass with the 2-loop
contribution included, are practically independent of~$\mu$.
For the tree- and 1-loop Higgs masses, the
$\mu$-dependence in varying $\mu$ in the range
$0.5\mu (\phi_b)\ldots 2.0\mu (\phi_b)$
is of the order of $0.03g_3^2$.

Attempts to describe the correlation lengths in the symmetric
phase can be found in \cite{bph,dkls}.

{\bf Parameters of the phase transition}
The characteristics of the phase transition, such as the critical
temperature, latent heat, bubble nucleation rate, surface tension, and
correlation lengths in the symmetric phase cannot be defined in the
perturbative framework only, because the symmetric phase is in
the strong coupling regime.
The failure of perturbation theory is clearly
seen in order by order computations. For example,
the vector 4-loop contribution has a logarithmic singularity at
$\phi=0$. Nevertheless, it is still interesting to compare 2-loop and
3-loop predictions with the results of lattice simulations. The
RG-improved 2-loop effective potential
is real in the vicinity of the critical temperature
for the whole range of~$\phi$ and is regular at the origin. Hence,
the computation of the above-mentioned characteristics of the phase
transition is not faced with formal mathematical difficulties.

The critical temperature of the phase transition derived by 2-loop
perturbation theory is presented in table~\ref{table:critical}. It is
somewhat larger than that derived on the lattice. The
vev of the scalar field at $T_c^*$ for $m_H^*=60,70$ GeV
coincides with the lattice value within errorbars. This perfect
agreement, however, is an incident, since the comparison of vevs is
done at different temperatures. The same is true also for the latent
heat, since it is mainly determined by the value of the scalar
condensate in the broken phase. We do not expect the same coincidence
at larger Higgs masses.

One of the crucial quantities for the computation of the bubble
nucleation rate is the surface tension. For $m_H^*=60$ GeV, its value on the
lattice is considerably (about 3 times)
smaller than that derived
from perturbation theory as
\be
\sigma=\int_0^{\phi_b(T_c)}\sqrt{2V(\phi)}d\phi.
\ee
Note that this equation is valid to leading order only; in
higher orders the wave function renormalization must be taken into
account. However, the huge discrepancy with the lattice
result shows that perturbation theory is not applicable at all
for the computation of the surface tension, at least for
$m_H^*\gsim 60$ GeV.

To summarize, the computation of the characteristics of the phase
transition from 2-loop perturbation theory
(for 35 GeV$<m_H^*<$70 GeV) is accurate for the critical temperature within
$0.8-1.6\%$ and for the ratio $v/T$ within $6\%$. Perturbation
theory fails to describe the surface tension (and, therefore, the
bubble nucleation rate) at least for $m_H^*= 60$ GeV.

Adding to the effective potential
the 3-loop linear term determined in Sec.~\ref{sec:broken}
does not improve the accuracy of perturbative
predictions. We take as an example $m_H^*=60$ GeV and $\beta=50$, as
follows from lattice simulations. The critical
temperature remains roughly as far from the lattice $T^*_c$
as at 2-loop level, being now {\em smaller} than the lattice value,
$T_c^\rmi{3-loop}=136.8$ GeV.
The 3-loop ratio $v^\rmi{3-loop}(T_c^\rmi{3-loop})/T_c^\rmi{3-loop}$
is about $20\%$ larger than the exact
value\footnote{The estimate of $v/T$ in the broken phase
at a {\em given} temperature of course does improve, since the 3-loop part
is determined by just this requirement.
The discrepancy is due to the fact that
the exact and 3-loop critical temperatures
are different, and $v/T$ is strongly temperature
dependent near $T_c$.}. The 3-loop perturbative latent heat
is away from the lattice results by almost $50\%$. The most drastic
deviation is in the surface tension. The 3-loop value of it
is $\sigma/(T_c^*)^3 \simeq 0.018$ -
a factor $7$ larger than the lattice number!
The behaviour of the effective potential in two different situations
is illustrated in \figs\ref{veff1},~\ref{veff2}.
It is clearly seen that perturbation theory
cannot quantitatively describe the phase transition.

\begin{figure}[tb]
\vspace*{0cm}
\hspace{1cm}
\epsfysize=11cm
\centerline{\epsffile{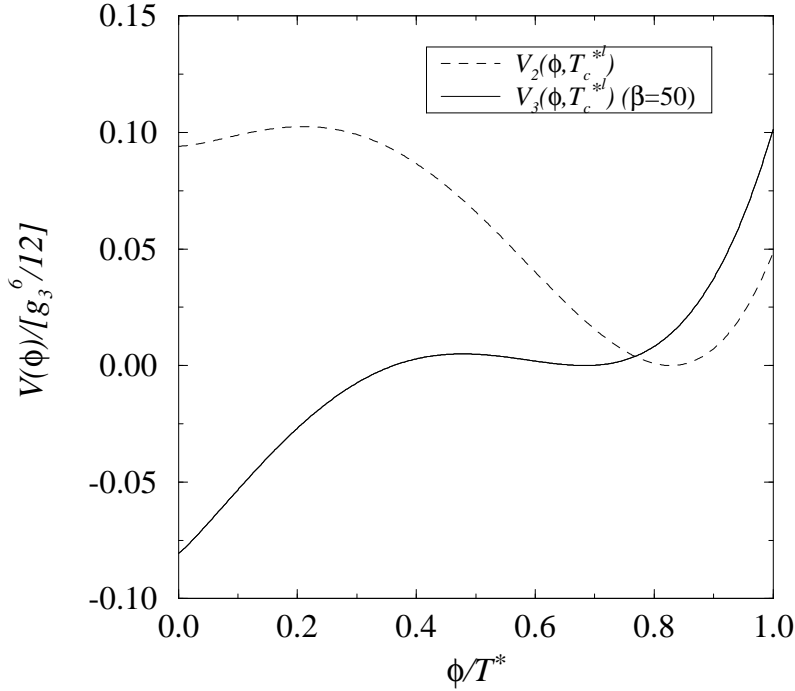}}
\vspace*{-3cm}
\caption[a]{\protect
The RG-improved 2-loop effective potential (dashed line)
and the 3-loop effective potential with $\beta=50$ (solid line)
for $m_H^*=60$ GeV
at the non-perturbative critical temperature $T_c^*=138.38$ GeV.
Note that in this figure we have expressed $\phi$ in
4d units [$\phi_{4d}=\phi_{3d}(T^*)^{1/2}$].}
\la{veff1}
\end{figure}
\begin{figure}[tb]
\vspace*{0cm}
\hspace{1cm}
\epsfysize=11cm
\centerline{\epsffile{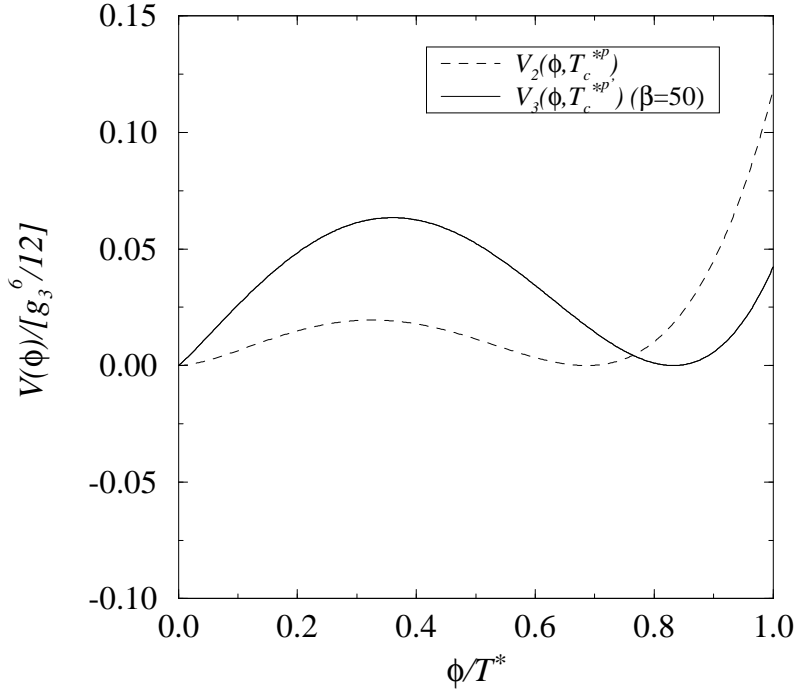}}
\vspace*{-3cm}
\caption[a]{\protect
The 2-loop and 3-loop effective potentials at the
corresponding perturbative critical temperatures,
$T_c^{*\rm p}=$ 140.25 GeV and 136.82 GeV, respectively.}
\la{veff2}\end{figure}

\subsection{Comparison with other non-perturbative approaches}

We were not able to make a detailed comparison of the results of
lattice simulations with all the non-perturbative approaches suggested in
literature. For example, in~\cite{ay} devoted to the application of
the $\epsilon$-expansion, the authors compare their results to
1-loop effective potential predictions. As we argued in~\cite{fkrs1},
1-loop computations in 3d have a considerable unphysical
scale dependence, and this fact makes a direct comparison
difficult. We could not make a comparison with the
renormalization group approach in~\cite{rw}, either,
nor with~\cite{dkls}.
Hence we discuss here~\cite{ms} and~\cite{bph,desy,eqz}.

In \cite{ms} it was suggested that the non-perturbative effects
may considerably modify the effective potential of the Higgs field
near the origin. Assuming that perturbation theory works in the
broken phase, a measure of the non-perturbative energy shift at
$\phi=0$ was introduced,
\be
A_F= \frac{12}{g_3^6}\Bigl[V(0,T_c^*)-V(\phi_b,T_c^*)\Bigr].
\ee
Here $T_c^*$ is the exact value of the critical temperature. It was
conjectured in~\cite{ms} that if $A_F$ is positive then the phase
transition is stronger than predicted by perturbation
theory. In particular, if $A_F$ is sufficiently large (say,
$A_F=0.4$), then the lower metastability temperature is considerably
smaller than the critical temperature, and the vev of the Higgs field
at $T_{-}$ is substantially larger than it is at the critical
temperature. The lattice simulations allow to check the validity of
this hypothesis. Let us take for definiteness $m_H^*=60$ GeV. Then,
$A_F$ is to be found from 3-loop perturbation theory (2-loop
perturbation theory gives a result off by $20\%$ for the vev of the
Higgs field and therefore is not to be used). We get $A_F\simeq
-0.08$, see \fig\ref{veff1}\footnote{In~\cite{fkrs2}
we had another estimate of the parameter $A_F$, $A_F=0.2$.
The difference is because in~\cite{fkrs2}
2-loop perturbation theory was used, the 2-loop relations of the lattice
and $\msbar$ schemes were not fully available, and the proper
extrapolation of the lattice results to the continuum limit was not
done due to the limited amount of data.}.
In other words, the phase transition is weaker than predicted
by 3-loop perturbation theory. Hence the assumption of the dominance of
the gluonic condensate contribution to the vacuum energy in the
symmetric phase, used in~\cite{ms}, appears not to be satisfied .

In~\cite{bph} it was suggested that a reasonable description of the
symmetric phase can be achieved with the 1-loop Schwinger-Dyson
equation. From the analysis of this equation it was anticipated that
the symmetric phase can be interpreted as a Higgs phase whose
parameters are determined non-perturbatively. One of the predictions
of this approach is that the vector boson mass in the symmetric phase
is smaller than  $0.27 g_3^2$, what is to be confronted with the
lattice value $m_W\sim 1.1 g_3^2$. We believe that the main reason why
this approach does not work is because it relies on an accurate
description of the symmetric phase by 1-loop perturbation theory.
In particular, it
was estimated in~\cite{bph} that the expansion parameter in the 3d
theory is $\frac{1}{6\pi}\frac{g_3^2}{m_T}$, while we find
it to be about 6 times larger.

In \cite{desy,eqz} it was suggested that the introduction
of a ``magnetic mass'' $m_M\sim g_3^2/3\pi$ in the propagator
of the gauge boson will cure the infrared problem and allow
one to estimate the magnitude of non-perturbative effects.
This recipe would effectively produce
a linear 3-loop term with a negative coefficient:
\be
V_1(\phi)\sim-\frac{1}{2\pi}(m_M^2+m_T^2)^{3/2}
=-\frac{1}{2\pi}\Bigl(m_T^3+\frac{3}{4}g_3\phi m_M^2+\ldots\Bigr),
\ee
where $m_T=g_3\phi/2$. This is clearly in contradiction to the
existence of the positive linear term we found in the simulations.

\section{Results for different physical 4d theories\la{sec:phys}}

So far we have given results for
the characteristics of the phase transition in the SU(2)+Higgs
model using the simplified relations of 4d and 3d
parameters given in eqs.~\nr{g3eff}-\nr{yeff}. Here we explain
how these results can be converted into characteristics
of finite-temperature phase transitions in physical
theories where the relations between 4d and 3d are more complicated.
We also give explicit results for the phase transition
in the 4d SU(2)+Higgs theory without fermions,
in the 4d SU(2)+Higgs theory with the fermionic
content of the Standard Model, and in the Standard Model.

To begin with, let us restate the observables relevant
for the phase transition in terms of properties of the
3d theory alone.
The parameters of the 3d theory are
$g_3^2,x,y$, defined in eqs.~\nr{g3eff}--\nr{yeff}. From
lattice measurements, one can derive
the expectation value of dimensionless gauge-invariant
observables, like
$\langle\phi^\dagger\phi(g_3^2)\rangle/g_3^2$ (one must use eq.~\nr{rl2}
to change scheme from lattice to $\msbar$
at $\mu=g_3^2$).
Let us denote this particular expectation value by $\ell_3$,
$\ell_3\equiv\langle\phi^\dagger\phi(g_3^2)\rangle/g_3^2$.
Let $\ell_3^b$ be the value of $\ell_3$
in the broken phase at the critical point $(x_c,y_c)$,
and $\Delta \ell_3$ be the difference
of the values of $\ell_3$ in the broken and symmetric phases
at the critical point, $\Delta \ell_3=\ell_3^b-\ell_3^s$.
The critical point $(x_c,y_c)$ is determined
as explained in Sec.~\ref{sec:critical}.
The observable $\Delta \ell_3$ is related to the
dimensionless quantity ${\epsilon}_3(x,y)$
defined by
\be
\exp[{-V_3g_3^6{\epsilon}_3(x,y)}]=\int{\cal D}\phi
{\cal D}A \exp({-S[\phi,A]}),
\ee
through
\be
\frac{\partial \Delta{\epsilon}_3(x_c,y_c) }{\partial y_c}=
\Delta \ell_3
\la{o3}
\ee
where
$\Delta{\epsilon}_3(x,y)=\epsilon_3^b(x,y)-\epsilon_3^s(x,y)$
and $V_3$ is the volume of the system. Finally, we define the
dimensionless quantity $\sigma_3$ related to the surface
tension by
\be
\sigma_3 = \lim_{V\to \infty}\frac{1}{2g_3^4A}\ln
\frac{P_{\rm max}}{P_{\rm min}},
\ee
in analogy with eq.~\nr{sigma}.
The observables $x_c$, $y_c$, $\ell_3^b$, $\Delta \ell_3$
and $\sigma_3$ are all dimensionless
quantities defined
strictly inside the 3d theory,
are independent of the parametrization used
in eqs.~\nr{g3eff}-\nr{yeff},
and are measurable, in principle, to arbitrary accuracy
with lattice simulations, since the lattice
counterterms of the 3d theory are known exactly.

Next, one needs the relations to 4d. The relations to 4d
consist of two parts:\\
(a) One needs the expressions of the parameters $g_3^2$, $x$ and $y$
of the 3d theory in terms of temperature and the physical
parameters of the 4d theory. For definiteness, let us take as one
of the physical parameters the pole mass of the Higgs field, $m_H$.
Then the parameters of the 3d theory are of the form
$g_3^2=g_3^2(m_H,T)$, $x=x(m_H,T)$ and $y=y(m_H,T)$. \\
(b) One needs the relation of 3d observables to 4d thermodynamics.
This relation is completely determined
by the equation $p(T)=-Tg_3^6{\epsilon}_3(x,y)$, which
holds apart from
inessential mass-independent terms.

Using the relations in (a) and (b),
one can relate $x_c$, $y_c$, $\ell_3^b$, $\Delta \ell_3$
and $\sigma_3$ to 4d quantities characterizing the
thermodynamics of the phase transition. First, from the equations
\be
\left\{
\begin{array}[c]{c}
x_c=x(m_H,T_c) \\
y_c=y(m_H,T_c),
\end{array}
\right.\la{mHTc}
\ee
one can solve for the Higgs mass $m_H$ and
critical temperature $T_c$ to which the phase
transition in the 3d theory at the
point $(x_c,y_c)$ corresponds. When
$T_c$ is known, one can calculate the
gauge coupling: $g_3^2=g_3^2(m_H,T_c)$.
Then one can determine $v^2(T_c)/T_c^2$, defined by
$v^2(T_c)/T_c^2=2\langle\phi_3^\dagger\phi_3(T_c)\rangle/T_c$
according to eq.~\nr{defvoT}, from
\be
\frac{v^2(T_c)}{T_c^2}=2
\frac{g_3^2}{T_c}
\biggl[\ell_3^b-\frac{3}{16\pi^2}\ln
\frac{g_3^2}{T_c}
\biggr]. \la{v2oTc2}
\ee
The surface tension $\sigma/T_c^3$ is obtained
from
\be
\frac{\sigma}{T_c^3}=\sigma_3
\frac{g_3^4}{T_c^2}. \la{soTc3}
\ee
For the latent heat $L=T_c[p_s'(T)-p_b'(T)]$, one gets
\ba
\frac{L}{T_c^4} & = & \frac{1}{T_c^3}
\frac{d}{dT} \left.\Bigl[Tg_3^6\Delta{\epsilon}_3(x,y)\Bigr]\right|_{T=T_c}
\nonumber \\
& = & \Delta \ell_3\frac{g_3^6}{T_c^2}
\biggl(\frac{dy}{dT}-\frac{dx}{dT}\frac{dy_c}{dx_c}\biggr). \la{LoTc4}
\ea
Here we utilized eq.~\nr{o3} together with the fact that
at the critical line, the following two equations hold:
\ba
\Delta{\epsilon}_3(x_c,y_c) & = & 0,
\\ 
\frac{\partial\Delta{\epsilon}_3(x,y) }{\partial x}
& = & -\frac{dy_c}{dx_c}
\frac{\partial\Delta{\epsilon}_3(x,y) }{\partial y}.
\la{clcl}
\ea
Note that both $dx/dT$  and $dy_c/dx_c$
in eq.~\nr{LoTc4} are non-zero through
loop corrections: $dx/dT$ is non-vanishing due to
logarithmic 1-loop corrections to the dimensional
reduction of the coupling constants $g_3^2$, $\lambda_3$,
and $dy_c/dx_c$ is non-zero due to loop corrections
inside the 3d theory (at tree-level, the phase transition
takes place at $y=0$ independent of $x$).
Numerically, $dy_c/dx_c\sim -1$ but $dx/dT\ll dy/dT$,
so that in realistic cases
\be
\frac{(dx/dT)(dy_c/dx_c)}{dy/dT}
\sim
0.02.
\ee
Hence one does not need to determine the derivative of the critical
line $y=y_c(x)$ with as good a relative accuracy as the jump
$\Delta \ell_3$ of the order parameter at the point $(x_c,y_c)$. If one uses
the parametrization of eqs.~\nr{g3eff}-\nr{yeff} in eq.~\nr{LoTc4},
then $dx/dT^*=0$ and one gets the expression in eq.~\nr{Lwithpdp}.

To get the values of $\ell_3^b$, $\Delta \ell_3$
and $\sigma_3$ from the results given in Sec.~\ref{pt},
one has to use eqs.~\nr{v2oTc2}-\nr{LoTc4} in the inverse direction,
employing the parametrization of eqs.~\nr{g3eff}-\nr{yeff}.
For instance, the surface tension $\sigma/T_c^{*3}=0.0023$
for $m_H^*=60$ GeV given in table~\ref{table:sigma}
corresponds to $\sigma_3=0.0023/0.44015^2=0.0119$
according to eqs.~\nr{soTc3} and~\nr{g3eff}.
The values of $x_c$, $y_c$, $\ell_3^b$, $\Delta \ell_3$
and $\sigma_3$ obtained this way are shown in the second
block in table~\ref{tab:phys}.
To go back to 4d units for different physical theories,
one needs the true values of $g_3^2/T_c$, $x'(T_c)$ and
$y'(T_c)$. In general, the value of $g_3^2/T_c$ differs
from 0.44015, so that $\sigma/T_c^3$ differs from
the value in table~\ref{table:sigma}, see table~\ref{tab:phys}.

The explicit form of the expressions
for $g_3^2$, $x$ and $y$ in terms of temperature
and the physical 4d parameters of the Standard Model
to order $g^4$ have been given in~\cite{klrs}, using
the approximation $g'^2\sim g^3$. With these relations, we can
give results for the thermodynamical properties
of the EW phase transition. To be more precise,
we will consider three different theories reminiscent
of the EW sector of the Standard Model.
As a starting point, we consider
the case $m_Z=m_W$ so that $g'=0$.
This is a SU(2)+Higgs theory with the fermionic
content of the Standard Model. The functions $x(m_H,T)$, $y(m_H,T)$
and $g_3^2(m_H,T)$ in this theory, with $m_{\rm top}=175$ GeV,
have been given in figs.~7-8 of~\cite{klrs}.
The results for the phase transition,
obtained from the values in the second block in table~\ref{tab:phys}
using eqs.~\nr{mHTc}-\nr{LoTc4},
are shown in the fourth block in table~\ref{tab:phys}.
Results for the full Standard Model can be
obtained from the
SU(2)+Higgs+fermions model
by taking into account the U(1)-subgroup perturbatively
with the help of fig.~9 in~\cite{klrs}.
The results are in the bottom block in table~\ref{tab:phys}.
Finally, we give results also for the SU(2)+Higgs model
without fermions. This case is
obtained from the SU(2)+Higgs+fermions model by
putting $g_Y=n_F=\alpha_S=0$ and fixing $m_W=80.22$ GeV.
The results are shown in the third
block in table~\ref{tab:phys}.
In principle, the results for the SU(2)+Higgs theory
should allow a comparison with the 4d lattice
simulations in~\cite{desylattice1,desylattice2,desylattice3}.
At present such a comparison is not rigorously possible,
however, since the relation of the gauge coupling $g_R^2$
used in 4d simulations to continuum physics is not known.

\newcommand{\none}{--\hspace*{0.25cm}}

\begin{table}[p]
\centering

\begin{tabular}{|l@{\hspace{1.0cm}}l@{\hspace{0.5cm}}|l|l|l|}
\hline
            &  & $m_H^*=35$ GeV & $m_H^*=60$ GeV & $m_H^*=70$ GeV\\ \hline
\mbox{SU(2)+Higgs in 3d} & $x_c$  & 0.01830 & 0.06444 & 0.08970 \\ \cline{3-5}
(perturbative)& $y_c$ & 0.0644(2) & 0.0114(15) & 0.0028(26)\\ \cline{3-5}
               & $\Delta \ell_3$ & 3.48(1) & 0.493(5) & 0.312(7) \\ \cline{3-5}
               & $\ell_3^b$  & 3.47(1) & 0.508(17) & 0.330(20) \\ \cline{3-5}
               & $\sigma_3$  & 0.339(1) & 0.0401(7) & 0.0253(9) \\ \hline
\mbox{SU(2)+Higgs in 3d} & $x_c$  & 0.01830 & 0.06444 & 0.08970 \\ \cline{3-5}
(lattice) & $y_c$ & 0.05904(56) &-0.00146(35) &-0.01531(69) \\ \cline{3-5}
          & $\Delta \ell_3$ & 4.07(13) & 0.491(8) & 0.302(18) \\ \cline{3-5}
          & $\ell_3^b$      & 3.91(13) & 0.500(12) & 0.353(26) \\ \cline{3-5}
      & $\sigma_3$   & [0.47(1)] & 0.0119(26) & ?\hspace*{2.5mm}\\ \hline
\mbox{SU(2)+Higgs in 4d}  & $m_H/$GeV  & 29.1 & 54.4  & 64.3  \\ \cline{3-5}
                    & $T_c/$GeV    & 76.85 & 132.6 & 151.2 \\ \cline{3-5}
                    & $L/T_c^4$      & 0.200 & 0.0294 & 0.0194 \\ \cline{3-5}
                    & $v/T_c$    & 1.74 & 0.626 & 0.529 \\ \cline{3-5}
               & $\sigma/T_c^3$ & [0.071] & 0.0017 & ?\hspace*{2.5mm}\\ \hline
\mbox{SU(2)+Higgs}  & $m_H/$GeV      & \none & 51.2 & 68.0  \\ \cline{3-5}
\mbox{+fermions}    & $T_c/$GeV     & \none  & 89.79 & 105.8 \\ \cline{3-5}
                    & $L/T_c^4$      & \none & 0.103 & 0.0651 \\ \cline{3-5}
                    & $v/T_c$    & \none & 0.642 & 0.542 \\ \cline{3-5}
               & $\sigma/T_c^3$ & \none & 0.0019 &  ?\hspace*{2.5mm}\\ \hline
\mbox{The Standard} & $m_H/$GeV      & \none & 51.2 & 68.0  \\ \cline{3-5}
\mbox{Model}        & $T_c/$GeV       & \none & 88.93 & 104.8 \\ \cline{3-5}
                    & $L/T_c^4$      & \none & 0.124 & 0.0769 \\ \cline{3-5}
                    & $v/T_c$    & \none & 0.689 & 0.575 \\ \cline{3-5}
               & $\sigma/T_c^3$ & \none & 0.0023 & ?\hspace*{2.5mm} \\ \hline
\end{tabular}
\vspace*{1mm}

\caption[a]{\protect
Properties of the phase transition in different physical 4d theories.
The simulations with $m_H^*=35$ GeV do not correspond to any physical
pole Higgs mass in the SU(2)+Higgs+fermions model, at least according
to the 1-loop formulas for the couplings used in~\cite{klrs}.
The errors for the 2-loop
perturbative results in the first block
indicate the effect of varying $\mu$ in the range
$0.5\mu_{\rm opt}\ldots 2.0\mu_{\rm opt}$.
The error estimates for the 3d lattice SU(2)+Higgs model
follow from the error estimates in
Secs.~\ref{sec:critical}-\ref{sec:tension}.
The surface tension measurements for $m_H^*=35$ GeV
are in parentheses for reasons explained in Sec.~\ref{sec:tension}.
In the 4d results, we only show the central value.
For the 4d SU(2)+Higgs model,
the additional relative error from the relations to 4d should be below 1\%,
and for the SU(2)+Higgs+fermions model, it may be
a few percent (in the critical temperature, the error is
an order of magnitude smaller)~\cite{klrs}.
The last block concerning the
Standard Model is based on a purely perturbative estimate of
the effect of the U(1)-subgroup~\cite[fig.~9]{klrs},
and the errors may be large
if the non-perturbative effects related to the U(1)-sector
are significant.\la{tab:phys}}
\end{table}

\section{Applications to cosmology}
\la{sec:appl}

\subsection{The phase transition}

The results of non-perturbative lattice
MC simulations allow one to considerably
reduce the uncertainties associated with the dynamics of the electroweak
phase transition. Strictly speaking, they still refer to a somewhat
unphysical situation, since the U(1) subgroup present in the
electroweak theory is omitted. As we discussed in the previous
section, the perturbative effects associated with the U(1) subgroup are
rather small; the estimate of the non-perturbative effects requires
lattice simulations in the complete 3d SU(2)$\times$U(1)+Higgs model.
All specific estimates in this section are based on the assumption
(quite reasonable, though) that all perturbative and
non-perturbative effects of the U(1) subgroup are small.

Let us reconstruct the picture of the phase transition for $m_H^*=60$
GeV, for which we have the best lattice data. For the MSM with the
top mass  $m_t=175$ GeV this corresponds to the pole Higgs mass of
$m_H=51.2$ GeV, excluded experimentally, but for a number of extensions
of the Standard Model this may be quite realistic. To make the
discussion less model dependent we will use the variable $T^*$; with
the help of the results of~\cite{klrs} and Sec.~\ref{sec:phys} everything
can be re-computed for any specific model. We shall omit here the
errorbars from the numbers.

The critical temperature of the phase transition is $T^*_c=138.4$ GeV, and the
vev-temperature ratio at $T_c^*$ is $v/T^*_c=0.67$. The scalar
correlation length in the symmetric phase is $\xi_s \sim 6/T^*_c$, and
in the broken phase $\xi_b\sim 8/T^*_c$. The corresponding vector
correlation lengths are a factor of three shorter. The domain wall separating
the broken and symmetric phases has the surface tension
$\sigma\sim 0.002 T_c^{*3}$, and the profile of the scalar field has
asymmetric tails on the different sides of the domain wall,
$\delta\phi(x) \sim \exp(-|x|/\xi_{s,b})$. At $T^*>T_{+}^*= 138.7$ GeV only
the symmetric phase is stable, and at $T^*<T_{-}^*=137.4$ only the broken
phase is stable, while in between both phases can exist
simultaneously\footnote{We use here the results of
subsection~\ref{sec:met-reweight} which are the most accurate.}. The bubble
nucleation temperature $T_{\rm bubble}$ lies
somewhere between $T_c$ and $T_{-}$ and may be estimated with the use
of the surface tension and latent heat found on the lattice.

The thin wall bubble nucleation rate is given by
\be
\Gamma= \kappa T_c^4 \exp(-\frac{4\pi \sigma R_c^2}{3T_c})
\ee
where $R_c=2\sigma/\epsilon$ is the radius of the critical bubble,
$\epsilon=L\frac{\Delta T}{T_c}$ is the pressure difference
between the broken and symmetric phases, $\Delta T=T_c-T_{\rm bubble}$,
$L$ is the latent heat of the transition, and $\kappa$ is the prefactor.
Estimates for $\kappa$ in different models and approximations can be
found in~\cite{helbig,fluct}. The bubble nucleation
temperature is roughly determined from the condition
\be
\Gamma \frac{M_{Pl}^4}{T_c^8} \simeq 1.
\ee
Inserting the lattice numbers to these relations gives an estimate
$\frac{\Delta T^*}{T_c^*}\simeq 0.001$, i.e., the bubble nucleation
temperature is very close to the critical one,
$T^*_{\rm bubble}=138.3$ GeV.
The smallness of ${\Delta T^*}/{T_c^*}$ is due to fact that the ratio
$\sigma^3/L^2T_c^*\sim 10^{-5}$ is so small~\cite{ikkl}.
Since $\Delta T^*/(T_c-T_{-})\simeq 0.1$ is also small,
one is in the thin-wall regime; indeed,
the size of the bubbles when they nucleate is at least $R_c \simeq
110/T_c^*$ which is much larger than the scalar correlation lengths in
the broken and symmetric phases at $T_{\rm bubble}$.
Since $T_{\rm bubble}$ is very close to the critical
temperature, the vev of the Higgs field at $T_{\rm bubble}$ is almost the
same as at $T_c^*$.

\subsection{The out of equilibrium condition for electroweak
baryogenesis}
One of the motivations for the study of the electroweak phase transition
is its application to electroweak baryogenesis. The rate of the
anomalous baryon number non-conserving processes is high in the
symmetric phase~\cite{kurs}, but is suppressed by the Boltzmann
exponent in the broken phase~\cite{kurs}. Baryogenesis occurs at
the 1st order electroweak phase transition, and the
mechanism-independent constraint on the strength of the phase
transition is that the rate of fermion number non-conservation
in the broken phase at the bubble nucleation temperature be smaller
than the rate of universe expansion~\cite{mes}.

We parametrize the rate of sphaleron transitions in the broken phase
as
\be
\Gamma = T^4 \left(\frac{\alpha_W}{4\pi}\right)^4
N_\rmi{tr} N_\rmi{rot} \left(\frac{2E_\rmi{sph}(T)}{\pi T}\right)^7 \exp
\left(-\frac{E_\rmi{sph}(T)}{T}\right)
\label{sph_rate}
\ee
where the factors $N_\rmi{tr} \simeq 26$ and $N_\rmi{rot}
\simeq 5.3\times10^3$
are zero mode normalizations \cite{armc} and $E_\rmi{sph}(T)$
is the effective sphaleron mass at temperature $T$.  Then the out of
equilibrium constraint reads \cite{mes}:
\be
E_\rmi{sph}(T_{\rm bubble})/T_{\rm bubble} > 45.
\label{Esphlim}
\ee
In principle, this bound can be converted into an upper bound on the
parameter $\lambda_3/g_3^2$, completely defining the dynamics of the
3d theory. An exact determination of this bound would require a
non-perturbative evaluation of the sphaleron rate in the broken phase
(for lattice simulations of topology changing processes in
the symmetric phase, see \cite{amb1,amb2}). Unfortunately, this
problem has not been solved yet. Below we will estimate the critical
value of $\lambda_3/g_3^2$ assuming that the expansion parameter in
the broken phase is $g_3^2/(\pi m_T)=2 g_3/(\pi \phi_b)$ -- a value
inspired by the higher order computations and lattice simulations.

The best estimate of the sphaleron rate available now is the 1-loop
computation in~\cite{baacke}, where the determinant of small
fluctuations was computed numerically in a bosonic
theory\footnote{Recently the fermionic determinant in the background
of a sphaleron was computed in~\cite{diakonov}. The authors
concluded that the fermionic contribution suppresses
the rate and is numerically very important. However,
the effect of fermions can be absorbed into the definition of the
3d coupling constants; after this the fermionic contribution is
negligible, see~\cite{guy}.}. According to~\cite{baacke}, the
rate in 1-loop approximation is just given by eq. (\ref{sph_rate}),
where
\be
\frac{E_\rmi{sph}(T)}{T}=
B(\frac{\lambda_3}{g_3^2})\frac{2 \pi T^{1/2}}{g_3}
\frac{\phi_b}{T}
\label{sph1loop}
\ee
and $\phi_b$ is the vev to be determined
from the 1-loop effective potential (measured here in 4d units).
Arguments in favour of absorption of the 1-loop effects into the
vev of the scalar field have been presented in~\cite{bosh}.
Some numerical values of the function $B$ are: for
$m_H^*=0,40,45,50$, $B=3.04,3.41,3.44,3.48$, correspondingly
\cite{brihaye}. We assume then that the exact value of the sphaleron
mass is given by (\ref{sph1loop}) plus corrections, and that $\phi_b/T$
is to be replaced by the
exact gauge-invariant value $v/T$ determined on lattice. Then the first
correction is of the order $A(g_3^2/(\pi m_T))^2$, where $A$ is a
number of the order of unity. A conservative limit on the ratio
$v/T$ is obtained when $A$ is positive, so that $v/T>1.22$ (we take
$B$ corresponding to $m_H^*=50$ GeV).  At this value the 2-loop
correction is about $10\%$ and the 3-loop correction is expected to
be of the order of $4\%$ and can be neglected. If, in the contrary,
$A$ is negative, we get $v/T>1.49$.

Now, we choose the weaker constraint $v/T>1.22$ and  convert it
into an upper limit on $m_H^*$. Since we do not have lattice simulations
for the whole range of $m_H^*$ values from $35$ GeV to $60$ GeV,
we take the 2-loop predictions for $v/T$; as we discussed, this is
accurate within a few percent. We find that if $m_H^*\simeq 42$ GeV then
$v/T_c=1.22$. The bubble nucleation temperature is somewhat smaller
than the critical temperature. The thin wall approximation for the
tunneling rate is not applicable here. Assuming that the perturbative
description of bubble nucleation is valid
in this region of the Higgs masses, we estimate that
$v/T$ at the nucleation temperature is about $20\%$
larger than at the critical temperature.
This is derived by defining the bounce for the action
\be
\fr12(\partial\phi)^2 + V_\rmi{2-loop}(\phi)
\ee
numerically, and requiring that the value of the bounce action
is $\simeq 140$ (see, e.g., \cite{linde}). For example,
for $m_H^*=43$ GeV, $T_c^*=108.6$ GeV, $v(T_c^*)/T_c^*=1.16$,
$T_\rmi{bubble}=106.3$ GeV,
$v(T_\rmi{bubble})/T_{\rm bubble}=1.43$.
Then the requirement $v/T>1.22$ gives $m_H^*<46$ GeV. 
The perturbative account of the U(1) factor makes the phase
transition stronger first order, correcting this bound by a factor
\be
\sqrt{\fr13(2+\frac{1}{\cos^3\theta_W})}
\ee
following from the 1-loop effective potential.
We then finally get  $m_H^*\lsim 50$ GeV.
If the correction to the sphaleron rate is in fact negative,
then the number is smaller, about $45$ GeV. It is interesting to note that
this bound is very close to the initial 1-loop computation
in~\cite{mes}. We stress, however, that these estimates are subject to
verification by the future lattice simulations of the theory with
the U(1) subgroup and to non-perturbative evaluation of the sphaleron rate.

{}From eq.~(\ref{lameff}), $m_H^*< 50$ GeV corresponds to $x<0.043$ and
$m_H^*< 45$ GeV to $x<0.034$.
To summarize, the upper limit to the parameter $\lambda_3/g_3^2$ in
the 3d SU(2)$\times$U(1)+ Higgs theory is likely to be
\be
\lambda_3/g_3^2< 0.04.
\label{bound}
\ee

In order to define the constraints following from this requirement on
the particle spectrum of the underlying 4d theory, one has to express
this ratio through the physical parameters of
the 4d theory at the critical temperature.
This computation may be quite involved~\cite{klrs}, but it is very clean from
the physics point of view and does not contain any infrared
divergencies. An essential point is that
only {\em 1-loop graphs need be computed}.
Indeed, a 1-loop computation
provides $O(\alpha^2)$ accuracy in
the coupling constants of the effective theory.
Moreover, the critical temperature enters to the ratio (\ref{bound})
only through logarithms, so that even a 1-loop estimate of
it will give sufficient accuracy.

The application of the constraint of eq.~\nr{bound} to the case of the
Minimal Standard Model follows immediately from \fig8 of~\cite{klrs}.
Indeed, if $m_t=175$ GeV, then {\em no Higgs mass} can ensure the
necessary requirement of eq.~(\ref{bound})\footnote{To be more
precise, the computations in~\cite{klrs} relating the physical masses
of the W boson and Higgs particle to the parameters of $\msbar$ scheme
break down if the physical Higgs mass is close to the Coleman-Weinberg
limit. In this limit the higher order Yukawa corrections start to be
important in the procedure of dimensional reduction as well. So, it is
not excluded that Higgs masses close to the Coleman-Weinberg limit are
still possible.}. If the top quark were lighter, then some low Higgs
mass value might be possible, see \fig\ref{xmHmt}.

\begin{figure}[tb]
\vspace*{0cm}
\hspace{1cm}
\epsfysize=11cm
\centerline{\epsffile{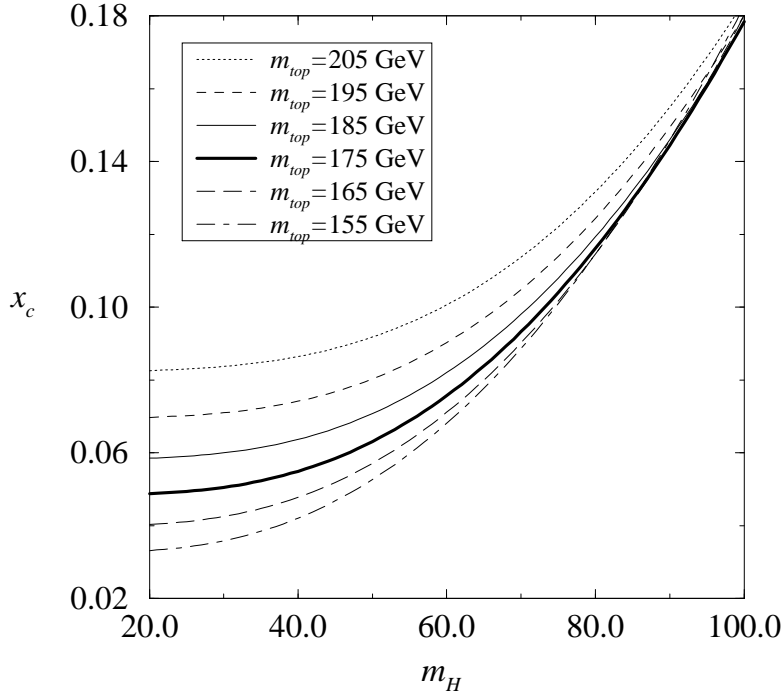}}
\vspace*{-3cm}
\caption[a]{\protect
The critical value $x_c=\lambda_3/g_3^2$ as a function of the physical Higgs
mass $m_H$ and the top quark mass $m_{\rm top}$. In general,
$x$ depends on the Higgs mass, the top mass and logarithmically
on the temperature. In calculating $x_c$, we have solved the temperature
from the equation $y=0$, which according to table~\ref{tab:phys}
is very close to the true critical value of $y_c$. The small error
in $T_c$ does not affect $x_c$ much due to the logarithmic dependence.
The value of the dimensionless U(1)-coupling $g'^2_3/g_3^2$
is not shown in this figure; it is $g'^2_3/g_3^2\approx 0.3$.}
\la{xmHmt}
\end{figure}

According to \cite{mssm} the phase transition in the MSSM occurs in
the same way as
it does in the MSM. If true, then MSSM also fails in generating a
sufficiently strong first order phase transition. The two Higgs doublet
model has more freedom, and the results of \cite{2doubl} indicate that the
constraint (\ref{bound}) can be satisfied there.

\section{Conclusions}
\la{sec:conclusions}

The 3d formalism, developed in the series of papers
\cite{krs}-\cite{klrs},\cite{mlaine2} provides a powerful tool for the
study of phase transitions in weakly coupled gauge theories. It
factorizes the perturbative and non-perturbative physics and allows
one to construct effective 3d theories, describing in a universal way
phase transitions in a large class of underlying 4d theories. The
effective theories in 3d contain bosons only, and may be used for high
precision lattice Monte Carlo simulations.

In this paper we reported on lattice simulations in the 3d SU(2)+Higgs
model, which is an effective theory for the SU(2) sector of the MSM
and its extensions.  The nature of the phase transition at moderate
Higgs masses $m_H<m_W$ is clarified, and the results presented here
form an ``experimental'' basis for different theoretical schemes
attempting to describe strong coupling phenomena at $T\sim T_c$.

{}From the phenomenological point of view, to our mind, the most
interesting further problems to be solved are the role of the U(1) factor in
the phase transition, and the rate of the sphaleron transitions in
the broken phase near $T_c$. The solution of the first problem on the
lattice is a straightforward generalization of the SU(2) case
considered in this paper. The non-perturbative estimate of the
sphaleron rate is a much more complicated problem, and it is even
not clear how it can be solved in principle.

\section*{Acknowledgements}
We thank K. Farakos for collaboration during the early stages of the work.
We are grateful to the Finnish Center for Scientific Computing
and Indiana University  for computational facilities.
M.S. thanks Glennys Farrar for helpful comments and ITP, Santa Barbara
for hospitality. K.R. is supported by the DOE grant \# DE--FG02--91ER40661.
This research was supported in part by the National Science Foundation
under Grant No. PHY94-07194.

\appendix
\renewcommand{\thesection}{Appendix~~\Alph{section}}
\renewcommand{\theequation}{\Alph{section}.\arabic{equation}}

\section{}

In this Appendix we rederive the result (II.76-78) for $\pdp$ in
the $\hbar$ (loop) expansion in the broken phase of the 3d SU(2)+Higgs
theory writing it in a more explicit and complete form,
and discuss the parametrisation of the 3-loop terms
(the statistical uncertainties do not allow
a determination of the 4-loop terms). 
The computation takes place in three steps:
first find the potential (gauge and $\mu$ dependent), then
find the value $V(v)$ of the potential at its broken minimum
(gauge independent but $\mu$ dependent)\footnote{
Note that we have simplified notation here: by $v$ we mean the
the location of the minimum of the effective potential, previously
denoted by $\phi_b$. The $v$ used here should not
be confused with the $v(T)$ used previously and defined as
$v^2(T)/T^2=2\langle\phi^\dagger\phi(T)\rangle/T$.},
and finally calculate the condensate from
\be
\pp{\mu}={dV(v)\over dm_3^2(\mu)}.
\la{condformula}
\ee

The starting point is the potential up to 3 loops ($\hbar$ is a loop
counting parameter):
\be
V=V_0+\hbar V_1+\hbar^2 V_2+\hbar^3V_3.
\la{potexp}
\ee
Perturbative computation gives $V$ in the form
$V=V(m_T,m_1,m_2,\phi;\mu)$, where
\be
m_T=\fr12 g_3\phi,\quad m_1^2=m_3^2(\mu)+3\lambda_3\phi^2,\quad
 m_2^2=m_3^2(\mu)+\lambda_3\phi^2. \la{masses}
\ee
Thus, equivalently $V=V(m_3^2(\mu),\phi;\mu)$.
Here, in contrast to eq.~\nr{condformula},
$\phi$ denotes the real Higgs field.

The tree and 1-loop potentials are
\begin{eqnarray}
V_0 &=& \fr12 m_3^2(\mu)\phi^2
+\fr14\lambda_3\phi^4,\la{vtree} \\
V_1&=& -{1\over12\pi} \bigl(6m_T^3 + m_1^3+3m_2^3 \bigr),
\la{1looppot}
\end{eqnarray}
and the 2-loop part in the Landau gauge is in eq.~(I.33),
in the general covariant gauge in~\cite{mlaine} and in the $R(\xi)$
gauge in~\cite{kls}. The 3-loop potential 
is near the classical broken minimum of the
general form
\be
V_3={\partial V_1\over\partial m_3^2}{f_{2m}\over16\pi^2}
\biggl(\log{\mu\over m_T}+\fr12\biggr)
-{27\over128}\biggl(2+\fr12 h^3\biggr)^2{g_3^4\over(4\pi)^3}
{m_T^2\over m_2}
+{\beta\over(4\pi)^3}g_3^4m_T,
\la{V3}
\ee
where
\be
{\partial V_1\over\partial m_3^2}=-{1\over8\pi}(m_1+3m_2),\quad
h^2\equiv {8\lambda_3\over g_3^2}, \la{defofh}
\ee
and $\beta$
is a complicated function of the masses,
which we only need near the broken minimum.
Note that $h$ in eq.~\nr{defofh} differs from the definition
in eq.~\nr{def1h}.

The $\mu$ dependent terms of $V_3$ 
are known (eq.~(I.70)) since they should cancel the
$\mu$~dependence of $V_1$ 
so as to make
the $\mu$ dependence of $V_1+V_3$ to be of order $\hbar^5$.

The terms proportional to
powers of $1/m_2$ are singular at the tree
minimum of the potential. They identically cancel when the value
of the potential at the minimum is computed
perturbatively, see below.

The effective
potential itself is gauge dependent but its value in the broken
minimum is gauge independent (but dependent on the scale $\mu$).
The minimum is defined by
\be
{\partial V(m_3^2(\mu),\phi;\mu) \over \partial\phi}=0.
\ee
Denoting the solution of this equation by $\phi=v$ one can solve for it
in the loop ($\hbar$) expansion:
\be
v^2=v_{(0)}^2+\hbar v_{(1)}^2+\hbar^2 v_{(2)}^2 +
{\cal O}(\hbar^3), \la{vexp}
\ee
with the result
\ba
v_{(0)}^2&=&{-m_3^2(\mu)\over\lambda_3}, \la{0loopv}\\
v_{(1)}^2&=&-{2\over\lambda_3}V_1'(v_{(0)}^2)=
{3\over4\pi}\biggl(h+{4\over h^2}\biggr)\bar m_T,
\\
v_{(2)}^2&=&{4\over\lambda_3^2}V_1''(v_{(0)}^2)V_1'(v_{(0)}^2)
-{2\over\lambda_3}V_2'(v_{(0)}^2) \la{2loopv}\\
&=&{9\lambda_3\over32\pi^2}
\biggl(3+{8\over h^3}+{\bar m_1\over m_2}\biggr)
\biggl(1+{4\over h^3}\biggr)
-{2\over\lambda_3}V_2'(v_{(0)}^2) \nonumber\\
&=&{3g_3^2\over16\pi^2} \biggl[
\fr3{16}h^2\biggl(3+{8\over h^3}\biggr)\biggl(1+{4\over h^3}\biggr)
+1-\fr{5h}4  -h^2+
\fr{h^3}2-\fr{3h^4}8+{1\over 2+h} \biggr].\nonumber
\ea
Here $V_n'(v_{(0)}^2)$ etc. mean writing $V_n=V_n(m_3^2,\phi)$, taking
partial derivative with respect to $\phi^2$ and evaluating the result
at the saddle point value (\ref{0loopv}). Hereby one will need the saddle
point values of the masses:
\be
\bar m_1=h\bar m_T=\sqrt{-2m_3^2(\mu)},\quad
\bar m_2=0.
\la{saddleptvalues}
\ee
The saddle point values satisfy the equations
\be
{d\bar m_T\over dm_3^2}= -{1\over h^2\bar m_T},\quad
{d\bar m_1\over dm_3^2}= -{1\over \bar m_1}.
\ee
The pole terms $1/m_2$ in (\ref{2loopv}) cancel before taking the
limit $m_2\to0$. These arise when taking derivatives of $V_1\sim
m_2^3,\,V_2\sim m_2, V_3\sim 1/m_2$, etc.
In gauge invariant quantities calculated below
they cancel and will not be explicitly written down.


Inserting (\ref{vexp}) to (\ref{potexp}) and expanding one gets
\ba
V(v)&=&V_0+\hbar V_1 + \hbar^2 \Bigl[V_2
-\lambda_3^{-1}(V_1')^2\Bigr]
+\hbar^3 \Bigl[V_3+(V_2'+\fr12 V_1''v_{(1)}^2)v_{(1)}^2 \Bigr],
\ea
where all quantities on the RHS should be evaluated at the saddle
point values (\ref{saddleptvalues}).
Note that $v^2$ is only needed up
to 1 loop.

The potential at the minimum thus is the potential evaluated at the
classical minimum corrected by some terms, related to one-particle
reducible diagrams and
calculable in terms of potentials of lower order. The result for the
3-loop one-particle reducible contribution is
\ba
(V_2'+\fr12 V_1''v_{(1)}^2)v_{(1)}^2&=&
\fr32\biggl(h+\fr4{h^2}\biggr)\bar m_T{f_{2m}\over(4\pi)^3}
\biggl(\log{\mu\over\bar m_T}+\fr12\biggr)+ \nonumber\\
&&+\beta_\rmi{disc}(h)\bar m_T {g_3^4\over(4\pi)^3}+
\\&&
+{27\over128}\biggl(2+\fr12 h^3\biggr)^2{g_3^4\over(4\pi)^3}
{\bar m_T^2\over m_2},\nonumber
\ea
where
\ba
\beta_\rmi{disc}(h)&=&\fr9{32}\biggl(h+\fr4{h^2}\biggr)
\biggl[
(\fr12h^4-21)\log3+h^4\log h+
\nonumber\\&&+
(4-3h^2+h^4)\log(2+h)-h^2(1+h^2)\log(1+h)+
\nonumber\\&&
-\fr{21}2-\fr6{h^2}-\fr54 h-h^2-h^3+\fr{11}{16}h^4-
{2\over 2+h}\biggr].
\ea
Defining further
\be
f_2(h)=\fr{17}2+h^2-\fr12h^4,
\ee
\be
{f_{2m}\over g_3^4}=\fr38\biggl(\fr{17}2+3h^2-\fr12h^4\biggr),
\ee
\ba
q_2(h)&=&(2-h^2+\fr14 h^4)\log(2+h)-2h+\fr14h^2-\fr12h^3+
\nonumber\\
&&+\fr14h^4\log(3h)-\fr{21}2\log 3-\fr32-\fr6{h^2}-\fr14h^4,
\la{q2}
\ea
the vacuum energy density is
\ba
V(v)&=&-{m_3^4(\mu)\over\lambda_3}-{\hbar\over12\pi}\bar m_T^3(6+h^3)+
\nonumber\\
&&+\hbar^2\bar m_T^2{3g_3^2\over 64\pi^2} \biggl[f_2(h)
\biggl(\log{\mu\over\bar m_T}+\fr12\biggr)+q_2(h)\biggr]+
\nonumber\\&&
+\hbar^3\bar m_T\biggl[\biggl(h+\fr6{h^2}\biggr)
{f_{2m}\over(4\pi)^3}
\biggl(\log{\mu\over\bar m_T}+\fr12\biggr)+
\nonumber\\&&
\qquad+\Bigl[\beta_\rmi{disc}(h)+\beta(h)\Bigr]
{g_3^4\over(4\pi)^3}\biggr].
\la{v4min}
\ea

Computing $\partial V(v)/\partial m_3^2$ gives the final result
for the scalar condensate in loop expansion:
\ba
&&{\pp{\mu}\over g_3^2}={\pp{\mu_0}\over g_3^2}+{3\over
16\pi^2}\ln{\mu\over\mu_0}=
\label{sp_for_pdp}\\
&&{-m_3^2(\mu)\over2\lambda_3}
\nonumber\\
&&+{1\over4\pi}\biggl(h+\fr6{h^2}\biggr)
{\bar m_T(\mu)\over g_3^2}
\nonumber\\
&&-{1\over2\lambda_3}
{3\over8(4\pi)^2}g_3^4\biggl[f_2(h)
\log{\mu\over \bar m_T(\mu)}+q_2(h)
\biggr]
\nonumber\\
&&-{1\over(4\pi)^3}{g_3^2\over h^2\bar m_T(\mu)}\biggl[
{f_{2m}\over g_3^4}
\biggl(h+\fr6{h^2}\biggr)\biggl(\log{\mu\over\bar m_T}-\fr12
\biggr)+\beta_\rmi{disc}(h)+\beta(h) \biggr], \nonumber
\ea
where the different loop contributions are on different lines
and where $\beta(h)$ is the 3-loop quantity to be determined.
Note how the general $\mu$ dependence arises
from an incomplete cancellation
between the tree and 2-loop terms and
how the $\mu$ dependence of the 1-loop term is compensated to
order $\hbar^5$ by the 3-loop term.

\section{}
\label{oldB}

In this appendix we describe an ``improved'' version of the 2-loop
effective potential in Landau gauge used for the computation of the
scalar condensate at different temperatures by the
``Coleman-Weinberg'' method.

It was pointed out in \cite{fkrs3}
(see also Appendix~A)
that the effective potential in
the Landau gauge contains singularities at the classical
tree minimum, where the Goldstone mass
\be
m_2^2 = m_3^2 +\lambda_3\phi^2,
\ee
vanishes. The order of leading singularities on the $n$-loop level
can be found from power counting:
\be
V_n^\rmi{sing} = \frac{\hbar^n C_n}{(m_2^2)^{n-\fr52}},
\label{sing}
\ee
where the $C_n$ are some coefficients. These terms are not dangerous if
the location of the minimum of the effective
potential is far enough from the tree-level value (this is the
Coleman-Weinberg regime). However, deep in the broken phase the
quantum corrections are small, the Goldstone mass $m_2$ is close to
zero and terms like (\ref{sing}) may become important. As we
discussed in Sec.~5 of~\cite{fkrs3},
due to the singular terms
the computation of the ground
state energy by the CW method,
needed for the estimate of the value of the condensate $\pdp$,
differs from the straightforward $\hbar$ expansion
described in Appendix~A by fractional
powers of $\hbar$; at $n$ loops, the difference is of order
$\hbar^{(n+\fr12)}$. In particular, since the effective
potential is known up to two loops, the difference between the two
methods is of order $\hbar^{\fr52}$. This is unacceptable since
the $\hbar^3$-contribution
to the effective potential is to be determined.

Below we show how this disadvantage of the CW method can be removed,
so that the CW method can be used both near and far from the critical
temperature.

The main idea is to redefine the 2-loop potential, including in it
all leading singularities:
\be
V_\rmi{2-loop}^\rmi{improved}=V_\rmi{2-loop}+\sum_{n=3}^{\infty}
V_n^\rmi{sing}.
\ee
In this way,
the expression for the ground state energy becomes
analytic in $\hbar$, as it must be in the broken phase, where there are
no massless physical excitations. It is interesting that the
requirement that the $\hbar^{\fr52}$ correction be absent in the
expression for the ground state energy is powerful enough to determine
the structure of the leading singularities.

Let us rewrite the tree and 1-loop effective potentials in
the following form:
\be
V_0= \frac{1}{4\lambda_3}z^4,\quad
V_1=-\frac{\hbar}{12\pi}[6m_W^3+m_H^3 + 3z^3] -
\frac{1}{2\lambda_3}m_G^2 z^2 + O(z^4).
\ee
Here $m_W=\fr12g_3 \sqrt{(-m_3^2)/\lambda_3}$ and
$m_H^2=-2m_3^2$ are the tree-level values of the W and Higgs masses,
$z=m_2$ and
\be
m_G^2=\frac{3\hbar}{8\pi}g_3^2(m_W+\frac{2\lambda_3}{g_3^2}m_H).
\ee
Then the solution to the minimization equation $\frac{\partial
V_\rmi{eff}}{\partial z}=0$ has the form
\be
z^2=m_G^2 +O(\hbar^{\fr32}).
\ee
Now, if this solution is used in computing
the values of the singular parts of the
effective potential in eq.~(\ref{sing}), one can see that any of
the $n$-loop contributions is of order $\hbar^{\fr52}$. Since
such a term must be absent, we get
\be
\sum_{n=1}^{\infty} V_n^\rmi{sing}(z^2=m_G^2)=0.
\ee
Therefore, the whole sum must have the form
\be
\sum_{n=1}^{\infty}
V_n^\rmi{sing}(z)=-\frac{\hbar}{4\pi}(z^2-m_G^2)^{\fr32}
\label{singsum}
\ee
in order to reproduce correctly the known 1-loop
term $\sim z^3$, which is non-analytic in~$z^2$.

In this derivation the explicit form of the 2-loop potential was not
used, so that the prediction of the term
non-analytic in $\hbar^2$ may be
compared with the direct 2-loop computation. From \cite{fkrs1},
the non-analytic terms on the 2-loop level are
\be
V_2^\rmi{sing}
=\frac{\hbar^2}{(4\pi)^2}\fr94g_3^2(m_W+\frac{2\lambda_3}{g_3^2}m_H)m_2
=\frac{3\hbar}{8\pi}m_G^2 z,
\ee
indeed coinciding with the $\hbar$ expansion of eq.~(\ref{singsum}).

To conclude, the improved form of the effective potential, which
reproduces the correct structure of the $\hbar$ expansion of the ground
state energy, has the form
\be
V_\rmi{2-loop}^\rmi{improved}=V_\rmi{2-loop}(m_2^2\rightarrow m_2^2-m_G^2)
-\frac{3}{8\pi}m_G^2 m_2,
\ee
where the last term must be subtracted in order to avoid
double-counting. It is this potential which was used for the
computation of the scalar condensate at fixed temperature by the CW
method in Sec.~\ref{sec:broken}.

\section{}
\label{oldC}

Here we calculate the W and Higgs correlators in the 3d theory of
eq.~\nr{lagr} at 1-loop order.
The Lagrangian masses of the vector, Higgs and Goldstone fields are
in eq.~\nr{masses}. With these masses,
the radiatively corrected propagators of the
vector and Higgs fields are of the form
\ba
\langle A^a_i(-p)A^b_j(p)\rangle & = &
\delta^{ab}\frac{\delta_{ij}-p_i p_j/p^2}
{p^2+m_T^2-\Pi^W(p^2)}
+ {\rm \quad longitudinal\; part},\nonumber \\
\langle\phi_1(-p)\phi_1(p)\rangle & = & \frac{1}{p^2+m_1^2-\Pi^H(p^2)}.
\la{prop}
\ea
To calculate the self-energies $\Pi^W$, $\Pi^H$, one needs
the basic integrals
\ba
A_0(m^2) & = & \int dp \frac{1}{p^2+m^2}=-\frac{m}{4\pi}, \\
B_0(k^2;m_1^2,m_2^2) & = &
\int dp\frac{1}{[p^2+m_1^2][(p+k)^2+m_2^2]}\nonumber \\
& = &
\frac{i}{8\pi (k^2)^{1/2}}
\ln\frac{m_1+m_2-i(k^2)^{1/2}}{m_1+m_2+i(k^2)^{1/2}}.
\ea
The integration measure here is
\be
\int dp \equiv \int\frac{d^dp}{(2\pi)^d},
\ee
where $d=3-2\epsilon$.

The contributions of the diagrams in \fig\ref{pi3d}.a
to the vector self-energy $\Pi^W$ in eq.~\nr{prop} are
(V is a vector, S a scalar, and
$\eta$ a ghost propagator; $k^2$ is the
Euclidian external momentum)
\ba
\Pi^W_{\rm SS} & = &
-\frac{g_3^2}{8}\biggl\{
B_0(k^2;m_1^2,m_2^2)\Bigl[
k^2+2(m_1^2+m_2^2)+\frac{(m_1^2-m_2^2)^2}{k^2}
\Bigr] \nonumber \\
& & +B_0(k^2;m_2^2,m_2^2)
(k^2+4m_2^2) \nonumber \\
& & +\Bigl[A_0(m_1^2)-A_0(m_2^2)\Bigr]\Bigl(
\frac{m_1^2-m_2^2}{k^2}\Bigr)
-A_0(m_1^2)-3A_0(m_2^2)\biggr\},
\la{piwss}\\
& & \nonumber \\
\Pi^W_{\rm VV} & = &
g_3^2\biggl\{
B_0(k^2;m_T^2,m_T^2)\Bigl[5k^2-4m_T^2+\frac{k^4}{m_T^2}-
\frac{k^6}{8m_T^4}\Bigr] \nonumber \\
& & +B_0(k^2;m_T^2,0)\Bigl[\frac{(k^2+m_T^2)^2}{4m_T^4k^2}
(k^4-6k^2m_T^2+m_T^4)\Bigr] \nonumber \\
& & +B_0(k^2;0,0)\Bigl(-\frac{k^6}{8m_T^4}\Bigr)
+A_0(m_T^2)\Bigl[\frac{5}{12}-\frac{3}{2}\frac{k^2}{m_T^2}
+\frac{(k^2+m_T^2)^2}{4k^2m_T^2}\Bigr]
\biggr\},
\la{piwww}\\
& & \nonumber \\
\Pi^W_{\rm SV} & = &
\frac{g_3^2}{8}\biggl\{
B_0(k^2;m_1^2,m_T^2)\Bigl[
-k^2-2m_1^2+6m_T^2-\frac{(m_1^2-m_T^2)^2}{k^2}
\Bigr] \nonumber \\
& & +B_0(k^2;m_1^2,0)
\frac{(k^2+m_1^2)^2}{k^2}
\nonumber \\
& & +
A_0(m_T^2)\Bigl[1+\frac{m_1^2-m_T^2}{k^2}\Bigr]
+A_0(m_1^2)\frac{m_T^2}{k^2}
\biggr\},
\la{piwsv}\\
& & \nonumber \\
\Pi^W_{\eta\eta} & = &
\frac{g_3^2}{4}k^2B_0(k^2;0,0),
\la{piwee}\\
& & \nonumber \\
\Pi^W_{\rm S} & = &
-\frac{g_3^2}{4}\Bigl[A_0(m_1^2)+3A_0(m_2^2)\Bigr],
\la{piws}\\
& & \nonumber \\
\Pi^W_{\rm V} & = &
-\frac{8}{3}g_3^2A_0(m_T^2).
\la{piwv}
\ea
For the 1-loop contributions
to the Higgs self-energy $\Pi^H$ in eq.~\nr{prop},
one gets from the diagrams in \fig\ref{pi3d}.b the results
\ba
\Pi^H_{\rm SS} & = &
6\lambda_3^2\phi^2
\Bigl[3B_0(k^2;m_1^2,m_1^2)+
B_0(k^2;m_2^2,m_2^2)\Bigr],
\la{pihss}\\
& & \nonumber \\
\Pi^H_{\rm VV} & = &
\frac{3}{8}\frac{g_3^2}{m_T^2}\biggl\{
B_0(k^2;m_T^2,m_T^2)\Bigl[
k^4+4k^2m_T^2+8m_T^4
\Bigr] \nonumber \\
& & -2B_0(k^2;m_T^2,0)
(k^2+m_T^2)^2
\nonumber \\
& & +B_0(k^2;0,0)k^4-2A_0(m_T^2)m_T^2
\biggr\},
\la{pihww}\\
& & \nonumber \\
\Pi^H_{\rm SV} & = &
\frac{3}{4}\frac{g_3^2}{m_T^2}\biggl\{
B_0(k^2;m_T^2,m_2^2)\Bigl[
k^4+2k^2(m_T^2+m_2^2)+(m_T^2-m_2^2)^2
\Bigr] \nonumber \\
& & -B_0(k^2;m_2^2,0)(k^2+m_2^2)^2
\nonumber \\
& & +A_0(m_T^2)\Bigl[m_T^2-m_2^2-k^2\Bigr]
-A_0(m_2^2)m_T^2
\biggr\},
\la{pihsv}\\
& & \nonumber \\
\Pi^H_{\rm S} & = &
-3\lambda_3\Bigl[A_0(m_1^2)+A_0(m_2^2)\Bigr],
\la{pihs}\\
& & \nonumber \\
\Pi^H_{\rm V} & = &
-\frac{3}{2}g_3^2A_0(m_T^2).
\la{pihv}
\ea

\begin{figure}[tb]
\vspace*{1cm}

\hspace*{0.0cm}
\epsfysize=7cm
\epsffile[100 500 500 700]{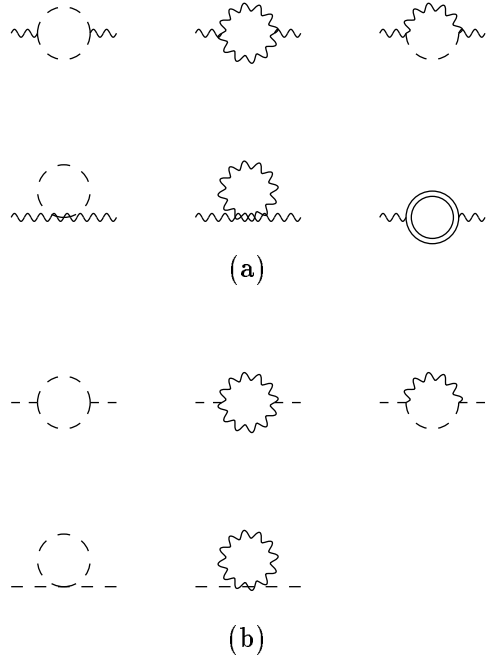}

\vspace*{2cm}
\caption[a]{\protect
The diagrams needed for calculating
(a) the self-energy of the W boson, and
(b) the self-energy of the Higgs particle.
Dashed line is a scalar propagator,
wiggly line a vector propagator, and
double line a ghost propagator.}
\la{pi3d}
\end{figure}

\end{document}